\documentclass[12pt, a4paper]{JHEP3}
\usepackage{multicol,bbm}
\usepackage[dvips]{epsfig, graphicx}
\usepackage{amsmath,amssymb,eucal,mathrsfs}
\usepackage{placeins}

\newcommand{\be}{\begin{equation}}
\newcommand{\ee}{\end{equation}}
\newcommand{\bea}{\begin{eqnarray}}
\newcommand{\eea}{\end{eqnarray}}

\newcommand{\mc}{\mathcal}

\newcommand{\half}{\frac{1}{2}}

 \newlength{\wth}
 \newcommand{\twographshuge}[2]{%
 \unitlength=1.1in
 \begin{picture}(5.8,2.6)
 \put(0,0){\epsfig{file=#1.eps, width=0.9\wth, height=0.5\wth}}
 \put(2.85,0){\epsfig{file=#2.eps, width=0.9\wth, height=0.5\wth}}
 \put(0,2.25){(a)}
 \put(2.65,2.25){(b)}
 \end{picture}
}

\preprint{DAMTP-2009-28\\ CERN-PH-TH/2009-037}

\title{Fitting the Phenomenological MSSM}

\author{Shehu S. AbdusSalam$^1$, Benjamin C. Allanach$^1$, Fernando Quevedo$^{1,2}$ \\
$^1$DAMTP, Centre for Mathematical Sciences, \\ $ $ Wilberforce Road, Cambridge
CB3 0WA, UK.\\ 
$^2$CERN, PH-TH, CH-1211, Geneva 23, Switzerland\\
$ $ E-mails: \email{s.s.abdussalam, b.c.allanach, f.quevedo@damtp.cam.ac.uk}}%,
\author{Farhan Feroz$^3$, Mike Hobson$^3$\\
$^3$Cavendish Laboratory, JJ Thomson Avenue, Cambridge CB3 0HE, UK \\
$ $ E-mails: \email{f.feroz, mph@mrao.cam.ac.uk}}%, \email{mph@mrao.cam.ac.uk}}

\abstract{
We perform a global Bayesian fit of the phenomenological minimal
supersymmetric standard model (pMSSM) to current indirect collider and
dark matter data. The pMSSM contains the most relevant 25 weak-scale
MSSM parameters, which are simultaneously fit using `nested sampling'
Monte Carlo techniques in more than 15 years of CPU time. We calculate
the Bayesian evidence for the pMSSM and constrain its parameters and
observables in the context of  two widely different, but reasonable,
priors to determine which inferences are robust. We make inferences
about sparticle masses, the sign of the $\mu$ parameter, the amount of
fine tuning, dark matter properties and the prospects for direct dark
matter detection without assuming a restrictive high-scale
supersymmetry breaking model. We find the inferred lightest CP-even
Higgs boson mass as an example of an approximately prior independent
observable. This analysis constitutes the first statistically
convergent pMSSM global fit to all current data. 
}

\keywords{Supersymmetry, Dark Matter phenomenology, Supersymmetric
  Standard Model}  

\begin{document}

%SSSSSSSSSSSSSSSSSSSSS
\section{Introduction}  \label{sec.intro}
%SSSSSSSSSSSSSSSSSSSSS

\paragraph{}
There is  currently an expectation that with the start of the Large
Hadron Collider (LHC),
high energy physics will
soon enter a new phase highly dominated by new data that could imply
physics beyond the Standard Model (SM). Over the past years, low energy
supersymmetry (SUSY) has become  the standard approach to study the
potential physics beyond the SM,
mostly because of its natural power to address the hierarchy problem.
The $124$ Lagrangian parameters of the minimal supersymmetric
extension of the SM (for a recent review, see~\cite{hep-ph/0312378}) makes
its phenomenological study impractical. It may well be that the mechanism that
mediates SUSY breaking to the observable sector provides
relations between many of these parameters. Unfortunately, however, there are
many different mediation mechanisms in the literature, with no one clearly
preferred. 

\paragraph{}
In order to extract computable information, many  works have reduced
the number of parameters by truncating to a handful of soft-breaking parameters
at a high energy scale.
The remaining set of parameters
are used as boundary conditions for renormalisation
group equations (RGE), which are run down to the weak-scale.
A large amount of minimal supersymmetric standard model (MSSM) based studies
have been 
carried out in the way described above. Most of them 
 were performed in the context of the constrained MSSM (CMSSM, also sometimes
 called mSUGRA for minimal supergravity) set-up which
 have only 
$4$ independent non-SM parameters (and a sign choice). Many groups have
 been pursuing a programme  
to fit this model and identify regions in parameter space that might
be of interest with the forthcoming LHC data. See for instance
\cite{Arnowitt:1993mg,Ellis:2003si,Profumo:2004at,Baltz:2004aw,Ellis:2004tc,Stark:2005mp}. 
Complete scans over up to $8$ free parameters of CMSSM
with a combined treatment of likelihoods from different experimental
constraints were possible with Markov Chain Monte Carlo (MCMC)
sampling techniques~\cite{Baltz:2004aw, Allanach:2005kz, hep-ph/0602028,
hep-ph/0609295, Allanach:2006jc,0705.2012, 0705.0487,
Buchmueller:2008qe}. However, 
the truncation to a handful of parameters in 
the CMSSM, is at best a very strong assumption, and most likely
over-restrictive. 

\paragraph{}
There are two directions that can be followed to properly study low-energy
supersymmetric models. The top-down approach has been tried over the years by deriving the
otherwise free parameters from an ultraviolet extension of the MSSM\@.
Models of unification, different sources of SUSY breaking,
such as gravity and gauge mediation and classes of string
compactifications
\cite{Phys.Lett.B306.269,Brignole:1997dp,Allanach:2005pv,Choi:2005ge,AbdusSalam:2007pm,Acharya:2008zi,Aparicio:2008wh,Krippendorf:2009zz,Allanach:2008tu} 
have been used to provide high energy expressions for the soft
breaking parameters. 
 Soft SUSY breaking terms can be computed at  energies as high 
as the GUT scale of $\sim 10^{16}$ GeV, and renormalisation group running
to the TeV scale allows contact to be made with potential quantities of
interest such as 
sparticle masses. Recent progress in moduli stabilisation in string theory 
has made
this approach more concrete and calculable with explicit results for some
classes 
of models. This is very encouraging but usually the string derived models
fill only a small subset of the full CMSSM parameter space, which could make
them impossible to differentiate from the CMSSM\@.
The proliferation of SUSY breaking mechanism set-ups mean that analyses where
only one is picked tend to be very specific, with a rather limited range of
applicability. It is desirable to side-step such extreme model dependence with
a different approach.

\paragraph{}
Alternatively, one can use a bottom-up approach to low-energy
SUSY. In 
this case the soft breaking parameters are considered at the SUSY
scale without referring to their high
energy origin. All the parameters can in principle be varied over the
experimentally allowed range and compared with potential observations at
the LHC and other experiments. It is a formidable task to consider all the
124 parameters, due mostly to computing limitations. An interesting
compromise is 
the pMSSM~\cite{Djouadi:1998di}  which we consider here. 
In this model, the number of free parameters is
20 soft-breaking parameters (and a $\pm 1$ parameter) plus 5 SM ones
and are selected by the 
requirements of consistency with unobserved flavour changing and CP
violating processes. However, even this simplified version of the MSSM
requires a lot of computer power to be analysed in 
and details. The ability of 
future collider measurements to constrain the pMSSM has been estimated using
MCMC methods in Refs.~\cite{Baltz:2006fm,Lafaye:2007vs}.%Lafaye:2004cn}.

\paragraph{}
At present, SUSY forecasts for the LHC necessarily contain large
uncertainties. In particular, there is a strong model dependence on
the mechanism for SUSY breaking. Realistic predictions need guidance
from direct and precise (collider and other related) experimental
data. Interestingly, the converse is also very important: the
experiments need unbiased phenomenological guidance about the
expected nature or properties of SUSY. This is what we aim to
accomplish, eventually.

\paragraph{}
The purpose of this article is to perform a global fit of the pMSSM
and make SUSY forecasts for collider and dark matter search
experiments using  Bayesian statistics methods. For given 
prior probability and likelihood densities, Bayes theorem provides the way to
extract the 
posterior probability density for the parameters. It can also be used for model
comparison when enough data is available. This formalism has been
used in many fields of science, including cosmology (see
\cite{0803.4089,0903.4210} for recent reviews).
Markov Chain Monte Carlo
(MCMC) and related techniques have recently begun to be used to perform
Bayesian inference on supersymmetric models. The increasing access to large
scale 
computing power and improved methods of calculation are making these
techniques more
manageable with time and we have been able to tackle the relevant
parameters of the pMSSM\@. 

\paragraph{}
The complete and simultaneous scan of the 25
parameters and a sign\footnote{Where we refer to 25 parameters, 
we shall really mean 25 continuously varying parameters plus the one discrete
sign choice.} for the pMSSM construction was performed
using the {\sc MultiNest} program~\cite{Feroz:2007kg,
  Feroz:2008xx}. At the heart of the algorithm is the 
nested sampling 
technique~\cite{Skilling} that revolutionised computational Bayesian
inference by prioritising a computation of the Bayesian evidence rather than
solely on
computing the posterior probability distribution function (PDF) of model
parameters (although 
the latter is obtained at no additional cost), as is the case in traditional
Monte Carlo algorithms (e.g.\ MCMC). We will review the nested sampling method
in an appendix.  
In simple terms, similar to the MCMC, nested sampling is an iterative Monte
Carlo method that, starting with a relatively small number of points 
(a few thousand in our case), it produces a list of a large number of points
($10^7-10^8$ in our case) ordered in increasing likelihood. 
We used it here because it computes both the evidence {\em and}\/ the
posterior PDF, as opposed to traditional MCMC methods which only
calculate the posterior PDFs. Further the {\sc MultiNest} algorithm is
efficient in handling complicated problems with multimodal/degenerate
posterior distributions. 

\paragraph{}
We emphasise that the current situation with no direct sparticle
measurement data yet from LHC makes the issue of prior dependence
critical. For this reason, it is expected that extraction of prior
independent information of our analysis will be
difficult. Interestingly, however, we find some results with
approximate prior independence. In order to illustrate the issue of
prior dependence of results we consider priors that are flat in the
parameters themselves (`linear') and flat in the logarithm of the
parameters (`log') priors. As usual in Bayesian statistics prior
dependence should not be understood as a drawback but as a positive
feature that can be used to determine when enough data is available to
unambiguously make inferences. It is expected that, if SUSY is
discovered, the addition of LHC sparticle mass data will relax any
prior dependence, and so an analysis along the lines of ours could be
used to extract prior independent information.

\paragraph{}
The amount of work related to this project required very efficient
algorithms and access to high performance computing. We used the
University of Cambridge supercomputers: COSMOS from the Department of
Applied Mathematics and Theoretical Physics (DAMTP)  and the {\it 
Darwin}\/ cluster from the High  Performance Computing Service      
(HPC). The final run was made in terms of 60 12-hour jobs, each      
corresponding to a cluster of 128 CPUs on HPC and 40 8-hours jobs,   
each corresponding to a cluster of 64 CPUs, on COSMOS (making a total
of more than 15-year standard CPU time). Some results of a complete
and independent (from the one we present here) runs with fewer
experimental constraints were presented by one of us (SSA) at the
SUSY 2008 conference and reported in Ref.~\cite{AbdusSalam:2008uv}.
While we were upgrading our analysis, a study of randomly scanned
pMSSM points appeared~\cite{Berger:2008cq}, similar in philosophy 
to~Ref.~\cite{Profumo:2004at}. In the conclusions, we contrast
the aims and methodologies of our work with these.

\paragraph{}
In order to make a self-contained presentation we briefly describe
Bayesian inference and relevant terminologies in
Section~\ref{sec.inference}. We construct the elements needed for 
inference in the context of the pMSSM in
Section~\ref{sec.phenomssm}. The experimental constraints or
observables used are described in Subsection~\ref{sub.phenodata}. The
sampling procedure and the different high energy physics software used
to predict the observables are presented in
Subsection~\ref{sub.phenoprocedure}. In Sections~\ref{sec.results}
and~\ref{sec.dm} we analyse our
results and then conclude. In the Appendix we briefly
review the nested sampling method and how the {\sc MultiNest} program
works.

%SSSSSSSSSSSSSSSSSSSSSSSSSSS
\section{Bayesian Inference} \label{sec.inference}
%SSSSSSSSSSSSSSSSSSSSSSSSSSS

\paragraph{}
Bayesian inference fits and plays an important role in the scientific
process of data collection and modelling. It particularly deals with
the steps that involve model fitting to data and the technique of
assigning preferences to alternative models (model comparison). This
subject is very important especially with the imminent start of the
LHC experiments. Here we will
give a short review of the basics of Bayesian statistics that are
useful in our work. 

%%%%%%%%%%%%%%%%%%%%%%%%%%%
\subsection{Bayes' theorem} \label{sub.bayes}
%%%%%%%%%%%%%%%%%%%%%%%%%%%

\paragraph{}
Consider a given model or hypothesis $H$ (we shall take $H$ to be the pMSSM)
defined 
by some set of parameters (in our case, 25 parameters) $\mathbf{\Theta}$.
We wish to know the PDF
$P(\mathbf{\Theta}| \mathbf{D}, H)$ of the parameters $\mathbf{\Theta}$ given
the data $\mathbf{D}$ and the model set-up, $H$. $P(\mathbf{\Theta}|
\mathbf{D}, H)$, being the parameters' PDF {\em after}\/ confrontation with
data, is called 
the {\em posterior}\/ PDF.
The likelihood,
$P(\mathbf{D}|\mathbf{\Theta}, H) \equiv
{L}(\mathbf{\Theta})$ is a measure of how well a model point
$\mathbf{\Theta}$ predicts data set $\mathbf{D}$.
In order to calculate the posterior from the likelihood, 
one must assign some {\em prior}\/
PDF to the parameters $P(\mathbf{\Theta} | H) \equiv
\pi(\mathbf{\Theta})$ to parametrise our uncertainty in them before the model
is confronted with data. 
Bayes' theorem then describes how one may obtain the posterior from the other
two PDFs and 
a normalisation constant $P(\mathbf{D}|H)
\equiv {Z}$, the {\em Bayesian evidence}\/ for the model in light of the
data:
\be \label{posterior} P(\mathbf{\Theta}|\mathbf{D}, H) =
\frac{P(\mathbf{D}|\,\mathbf{\Theta},H)P(\mathbf{\Theta}|H)}
{P(\mathbf{D}|H)}\equiv \frac{{L}(\mathbf{\Theta})\, \pi(\mathbf{\Theta})}{{Z}}. \ee 

The Bayesian evidence is given by
\be \label{eq:3}
{Z} =
\int{{L}(\mathbf{\Theta})\pi(\mathbf{\Theta})}d^N\mathbf{\Theta}.
\ee
Here $N$ is the dimensionality of the parameter space; $N=25$ for the
pMSSM. Since the Bayesian evidence does not
depend on the parameter values $\mathbf{\Theta}$, it is usually
ignored in parameter estimation problems and  posterior inferences
are obtained by exploring the un--normalised posterior using standard
MCMC sampling methods. However, the evidence plays a central role in our
discussion. 

\paragraph{}
A useful feature of Bayesian parameter estimation is that one can
easily obtain the posterior PDF of any function, $f$, of the
model parameters $\mathbf{\Theta}$. Since, 
\bea \label{eq:derived_post}
P(f|\mathbf{D}) = \int{P(f,\mathbf{\Theta} | \mathbf{D}) d
  \mathbf{\Theta}} &=& 
\int{P(f|\mathbf{\Theta},\mathbf{D}) P(\mathbf{\Theta} | \mathbf{D})
  d \mathbf{\Theta}} \nonumber \\\ 
&=& \int{\delta(f(\mathbf{\Theta})-f) P(\mathbf{\Theta}|\mathbf{D}) d
  \mathbf{\Theta}} 
\eea
where the probability chain rule is employed for the second equality
and $\delta$ is the Dirac delta function. Thus one simply needs to
compute $f(\mathbf{\Theta})$ for every Monte Carlo sample and the
resulting sample will be drawn from $P(f|\mathbf{D})$. We make use
of this feature in Sections~\ref{sec.results} and \ref{sec.dm} where
we present the posterior probability PDFs of various observables used
in the analysis of the pMSSM\@.

\paragraph{}
Before proceeding to discuss prior distributions, we first briefly 
address~\footnote{We follow the description in:\protect\par
\texttt{http://www.dsg.port.ac.uk/~valiviitaj/Lectures2006/CrittendenCMB2003.pdf}},  
the difference between the Bayesian and the frequentist approaches to 
inference. Bayesian inference (using Bayes' theorem) is a robust technique for
updating prior knowledge or belief based on new data. It is unlike the
frequentist approach where observations are viewed as random draws
from some pool of possible observations such that the probability of
an observation is the frequency observed with a large number of repeated
measurements. Frequentists usually focus on the likelihood and
argue that the Bayesian approach is too subjective because of the use
of priors. Bayesians reply that when many examine frequentist
statistics, they are actually implicitly using
priors anyway. If one thinks of a region of parameter space with the lowest
chi-squared values as being more likely than a region of parameter space with 
much higher chi-squared values, one is implicitly using some vague prior.
The two approaches are asking different
questions. Bayesians ask, ``how likely is a given parameter value
given the data?'', while frequentists ask, ``how probable is the data,
given certain parameters?'' In situations where the data is very
informative the two approaches give the same results. 
We are interested in PDFs of parameters and so we use Bayesian statistics.
By comparing results from different but reasonable priors, we obtain an
estimate of how robust an inference is given current data. 

%%%%%%%%%%%%%%%%%%%
\subsection{Priors} \label{sub.priors}
%%%%%%%%%%%%%%%%%%%

\paragraph{}
The prior probability of the model parameters 
is a
PDF that gives a subjective measure of our
initial knowledge/ignorance about the values of the parameters of the model
before the data are taken. 
Symmetries and physical observations or expectations are usually a good guide
to which priors to take.
Information theory also provides a way of selecting priors by favouring
those that maximise the entropy of the distributions. 
Some commonly used prior PDFs $P(\mathbf{\Theta} | H)$
are:
\bea \label{eg.priors}
P(\mathbf{\Theta} | H) & \propto & constant \textrm{ -- the linear
  prior, flat in }\mathbf{\Theta} \\
P(\mathbf{\Theta} | H) & \propto & \frac{1}{\mathbf{\Theta}} \textrm{
  -- the Jeffreys prior, flat in } \log(\mathbf{\Theta}) \\
P(\mathbf{\Theta} | H) & \propto & e^{-(\mathbf{\Theta} -
  \overline{\mathbf{\Theta}})^2/2\sigma^2} \textrm{  -- the Gaussian prior.}
\eea
The linear priors are often used for translational (like time and
location) parameters, where there is no information to suggest that one
value is preferred over others. The Jeffreys prior, also referred to as
logarithmic ({\em log}\/)
prior, is uniform in the logarithm of the parameter. These two priors are
improper since they diverge when integrated over an infinite
range. Our log and linear priors will be bounded by the requirements
of perturbativity of the 
model, by passing previous direct sparticle search constraints and by the
requirement of not too large fine-tuning in the Higgs potential
parameters. These three criteria are sufficient to bound all 20 non-SM input
parameters to a finite range. 
The Gaussian
prior, on the other hand, is proper and integrable but
requires previous experimental knowledge on $\sigma$ and $\mathbf{\bar
  \Theta}$. This is indeed the case for our 5 SM input parameters, and we
use Gaussian priors for them (see Section~\ref{sub.pars}).

\paragraph{}
Assuming the parameters are independent,
the resultant prior is obtained by the
product of all the prior probability densities for each of the
individual parameters. For instance, in the case of pMSSM with $25$
parameters, $\theta_1, \theta_2, \dots, \theta_{25}$, 
\be \label{pro}
P(\mathbf{\Theta} | H) \equiv \pi(\mathbf{\Theta}) = \prod_{i=1}^{25}
\pi(\theta_i). 
\ee  
For our construction and analysis we are going to choose a linear prior
measure for the pMSSM parameters described in Section ~\ref{sub.pars}.
This is because there are no observational evidence that hint to giving
preference of some parameter region over others. We are going to
determine bounds on the parameters from the fact that the parameter
values have to be not far away from the TeV-scale in order avoid
the little hierarchy problem. We are going to call such case/scenario
as the {\it linear}\/ prior. We then check the dependence of the
results in our analysis on prior change by performing another analysis
with a log prior. 
When we refer to a log prior, in fact the Jeffreys prior is only used for all
parameters which only have positive bounds. Parameters which are allowed to
take either sign present a problem with Jeffreys priors since the prior
diverges at the origin. Therefore, the priors of such parameters are always
taken to be linear. 

\paragraph{}
For parameter estimation, the priors become irrelevant once the
data employed are powerful enough. This is already in evidence by comparing
Bayesian CMSSM fits~\cite{0807.4512} with those in similar supersymmetric
models which have a 
lower number of free parameters, for example the large volume string
compactification (LVS) scenario~\cite{Allanach:2008tu}. The LVS scenario has two less
free parameters than the CMSSM and current indirect data are already enough to
make the result approximately prior independent. We may expect the addition of
two precise, constraining, non-degenerate measurements (such as sparticle mass
measurements from the LHC) to have the same effect upon the CMSSM\@.

\paragraph{}

For model comparison, the dependence
on priors always remains (although with more informative data the
degree of dependence on the priors is expected to decrease, see
e.g.~\cite{0803.4089}). Indeed this explicit dependence on priors is
one of the most attractive features of Bayesian model
selection. 
Refs.~\cite{0705.0487,Cabrera:2008tj} identified prior distributions in
high-scale CMSSM Lagrangian parameters. In particular, a Jacobian was defined
to transform between derived parameters (such as $\tan \beta$) and more
fundamental Lagrangian parameters from which they are derived. It is not our
purpose here to find the `most natural' prior because any such choice is
necessarily subjective. Instead, we shall
check the robustness of any inference
under a reasonable variation of the priors. Such a check
is especially required in 
model comparison hypothesis tests, which may be particularly sensitive to 
the particular choice of prior and its associated
metric in parameter space~\cite{Cousins:2008gf}.

%%%%%%%%%%%%%%%%%%%%%%%%%%%%%
\subsection{Model comparison} \label{sub.comparison}
%%%%%%%%%%%%%%%%%%%%%%%%%%%%%
\paragraph{} 
In order to evaluate and rank two alternative models $H_{0}$ and
$H_{1}$ in the light of data $\mathbf{D}$ one needs to compare their
respective posterior probabilities given the observed data set
$\mathbf{D}$, as follows\footnote{Here we follow the description
  by~\cite{Mackay}.}: use Bayes' theorem to relate the 
plausibility of $H_1$ given the data, $P(H_1|\mathbf{D})$, to the
predictions made by the model about the data, $P(\mathbf{D}|H_1)$,
and the prior plausibility of $H_1$, $P(H_1)$. With this procedure
one could construct the following probability ratio
\be \label{eq:3.1}
\frac{P(H_{1}|\mathbf{D})}{P(H_{0}|\mathbf{D})}
=\frac{P(\mathbf{D}|H_{1})P(H_{1})}{P(\mathbf{D}|
H_{0})P(H_{0})}
=\frac{{Z}_1}{{Z}_0}\frac{P(H_{1})}{P(H_{0})}.
\ee 
$P(H_{1})/P(H_{0})$ is the prior probability ratio for the two
models, it measures how much our initial beliefs favour $H_1$ over
$H_2$. It is often set to unity but may occasionally require further
consideration. The other ratio, $P(\mathbf{D}|H_{1})/P(\mathbf{D}|
H_{0})$ measure how well the observed data were predicted by $H_1$
and $H_0$. It can be seen from~\ref{eq:3.1} that the Bayesian evidence
takes the centre stage in Bayesian model comparison. As the average of
likelihood over the prior, the Bayesian evidence is higher for a model
if more of its parameter space is likely over some region of significant 
integrated prior (`prior mass')
and smaller for a model with
highly peaked likelihood but has a large prior mass
associated with low likelihood values. Hence, Bayesian model selection
automatically implements Occam's razor, since the prior PDF decreases for 
models with higher numbers of parameters.
A more
complicated theory will only have a higher evidence if it is
significantly better at explaining the data than a simpler
theory with less parameters. This technique was applied in~\cite{0807.4512} to
compare two CMSSM models: $sign(\mu) > 0$ versus  $sign(\mu) < 0$. We
shall perform a similar comparison for the pMSSM in
Subsection~\ref{sub.signmucomp}. The comparison of different GUT scale
SUSY breaking models is also interesting~\cite{AbdusSalam:2009tr}.

\paragraph{}
Another perspective on model comparison is in
quantifying the consistency between two or more data sets or
constraints~\cite{astro-ph/0203259,Phys.Rev.D73.067302}. Different experimental
constraints may `pull' the model parameters to different directions
and consequently favour different regions of the parameter space. Any
obvious conflicts between the observables are likely to be noticed by
the `chi by eye' method commonly employed to-date, but it is imperative for
forthcoming high--quality data to have a method that can
quantify such discrepancies. The simplest method for analysing
different constraints on a particular model is to assume that all
constraints and data provide information on the same set of parameter
values. This can be considered as one hypothesis or model, $H_1$. This
is the assumption which underlies the joint analysis of the
constraints. However, if we are interested in accuracy as 
well as precision then any systematic differences between
constraints or data should also be taken into account. In the most
extreme case they could be in conflict to the extent
that each of them would require its own set of parameter
values, since they are in different regions of parameter space. This
scenario can be considered as another hypothesis or model, $H_0$, to be
compared with the other, $H_1$. Bayesian inference provides a
very easy method of distinguishing between the scenarios $H_0$ and
$H_1$. This technique was illustrated in Refs.~\cite{0807.4512,Feroz:2009dv} by
examining the 
compatibility between $B \rightarrow X_s \gamma$ branching ratios and the
anomalous magnetic moment of the muon in the CMSSM\@.
\paragraph{} 
The natural logarithm of the ratio of posterior model probabilities
provides a useful guide to what constitutes a significant difference
between two models: 
\be \label{eq:Jeffreys}
\log \Delta E = \log \left[ \frac{P(H_{1}|\mathbf{D})}{P(H_{0}|\mathbf{D})}\right]
=\log \left[ \frac{{Z}_1}{{Z}_0}\frac{P(H_{1})}{P(H_{0})}\right].
\ee In Tab.~\ref{tab:Jeffreys} we summarise the conventions we use in
this paper to separate between  different levels of evidence.

\TABULAR{|l|l|l|l|}{\hline
$|\log \Delta E|$ & Odds, $Z_1 / Z_0$ & Probability & Remark \\ \hline
$<1.0$ & $\lesssim 3:1$ & $<0.750$ & Inconclusive \\
$1.0$ & $\sim 3:1$ & $0.750$ & Weak Evidence \\
$2.5$ & $\sim 12:1$ & $0.923$ & Moderate Evidence \\
$5.0$ & $\sim 150:1$ & $0.993$ & Strong Evidence \\ \hline
}{\it The scale we use for the interpretation of model
  probabilities. Here the $\log$ represents the natural
  logarithm. \label{tab:Jeffreys}} 
\paragraph{} 
  The evaluation of the multi-dimensional
integral~\eqref{eq:3} is a challenging numerical task. Standard
techniques like thermodynamic integration~\cite{O'Ruanaidh} are
extremely computationally expensive which makes evidence evaluation
typically at least an order of magnitude more costly than parameter
estimation. Some fast approximate methods have been used for evidence
evaluation, such as treating the posterior as a multivariate Gaussian
centred at its peak (see e.g.~\cite{astro-ph/0203259}), but this approximation
is clearly a poor one for multi--modal posteriors (except perhaps if
one performs a separate Gaussian approximation at each mode). The
Savage--Dickey density ratio has also been
proposed~\cite{astro-ph/0504022} as an exact, and potentially faster,
means of evaluating evidences, but is restricted to the special case
of nested hypotheses and a separable prior on the model
parameters. 
Bridge sampling~\cite{Jnl.Comp.Phys.22245,Stat.Sci.13163,r.m.neal05} 
allows the evaluation of
the ratio of Bayesian evidence of two models and was implemented in fits to
the CMSSM in Ref.~\cite{hep-ph/0609295}. 
Bank sampling~\cite{Allanach:2007qj} (a particular implementation of
the MCMC technique)
also allows evidence ratios to be calculated with half the number of points
than required for bridge sampling, and was used in CMSSM fits in
Ref.~\cite{0705.0487}. 
It is not yet clear
how accurately bank sampling can calculate these evidence
ratios, and both bridge and bank sampling have the disadvantages of only being
able to calculate evidence ratios, not the evidence itself. When comparing
$N$ models, this introduces an inefficiency of a factor of $2(N-1)/N$ compared
to the latter case.
Various alternative information criteria for model selection
are discussed by~\cite{0903.4210}, but the evidence remains the preferred
method.  
\paragraph{} 
The nested sampling approach, introduced by Skilling~\cite{Skilling},
is a Monte Carlo method targeted at the efficient calculation of the
evidence. It also produces posterior inferences as a
by--product. Feroz \& Hobson~\cite{Feroz:2007kg,Feroz:2008xx} built on
this nested sampling framework and have recently introduced
the~{\sc MultiNest} algorithm which is efficient in sampling from
multi--modal posteriors exhibiting curving degeneracies, producing
posterior samples and calculating the evidence value and its
uncertainty. This technique has greatly reduced the computational cost
of model selection and the exploration of highly degenerate
multi--modal posterior PDFs. We employ this technique in this
paper.

%SSSSSSSSSSSSSSSSSS
\section{The pMSSM} \label{sec.phenomssm}
%SSSSSSSSSSSSSSSSSS

\paragraph{}
The MSSM Lagrangian has the form 
${\mc L} = {\mc L}_{\rm SUSY} + {\mc L}_{\rm soft}$
where ${\mc L}_{\rm SUSY}$ contains all of the kinetic terms, gauge
and Yukawa interactions while preserving SUSY
invariance. It is based on the gauge group $G = SU(3)_c \times SU(2)_L
\times U(1)_Y$ and superpotential, $W$, constructed with a
particle content in the following chiral superfields shown with their
corresponding $G$ charges:
\bea \label{fields}
L:&(1,2,-\half),\quad {\bar E}:&(1,1,1),\qquad\,
Q:\,(3,2,\frac{1}{6}),\quad
{\bar U}:\,(3,1,\frac{2}{3}),\nonumber\\ 
{\bar D}:&(3,1,-\frac{1}{3}),\quad
H_1:&(1,2,-\half),\quad  H_2:\,(1,2,\half).
\eea
The superpotential is given by
\be \label{superpot}
W = \epsilon_{ab} \left[ (Y_E)_{ij} L_i^b
H_1^a {\bar E}_j + (Y_D)_{ij} Q_i^{bx} H_1^a {\bar D}_{jx} +
(Y_U)_{ij} Q_i^{ax} H_2^b {\bar U}_{jx}  + \mu H_2^a H_1^b\right].
\ee Here we use the convention in
\cite{Allanach:2001kg} and denote the $SU(3)$ colour index of the
fundamental representation by $x, y = 1, 2, 3$; the $SU(2)_L$
fundamental representation indices by $a, b = 1, 2$ and the
generation indices by $i, j = 1, 2, 3.$ $\epsilon_{ab}=\epsilon^{ab}$
is the totally antisymmetric tensor, with $\epsilon_{12}=1$.

\paragraph{}
The soft SUSY-breaking part of the Lagrangian 
consists of different mass and coupling terms: 
\be \label{lsoft}
{\mc L}_{\rm soft} = {\mc L}_{\rm gauginos} + {\mc L}_{\rm sfermions} + {\mc
  L}_{\rm trilinear} + {\mc L}_{\rm higgs}
\ee
where the part including the SUSY breaking sfermion masses is   
\bea 
-{\cal L}_{\rm sfermion} &=&
{\tilde{Q}^*}_{ixa} (m_{\tilde Q}^2)_{ij} \tilde{Q}_j^{xa} +
{\tilde{L}^*}_{ia} (m_{\tilde L}^2)_{ij} \tilde{L}_j^{a}  + \nonumber
\\ &&
 \tilde{u}_i^{x} (m_{\tilde u}^2)_{ij}  {\tilde{u}^*}_{jx} +
\tilde{d}_i^{x} (m_{\tilde d}^2)_{ij}  {\tilde{d}^*}_{jx} +
\tilde{e}_i (m_{\tilde e}^2)_{ij} {\tilde{e}^*}_{j}. \label{msq}
\eea
Each mass parameter in Eq.~\ref{msq} is a
$3\times 3$ hermitian matrix in generation space.  
\be -{\cal L}_{\rm higgs} =
m_{H_1}^2 {{H_1}_a}^* {H_1^a} + m_{H_2}^2 {{H_2}_a}^* {H_2^a}
+ \epsilon_{ab} (m_3^2 H_2^a H_1^b + H.c.)
\ee gives the SUSY breaking higgs masses and bi-linear coupling terms. 
The SUSY breaking scalar trilinear couplings are
\be -{\cal L}_{\rm trilinear} = \epsilon_{ab}
\left[ \tilde{Q}_{i_L}^{xa} (U_A)_{ij}  \tilde{u}_{jx_R}^* H_2^b +
\tilde{Q}_{i_L}^{xb} (D_A)_{ij}  \tilde{d}_{jx_R}^* H_1^a +
\tilde{L}_{i_L}^{b} (E_A)_{ij}  \tilde{e}_{j_R}^* H_1^a \right] +H.c.,
\ee
where fields with a tilde represent the scalar components of the
corresponding capital letter superfield and the soft
SUSY-breaking 
A-terms, each a complex $3\times 3$ matrix in generation space, are
defined (no summation on $i,j$ is inferred) as
\be (A_{U,D,E})_{ij} = (U_A,D_A,E_A)_{ij} / (Y_{U,D,E})_{ij}.\ee 
Finally, writing the bino as ${\tilde b}$, ${\tilde w}^{A=1,2,3}$ as the
unbroken-$SU(2)_L$ gauginos and ${\tilde g}^{X=1,\ldots,8}$ as the
gluinos, then the gaugino-mass part of the Lagrangian is 
\be -{\cal L}_{\rm gaugino} = \half \left[M_1 {\tilde b} {\tilde b} +M_2
\sum_{A=1}^{3} {\tilde w}^A {\tilde w}^A + M_3 \sum_{X=1}^{8} {\tilde
  g}^X {\tilde g}^X +H.c. \right].
\ee The parameters together make a total of $105$ free 
parameters in $\mc L_{soft}$, before re-phasing and 
higgs potential minimisation \cite{Nucl.Phys.B452.496,Nucl.Phys.Proc.Suppl.62.469,Djouadi:1998di}. 
In the CMSSM, the SUSY breaking scalar masses 
the gaugino masses and trilinear couplings are
collapsed to the flavour independent parameters
$m_0$, $M_{1/2}$ and $A_0$ respectively, at
grand unification scales $M_{GUT} \sim 2 \times 10^{16}$ GeV. $m_3^2$
and $|\mu|$ are related to the $Z-$boson mass $m_Z$ through
higgs potential minimisation conditions. 
$sign(\mu) = \pm$ and
$\tan \beta$, the ratio of the Higgs vacuum expectation values (vevs)
remain as free parameters. 
However in this paper we instead explore the 
parameters at the weak-scale in its
phenomenologically most relevant directions and following
Ref.~\cite{Djouadi:2002ze} call the set-up 
the pMSSM\@. In Subsection~\ref{sub.pars} we describes how the $25$
parameters in the pMSSM set-up are derived from the much larger
parameter space of the parent MSSM\@.  

%%%%%%%%%%%%%%%%%%%%%%%
\subsection{Parameters} \label{sub.pars}
%%%%%%%%%%%%%%%%%%%%%%% 

\paragraph{}
Sources of CP-violation in the MSSM are tightly constrained by
experimental 
limits on the electron and neutron electric dipole moments and from
results on K-meson system experiments. Assuming that the MSSM soft
SUSY-breaking parameters are real is consistent with such tight
bounds, indeed significant departures from this assumption usually
require a specific cancellation or suppression mechanism in order to pass the
constraints. To suppress flavour  
changing neutral current (FCNC) processes, all 
off-diagonal elements in the sfermions masses and trilinear couplings
are set to zero and the first and second generation soft terms are
set to be equal. $A_t$, $A_b$, $A_\tau$ may all change the likelihood
significantly, and we also include $A_e = A_\mu$
because it is relevant for the computation of the anomalous magnetic moment of
the muon
\cite{Martin:2001st}. We set $A_u = A_c = A_d = A_s = 0$ since these 
are proportional to the SM Yukawa couplings which are very tiny and so they
will have negligible effect  on the likelihood.
All the other trilinear couplings are set to zero.
Our Higgs-sector parameters are specified by $(m^2_{H_1},
m^2_{H_2})$, and as discussed above, we must add $\tan \beta$ and $sign(\mu)$
to the list of parameters.

\paragraph{}
All of the parameters mentioned so far are purely non-SM\@. However,
some of the SM parameters 
significantly affect the likelihood. 
The relevant SM parameters include the electromagnetic   
coupling constant $\alpha_{em}(m_Z)^{\overline{MS}}$ and
the strong coupling constant
$\alpha_{s}(m_Z)^{\overline{MS}}$. The values of these two couplings are taken
at the $Z$-boson pole mass $m_Z$ energy scale evaluated in the
$\overline{MS}$ renormalisation scheme.  
The tau lepton mass and $G_F$, the Fermi constant have been so
precisely determined that their uncertainty has negligible error on the
likelihood 
and so they are fixed at their global average values:
$m_\tau = 1.777$ GeV
and $G_F = 1.16637 \times 10^{-5}$ GeV$^{-2}$\cite{J.Phys.G33.1}. The top
and bottom quark masses are not as precisely 
known and can have significant effects on predictions of supersymmetric
models. They are therefore included as parameters with, using the experimental
measurements of their central values and uncertainties, Gaussian priors.
Despite the fact that the $Z$-boson mass, $m_Z$, is precisely
determined, we include its uncertainty
because one of the observables used for the analysis (the total decay
width of the $Z$-boson $\Gamma_Z$) is proportional to $m_Z^3$ and the
pMSSM predicted values can fall outside the expected
experimentally determined range with a sigma variation in $m_Z$. So
adding the SM parameters, \be \theta_{SM} = \{m_Z, m_t,
m_b(m_b)^{\overline{MS}}, \alpha_{em}(m_Z)^{\overline{MS}},
\alpha_{s}(m_Z)^{\overline{MS}}\} \ee makes a total of $25$ continuously
varying parameters
in the pMSSM\@. These are listed together with their ranges or Gaussian prior
distributions in
Tab.~\ref{tab.param}. These pMSSM directions make up its
parameter-space. Our aim is to
eventually construct a detailed map of the parameter-space that could
be of help to or guide for collider and other SUSY-related
experiments. In Subsection~\ref{sub.consts} we briefly describe the
observables considered and summarise the experimental constraints
coming from each. 
\begin{table}[htbp!] 
\linespread{1.5}
\begin{center}{\begin{tabular}{|cll|}
\hline
Parameter & Description & Prior range\\
\hline
$M_1$               & Bino mass      & [$-4$ TeV, $4$ TeV]\\
$M_2$               & Wino mass        & [$-4$ TeV, $4$ TeV]\\
$M_3$               & Gluino mass       & [$-4$ TeV, $4$ TeV]\\
$m_{\tilde e_L} = m_{\tilde \mu_L}$    & 1st/2nd generation $L_L$
slepton & [$100$ GeV, $4$ TeV]\\  
$m_{\tilde \tau_L}$  & 3rd generation $L_L$ slepton & [$100$ GeV,
  $4$ TeV]\\   
$m_{\tilde e_R} = m_{\tilde \mu_R}$    & 1st/2nd generation $E_R$
    slepton & [$100$ GeV, $4$ TeV]\\  
$m_{\tilde \tau_R}$  & 3rd generation $E_R$ slepton & [$100$ GeV,
      $4$ TeV]\\   
$m_{\tilde u_L} = m_{\tilde d_L} = m_{\tilde c_L} = m_{\tilde s_L}$ &
    1st/2nd generation $Q_L$ squark   & [$100$ GeV, $4$
    TeV]\\   
$m_{\tilde t_L} = m_{\tilde b_L}$ & 3rd generation $Q_L$ squark
    & [$100$ GeV, $4$ TeV]\\
$m_{\tilde u_R} = m_{\tilde c_R}$ & 1st/2nd generation $U_R$ squark
    & [$100$ GeV, $4$ TeV]\\  
$m_{\tilde t_R}$ & 3rd generation $U_R$ squark   &
    [$100$ GeV, $4$ TeV]\\  
$m_{\tilde d_R} = m_{\tilde s_R}$ & 1st/2nd generation $D_R$ squark
    & [$100$ GeV, $4$ TeV]\\  
$m_{\tilde b_R}$ & 3rd generation $D_R$ squark  &
    [$100$ GeV, $4$ TeV]\\  
$A_t$ & Trilinear coupling for top quark  & [-$8$ TeV, $8$ TeV]\\ %% why 8?
$A_b$ & Trilinear coupling for b-quark  & [-$8$ TeV, $8$ TeV]\\
$A_\tau$ & Trilinear coupling for $\tau$-quark & [-$8$ TeV, $8$ TeV]\\
$A_e = A_\mu$ & Trilinear coupling for $\mu$-quark & [-$8$ TeV, $8$ TeV]\\
$m_{H_1}$ & up-type Higgs doublet mass & [$100$ GeV, $4$ TeV]\\
$m_{H_2}$ & down-type Higgs doublet mass &  [$100$ GeV, $4$ TeV]\\
$\tan \beta$ & scalar doublets vevs ratio & [2, 60] \\ \hline
$m_t$ & top quark mass~\cite{:2007bxa} &  172.6 $\pm$ 1.4 \\
$m_Z$ & Z-boson mass &  91.187 $\pm$ 0.021 \\
$m_b(m_b)^{\overline{MS}}$ & b-quark mass &  4.20 $\pm$ 0.07 \\
1/$\alpha_{em}(m_Z)^{\overline{MS}}$ & electromagnetic coupling
    constant& 127.918 $\pm$ 0.018 \\ 
$\alpha_{s}(m_Z)^{\overline{MS}}$ & strong coupling constant&
    0.117 $\pm$ 0.002 \\ 
\hline
\end{tabular}}\end{center}
\caption{\linespread{1.}\it{The 25 parameters of the
    pMSSM model. The first twenty non-SM parameters are listed with their
    corresponding prior range. Gaussian 
    priors are used for the SM parameters, which are the last five listed,
    along with their central values and standard deviations.
}}\label{tab.param} 
\end{table}

%%%%%%%%%%%%%%%%%%%%%%%%%%%%%%%%%%%%%%%%%%%%%%%%%%%%%%
\subsection{Observables and experimental constraints} \label{sub.consts}
%%%%%%%%%%%%%%%%%%%%%%%%%%%%%%%%%%%%%%%%%%%%%%%%%%%%%%

\paragraph{} \label{sub.phenodata}
The SM  fits high precision electroweak
data well \cite{Alcaraz:2006mx}. However, on the one hand,
there are some observables whose SM predicted values significantly
differ from the corresponding experimental indications.
The discrepancies could be explained by the direct or indirect
presence of supersymmetric particles  
(or sparticles) in the interactions. On the other hand the very precise
agreement between the SM prediction and the experimentally determined
values of some other set of observables could be altered by the
presence of non-SM particles. The absence of any significant such
deviation puts tight constraints on possible new physics beyond the SM
(SUSY in our case); see for instance  \cite{RamseyMusolf:2006vr,
  Ellis:2007fu} and references therein. The values of the sparticle
masses affects these tendencies. For instance, the effect of sparticles
in loop corrections to electroweak physics observables (EWPO)
decouple if their masses are much heavier than $m_Z$ (300 GeV and
above according to Ref.~\cite{chan-pok}). Lighter sparticles with masses
just above current experimental limits  will alter
the agreement between the electroweak data and SM predictions -- hence
the preference for low energy (weak-scale) SUSY.  
\paragraph{}
For the pMSSM set-up we use observables and constraints 
from  high precision collider measurements that are sensitive to
effects of new physics via virtual loops. These include five EWPO: the
$W$-boson mass, $m_W$, the effective leptonic weak mixing angle,
$\sin^2 \theta^{lep}_{eff}$, the total $Z$-boson decay
width, $\Gamma_Z$, the anomalous magnetic moment of the muon, $(g -
2)_\mu$ and the mass of the lightest MSSM Higgs boson, $m_h$; five
B-physics observables: branching ratios $BR(B \rightarrow X_s\gamma)$, $BR(B_s \rightarrow
\mu^+ \mu^-)$, $BR(B_{u^-} \rightarrow \tau^- \nu)$, $BR(B_u \rightarrow K^*
\gamma)$ and the $B_s$ mass-mixing parameter $\Delta M_{B_s}$; and the
cosmological observable, dark matter relic density from WMAP5
results. We next briefly describe each of these physical observables
and state the corresponding experimental constraints. We first discuss
constraints from EWPO and end the Section with discussion of sparticle
mass limits.   
\begin{itemize} 
\item{\bf $W$-boson mass, $m_W$:}\\
The CDF Run II electroweak public results cited the $W$-boson
mass measurement as the single most precise measurement to date and
quotes \cite{Alcaraz:2007ri} \be m_W = (80.399 \pm 0.025)
\, \textrm{GeV}. \ee   
Theoretically the mass can be calculated from  
\be
m_W = \frac{\pi \alpha_{em}}{\sqrt{2}G_F(1-m_W^2/m^2_Z)(1-\Delta r)},
\ee
where $\alpha_{em}$ is the fine structure constant at the $m_Z$
renormalisation energy scale, $G_F$ is the Fermi weak coupling
constant and $\Delta r$ includes all radiative corrections to the
mass (see e.g.~\cite{Phys.Rev.D22.971,Phys.Rev.D22.2695} and
references therein). The high precision in this measured quantity
constrain any radiative corrections from new physics effects. 
The experimental precision is very high to the
extent that measurements can be sensitive even to two-loop
effects involving sparticles. We include a theoretical uncertainty of
10 MeV on $m_W$ by adding it in quadrature with the experimental uncertainty.
We use 
\texttt{SUSYPOPE}~\cite{Heinemeyer:2007bw,Heinemeyer:2006px} to %%spope} to
calculate the $W$-boson mass $m_W$ and the other EWPO. The most
complete available SM two loop corrections and the dominant results for
two loop SUSY corrections as implemented in \texttt{SUSYPOPE} currently
give the most accurate predictions within the MSSM.

\item{\bf $Z$-boson decay width, $\Gamma_Z$:}\\
The partial $Z$-boson decay width in the massless fermion limit
($m_f^2/m_Z^2 \rightarrow 0$) is given by~\cite{yellow} %%:2005ema}
\be
\Gamma_{Z \rightarrow f \bar{f}} = N_c^f \, \frac{G_F
  m_Z^3}{6\sqrt{2}\pi}\, \delta_{QCD}(\bar{g}_v^{(f)2} +
\bar{g}_a^{(f)2}) + \Delta_{ew/QCD}
\ee
where $\bar{g}_{v,a}$ are the neutral weak coupling constants modified
to include electroweak (EW) radiative effects, $\delta_{QCD}$ parametrises the
QCD corrections and $\Delta_{ew/QCD}$ includes some
non-factorisable EW contributions. The colour factor $N_c^f$ is 1 for
leptons and 3 for quarks. The current experimental value for the total
decay width is \cite{Alcaraz:2007ri} \be \Gamma_{Z} = (2.4952 \pm
0.0023) \textrm{ GeV}.\ee  Theoretically
\be
\Gamma_{Z} = \Gamma_{l} + \Gamma_{h} + \Gamma_{inv}
\ee
where $\Gamma_{l,h}$ are the decay widths into SM leptons
and quarks. $\Gamma_{inv}$ is for the decays into invisible particles
(neutrinos and possibly, if they are light enough,
neutralinos). Supersymmetric contributions enter 
via virtual corrections to the partial decay widths into lepton and
quarks. 

\item{\bf Effective mixing angle, $\sin^2 \theta^{lep}_{eff}$:}\\
The effective electro-weak mixing angle depends only on the ratio of
the effective weak couplings 
\be Re(g_v/g_a) = 1 - 4\sin^2 \theta_{eff}^{lep}
\ee for the vertex that couples the $Z$-boson and leptons $l$ in the
Lagrangian: $i 
\overline{l} \gamma^\mu \, (g_v - g_a \gamma_5) \, Z_\mu l$. It is
determined from various asymmetry measurements around the $Z$-boson
peak from $e^+\,e^-$ colliders after removing QED effects
\cite{:2005ema}. We use the experimental estimate \cite{Alcaraz:2007ri}
\be \sin^2 \theta^{lep}_{eff} = 0.2324 \pm 0.0012. \ee

\item{\bf Z-pole asymmetry parameters from $e^+e^- \rightarrow f
  \bar{f}$ processes:}\\
The results from the LEP and SLC $e^+e^-$ colliders on
Z-boson properties (its mass, partial and total widths, and couplings
  to fermion pairs) are in good agreement with the SM 
predictions~\cite{:2005ema}. The precision is high enough to
probe loop-level predictions where both SM and beyond the SM corrections are
absorbed into effective coupling constants. The LEP data consist 
of hadronic and leptonic cross sections, leptonic
forward-backward asymmetries, $\tau$ polarisation asymmetries, 
$b\bar{b}$ and $c\bar c$ partial widths and forward-backward asymmetries. The
Z-boson parameters derived from the data which we 
employ for our analysis include the ratios $R_l$ (which we assume to
be the average of $R_e$, $R_\mu$ and $R_\tau$), $R_b$ and $R_c$. These
are defined as
\be R_b = \frac{\Gamma(Z \rightarrow b\bar{b})}{\Gamma(Z \rightarrow 
hadrons)}, \quad R_c = \frac{\Gamma(Z \rightarrow
c\bar{c})}{\Gamma(Z \rightarrow hadrons)}, \quad R_l =
\frac{\Gamma(Z \rightarrow l^+l^-)}{\Gamma(Z \rightarrow
hadrons)}\ee  
and are constrained to be
$$ R_b = 0.21629 \pm 0.00066, R_c = 0.1721
\pm 0.0030 \textrm{ and } R_l = 20.767 \pm 0.025 $$
\paragraph{}
The Z-boson interacts with fermions through a mixture of vector and
axial-vector couplings. This makes the strength of the interaction 
between left- and right-handed fermions unequal and leads to the
production of polarised Z-bosons at the $e^+e^- $ colliders. As result
there are measurable
asymmetries (such as a forward-backward asymmetry) in the angular 
distributions of the final-state fermions $f \bar{f}$. 
The forward-backward asymmetry is related to the probability that the
$\bar{f}$ travels in the 
same (forward) or opposite (backward) direction to the 
incident $e^-$
direction and is quantified by
\be A_{FB} = \frac{\sigma_F - \sigma_B}{\sigma_F + \sigma_B}\ee where 
$\sigma_F(\sigma_B)$ is the cross section in the forward(backward)
directions. In terms of the effective vector and axial-vector neutral
current couplings, $g_{Vf}$ and $g_{Af}$ respectively, other Z-pole
asymmetries are:  
\be A_{FB}^{0,f} = \frac{3}{4} {\mc{A}^e}  {\mc{A}^f}, \quad
{\mc{A}^f} \equiv \frac{2g_{Vf} g_{Af}}{g_{Vf}^2 + g_{Af}^2}.\ee 
Here $A^f$ gives a measure of the asymmetry for the different
possible final state fermions. At LEP the Z-pole forward-backward asymmetries
$A_{FB}^{0,b}$ and $A_{FB}^{0,c}$ were precisely measured for the
final states $b \bar{b}$ and $c \bar{c}$ respectively.  We impose
$A_{FB}^{0, b} = 0.0992 \pm 0.0016$ and $A_{FB}^{0, c} = 0.0707 \pm
0.0035$ for our analysis. 
\paragraph{}
Using polarised beams, the  SLD experiment made a direct and precise 
measurement of the parameter $A_e$ from the left-right asymmetry  
\be A_{LR}=\frac{\sigma_L - \sigma_R}{\sigma_L + \sigma_R}\ee were
$\sigma_L$ and $\sigma_R$ are the $e^+$ $e^-$ production cross
sections for Z bosons produced with left- and right-handed electrons
respectively. The same parameter, $A^e$, is also indirectly constrained by
LEP experiments. Using the measurements of $A^e$ the parameters
$A^\mu$ $A^\tau$ $A^b$ and $A^c$ can then be inferred from $A_{FB}$ 
measurements at LEP\@. Hence, the LEP and SLC results form a
complete set of the $A^f$ parameter measurements.
The asymmetry parameter constraints we use are 
$A^b = 0.923 \pm 0.020$, $A^c = 0.670 \pm 0.027$ and $A^l = 0.1513 \pm
0.0021 = A^e$. 

\item{\bf Muon anomalous magnetic moment, $\delta a_\mu$:}\\
The world average for the muon anomalous magnetic moment as determined
from $e^+e^-\rightarrow$ hadrons-based experiment at Brookhaven
\cite{Bennett:2006fi} is $a_\mu^{exp} \equiv \half(g-2)_\mu = 11 659
20.80 \pm 0.63 \times 10^{-9}$. Other results from experiments based
on the $\tau$ lepton decay to hadrons \cite{Davier:2007ua,
  Hagiwara:2006jt, Hertzog:2007hz} differ slightly and we are not
using those here\footnote{See \cite{Stockinger:2007pe} for a
recent review.}. The experimental results are around $3\sigma$ deviation
from the SM prediction \cite{Miller:2007kk} $a_\mu^{SM} = 11
659 17.85 \pm 0.61 \times 10^{-9}$ giving 
\be \delta a_\mu = a_\mu^{exp} - a_\mu^{SM} =  29.5 \pm 8.8 \times
10^{-10}.\ee 
SUSY is a good new physics candidate that can
explain this deviation. For sparticles all of mass
$M_{SUSY}$, their contribution to $a_\mu$ is of order
\cite{Stockinger:2006zn} \be a_\mu^{SUSY} \approx 13 \times 10^{-10}
\left( \frac{100 \textrm{ GeV}}{M_{SUSY}} \right) \tan \beta
\,\textrm{sign}(\mu). \ee
We use {\tt
  micrOMEGAs}~\cite{Belanger:2008sj,Belanger:2006is,Belanger:2004yn,Belanger:2001fz}
to predict $a_\mu$, correcting it  
with $\Delta_\mu$, which includes contributions from 
the dominant QED-logarithmic piece, some dominant two-loop 
corrections~\cite{Stockinger:2007pe} and the recently computed
$\tan\beta$-enhanced term 
(see \cite{Marchetti:2008hw} and references therein)
\be
a_\mu^{SUSY} = a_\mu^{SUSY, 1L} \left( 1 - \frac{4 \alpha}{\pi} \log
\frac{M_{SUSY}}{m_\mu} \right) \left( \frac{1}{1 + \Delta_\mu} \right)
+ \textrm{ 2-loop terms.} 
\ee
For our analysis we add in quadrature a theoretical error of $2.0
\times 10^{-10}$ to the $8.8 \times 10^{-10}$ error above and use the
average value $\delta a_\mu = (30.2 \pm 9.2) \times 10^{-10}$.

\item{\bf Lightest Higgs boson mass, $m_h$:}\\
The SM higgs mass is constrained to $m_h \geq 114.4$ GeV
by LEP direct search experiment (see e.g.\ \cite{Schael:2006cr} and
references therein). The predicted higgs mass can be parametrised by
$g_{hZZ}/g_{hZZ}^{SM}$, the ratio of the MSSM higgs coupling to two
neutral $Z$-bosons to the equivalent SM coupling. In the
MSSM $g_{hZZ}/g_{hZZ}^{SM} = \sin^2(\beta - \alpha)$ so we use the above LEP
mass limit for $\sin^2(\beta - \alpha) > 0.95$ and use the MSSM higgs
mass limit $m_h \geq 89.7$ for $\sin^2(\beta - \alpha) < 0.95$.

\item{\bf Branching ratio ($B\rightarrow X_s\gamma$):} \\
The experimental value of the decay rate of the flavour changing
process $B\rightarrow X_s\gamma$ agrees to high precision with the
SM prediction. This stringently constrains  new physics
models that may contribute to the process. For SUSY the leading
contributions come from loops with charged Higgs bosons (these
interfere
constructively with the SM contributions) and charginos. Loops with
neutralinos are small (see for instance,
\cite{Wick:2008sz}). A recent theoretical estimate for
the SM contribution to the branching ratio of $B\rightarrow
X_s\gamma$
which we call $BR(b \rightarrow s \gamma)$
at next-to-next-to leading order
(NNLO) in QCD
is \cite{Misiak:2006zs, Misiak:2006bw, Misiak:2006ab, Gambino:2008fj}
$BR(b\rightarrow s\gamma)_{SM} = (3.28 \pm 0.23) \times 10^{-4}$,
where a 1.6
GeV lower energy cut is applied to the photon.
The central value is $\sim 1\sigma$
lower than a world-average
experimental value from the Heavy Flavor Averaging Group
\cite{Barberio:2007cr} (HFAG)\footnote{We note that the most recent
central
 value has shifted slightly to 3.52: this small change would leave
no
 significant imprint on our fits.}
\be BR(B\rightarrow X_s\gamma)_{exp} = (3.55 \pm
0.22^{+0.09}_{-0.10} \pm
0.03) \times 10^{-4}. \ee
The values predicted for this observable at different parameter
points
are constrained by combining in quadrature the experimental and SM
prediction errors:
\be BR(B\rightarrow X_s\gamma)_{exp} = (3.55 \pm 0.42) \times
10^{-4}.
\label{expsg}
\ee
The branching ratio therefore constrains  non-SM contributions. We
use
SuperIso2.0~\cite{Mahmoudi:2007vz} to predict the MSSM
plus SM branching ratio\footnote{The program {\tt
SusyBSG}~\cite{Degrassi:2007kj}
now contains a more
accurate prediction of this branching ratio to two-loops in MSSM
parameters.
It could lead to an estimated shift in the predicted branching ratio
of $0.13
\times 10^{-4}$, which may be considered to be included within our
estimate of
theoretical error.}.
 
\item{\bf Branching ratio ($B_s \rightarrow \mu^+ \mu^-$):}\\
The branching fraction for the flavour changing process $B_s
\rightarrow \mu^+ \mu^-$ is predicted to be $(3.42 \pm 0.54)\times
10^{-9}$ within the SM \cite{Buras:2003td}. In the MSSM,
interactions involving
neutral Higgs bosons can enhance the branching fraction by several
orders of magnitude at high $\tan \beta$. The branching ratio is
experimentally
bounded from above by recent CDF II results, implying \be BR(B_s \rightarrow \mu^+
\mu^-) < 5.8 \times 10^{-8} \ee at $95\%$ CL \cite{:2007kv}. We apply a
continuous likelihood constraint derived from CDF II data upon 
the MSSM prediction~\cite{private2}.
The resulting penalty is shown
 in Fig.~\ref{fig.bsmumu}.
 \begin{figure}[h]
     \centering\includegraphics[width=.6\textwidth]{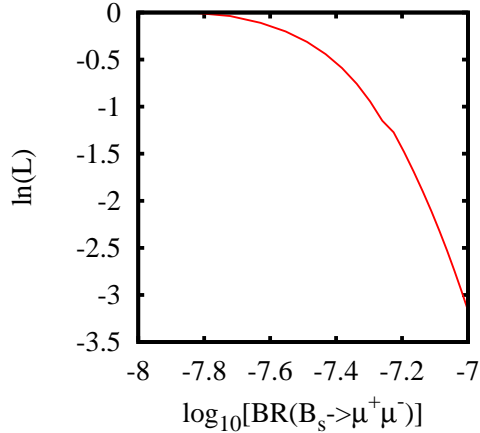}
   \caption{\linespread{1.} \it{Likelihood penalty on the
   predicted value of $BR(B_s \rightarrow \mu^+ \mu^-)$.}}  
  \label{fig.bsmumu} 
 \end{figure}

\item{\bf $B_s$-$\bar{B}_s$ mass difference, $\Delta M_{B_s}$:}\\
The neutral $B$-meson oscillates between particle and antiparticle
states via flavour changing processes. The frequency of oscillation is  
proportional to the mass difference $\Delta M_{B_s}$
It has been measured to be $(\Delta M_{B_s})_{exp} = 17.77 \pm 0.12
{\textrm{ ps}}^{-1}$ \cite{Abulencia:2006ze}. 
Its SM prediction can be obtained from an overall
unitarity triangle fit: 
$(\Delta M_{B_s})_{SM} = 20.9 \pm 2.6
{\textrm{ps}}^{-1}$ \cite{Bona:2006ah}. 
We use the ratio of the experimental constraint to the SM prediction
\be R_{\Delta M_{B_s}}^{exp} = \frac{(\Delta M_{B_s})_{exp}}{(\Delta
  M_{B_s})_{SM}} = 0.85 \pm 0.11, \ee  to constrain the
predicted frequencies. Our pMSSM predicted values
are based on the results in \cite{Isidori:2006pk} and
references therein: 
\bea
& &R_{\Delta M_{B_s}}^{MSSM} = \frac{(\Delta M_{B_s})_{MSSM}}{(\Delta
  M_{B_s})_{SM}} \\
&=& 1 - \frac{64 \pi \sin^2\theta_w}{\alpha_{em}M_A^2 S_0(m_t^2/m_W^2)}
\frac{m_b(\mu_b) m_s(\mu_b) \, (\epsilon_Y
  \tan^2\beta)^2}{[1+(\epsilon_0+\epsilon_Y)\tan \beta]^2[1+\epsilon_0
    \tan \beta]^2} \nonumber 
\eea
where $m_{b,s}$ are the bottom- and strange-quark masses evaluated
at the $\overline{MS}$ scale $\mu_b = m_b$; 
\be \label{eps0}
\epsilon_0 = -\frac{2\alpha_s \mu}{3 \pi m_{\tilde{g}}}\,H_2\left(
\frac{m_{\tilde{q}L}^2}{m_{\tilde{g}}^2},
\frac{m_{\tilde{d}R}^2}{m_{\tilde{g}}^2}\right), \quad
\epsilon_Y = -\frac{A_t y_t^2}{16 \pi^2 \mu}\,H_2\left(
\frac{m_{\tilde{q}L}^2}{\mu^2}, \frac{m_{\tilde{u}R}^2}{\mu^2}\right)
\ee
are the effective couplings that parametrise the correction to the
down-type Yukawa couplings,
\be
H_2(x,y)= \frac{x \log x}{(1-x)(x-y)} + \frac{y \log y}{(1-y)(y-x)};
\ee
$\mu$ is the supersymmetric Higgs mass terms and $A_t$ is the trilinear
soft breaking term involving the stops. $S_0$ is given by 
$$S_0(x) = \frac{4x-11x^2+x^3}{4(1-x)^2} - \frac{3x^3 \log
x}{2(1-x)^3}.$$

\item{\bf Branching ratio $(B_u \rightarrow \tau \nu)$:}\\
The purely leptonic decay $B_{u^-} \rightarrow \tau^- \nu$ proceeds via the
annihilation of $b$- and $\bar{u}$-quarks into $W^-$. The SM
prediction for the branching ratio of the process is given by  
\be
BR(B_u \rightarrow \tau \nu)_{SM} = \frac{G_F^2 m_B m_\tau^2}{8\pi} \left[1
  - \frac{m_\tau^2}{m_B^2}\right]^2 f_B^2 |V_{ub}|^2 \tau_B,
\ee
where $m_B$ and $m_\tau$ are the $B$ meson and $\tau$ pole masses,
respectively, and $\tau_B$ is the $B^-$-meson lifetime. 
For the SM prediction we use the average of the result from unitarity
triangle fits ($BR(B_u \rightarrow \tau \nu) = 0.85 \pm 0.14 \times
10^{-4}$) and the result obtained from the experimental determination
of $V_{ub}$ and $f_B \sqrt{B_{B_d}}$ ($BR(B_u \rightarrow \tau \nu) =
1.39 \pm 0.44 \times 10^{-4}$) \footnote{See  
  ``New Constraints from B and K rare decays'' at
  \texttt{http://www.utfit.org/}} adding the errors in quadrature to:  
\be BR(B_u \rightarrow \tau \nu)_{SM} =1.12 \pm 0.46 \times10^{-4} \label{one}\ee.
For the experimental constraint upon the branching ratio, we use the
average of the Belle and BaBar experiments, adding their errors in
quadrature: 
\be BR(B_u \rightarrow \tau \nu)_{exp} =1.41 \pm 0.43
\times10^{-4}. \label{two}\ee 
Eqs.~(\ref{one}), (\ref{two}) are then used to form the constraint
\be
R_{B\tau \nu}^{exp} = \frac{BR(B_u \rightarrow \tau \nu)_{exp}}{BR(B_u
  \rightarrow \tau \nu)_{SM}} = 1.26 \pm  0.41. 
\ee 
For the pMSSM predictions we follow \cite{Ellis:2007fu} and predict
\be
R_{B\tau \nu}^{MSSM} = \frac{BR(B_u \rightarrow \tau \nu)_{MSSM}}{BR(B_u
  \rightarrow \tau \nu)_{SM}} =
  \left[1-\left(\frac{m_{B_u}^2}{m_{H^{\pm}}^2} \right) \frac{\tan^2
  \beta}{1+\epsilon_0\tan \beta} \right]^2
\ee where $\epsilon_0$ is the effective coupling defined in
  Eq.~\ref{eps0}, $m_{B_u}$ is the the B-meson mass and
  $m_{H^{\pm}}$ the charged higgs boson mass.

\item{\bf $\Delta_{0-}$: $B \rightarrow K^* \gamma$ Isospin asymmetry:}
Isospin symmetry predicts the amplitudes for the decays $\bar{B}^0
\rightarrow \bar{K}^{*0} 
\gamma$ and $B^- \rightarrow K^{*-} \gamma$ to be equal at leading order
in perturbation theory. 
Isospin-breaking effects in the process $B \rightarrow K^* \gamma$
\cite{Ahmady:2006yr} 
may therefore provide a sensitive probe of physics beyond the SM\@.
The isospin asymmetry for the exclusive process $B \rightarrow K^*
\gamma$ is defined as 
\be
\Delta_{0-} = \frac{\Gamma(\bar{B}^0 \rightarrow \bar{K}^{*0} \gamma) -
  \Gamma(B^- \rightarrow K^{*-} \gamma)}{\Gamma(\bar{B}^0 \rightarrow
  \bar{K}^{*0} \gamma) + \Gamma(B^- \rightarrow K^{*-} \gamma)}. 
\ee
The world average experimental value is \cite{J.Phys.G33.1}
\be
\Delta_{0-} = 0.0375 \pm 0.0289.
\ee In order to fit this, we use the MSSM prediction from 
\texttt{SuperIso2.0}~\cite{Mahmoudi:2007vz} which
%. \texttt{SuperIso2.0} 
contains the 
available next-to-leading-order (NLO) contributions to $\Delta_{0-}$,
including the 
complete supersymmetric QCD corrections to Wilson coefficient
operators. It also includes some partial next-to-NLO (NNLO) SM QCD
corrections.
\item{\bf Dark Matter relic density}:\\% \label{dmrelic}
The Wilkinson Microwave Anisotropy Probe (WMAP) fits to a cosmological
constant plus cold dark matter model ($\Lambda$CDM) imply a dark
matter relic density of $ \Omega_{DM} h^2 = 0.1143 \pm 0.0034$, 
where $h$ is the reduced Hubble constant \cite{0803.0547}. 
We assume $R$-parity, resulting in a stable lightest supersymmetric particle
(LSP). 
The neutralino, $\tilde{\chi}^0_1$, LSP then has the correct properties to make up
the cold 
dark matter, being massive, stable and neutral and we constrain the prediction
of its relic density to lie on 
the WMAP5 central value, but inflate the uncertainties with an assumed error
on the theoretical prediction:
\be
\Omega_{DM} h^2 = 0.1143 \pm 0.02. \label{wmap5}
\ee
\end{itemize}
We summarise the experimental constraints used in our fits in 
Tab.~\ref{tab.observ}, listing the relevant references with each.
\begin{table}[!t]
\linespread{1.5}
\begin{center}{\begin{tabular}{|clll|}
\hline
Observable & Constraint & Th. Source & Ex. Source  \\
\hline
$m_W$ [GeV]& $80.399 \pm  0.027$& \cite{Heinemeyer:2006px}&
\cite{verzo}  \\
$\Gamma_Z$ [GeV]& $2.4952 \pm 0.0025$& \cite{Heinemeyer:2007bw} &  \cite{:2005ema}  \\
$\sin^2\, \theta_{eff}^{lep}$  & $0.2324 \pm 0.0012$& \cite{Heinemeyer:2006px} & \cite{:2005ema}\\
$\delta a_\mu $ & $(30.2 \pm 9.0) \times 10^{10}$&
\cite{Moroi:1995yh,Degrassi:1998es,Heinemeyer:2003dq,Heinemeyer:2004yq}
& \cite{Bennett:2006fi,Davier:2007ua, Hertzog:2007hz} \\ 
$R_l^0$ & $20.767 \pm 0.025$ & \cite{Heinemeyer:2007bw} & \cite{:2005ema}  \\ 
$R_b^0$ & $0.21629 \pm 0.00066$& \cite{Heinemeyer:2007bw} & \cite{:2005ema}       \\
$R_c^0$ & $0.1721 \pm 0.0030$& \cite{Heinemeyer:2007bw} & \cite{:2005ema}       \\
$A_{\textrm{FB}}^b$ & $0.0992 \pm 0.0016$& \cite{Heinemeyer:2007bw} & \cite{:2005ema}  \\
$A_{\textrm{FB}}^c$ & $0.0707 \pm 0.035$ & \cite{Heinemeyer:2007bw} & \cite{:2005ema}  \\
$A^l = A^e$& $0.1513 \pm 0.0021$  & \cite{Heinemeyer:2007bw} & \cite{:2005ema}   \\
$A^b$ & $0.923 \pm 0.020$ & \cite{Heinemeyer:2007bw} & \cite{:2005ema}            \\
$A^c$ & $0.670 \pm 0.027$ & \cite{Heinemeyer:2007bw} & \cite{:2005ema}            \\
$Br(B \rightarrow X_s \gamma)$ & $(3.55 \pm 0.42) \times 10^{4}$ &
\cite{Misiak:2006zs, Misiak:2006bw, Misiak:2006ab, Becher:2006pu} &
\cite{Barberio:2007cr}  \\ 
$Br(B_s \rightarrow \mu^+ \mu^-)$ & see Fig.~\protect~\ref{fig.bsmumu}&
\cite{Belanger:2008sj,Belanger:2006is,Belanger:2004yn,Belanger:2001fz}
& \cite{:2007kv} \\  
$R_{\Delta M_{B_s}}$ & $0.85 \pm 0.11$ & \cite{Bona:2006ah} & \cite{Abulencia:2006ze}  \\ 
$R_{Br(B_u \rightarrow \tau \nu)}$ & $1.26 \pm 0.41$&
\cite{Isidori:2006pk,Isidori:2007jw,Akeroyd:2003zr} &
\cite{Aubert:2004kz,paoti,hep-lat/0507015}  \\
$\Delta_{0-}$ & $0.0375 \pm 0.0289$& \cite{Mahmoudi:2007vz} & \cite{J.Phys.G33.1} \\ 
$\Omega_{CDM} h^2$ & $0.11 \pm 0.02 $&
\cite{Belanger:2008sj,Belanger:2006is,Belanger:2004yn,Belanger:2001fz}
& \cite{0803.0547} \\   
\hline
\end{tabular}}\end{center}
\caption{\linespread{1.}\it{Summary of constraints on predictions. 
Theoretical uncertainties have been added in quadrature to the
    experimental uncertainties quoted.}}\label{tab.observ}  
\end{table}

%%%%%%%%%%%%%%%%%%%%%%%%%%%%%%%%%%%%%%
\subsection{Direct search mass limits} \label{sub.mlimits}
%%%%%%%%%%%%%%%%%%%%%%%%%%%%%%%%%%%%%%

\paragraph{}
The absence of sparticle or Higgs boson production at current collider searches for
supersymmetric particles puts lower bounds on their possible masses
\cite{J.Phys.G33.1}. We veto any pMSSM 
points that violate the limits. The limits are derived from
various  experiments that usually {\it a priori}\/ assume the validity
of a chosen model (usually the CMSSM). Where possible, we use more appropriate
model independent limits upon sparticle masses coming from searches.
SUSY particles may be pair produced at colliders that have sufficient energy,
then undergo subsequent decay into SM particles and neutralino LSP. 
Hard jets or leptons associated with missing energy coming from the 
neutralino then constitute SUSY direct search signatures.
Constraints on sparticle of mass $m$ often dependent
upon the mass difference $\Delta m = m - m_{LSP}$ which is
correlated with the energy of visible sparticle decay products \cite{kane,
Djouadi:1998di}. The sparticle mass limits derived then depend on
this energy; depending upon whether $\Delta m$ is low (5 - 10 GeV) or
higher. 
The mass limits we impose on the sparticles and the 
lightest Higgs boson are summarised in Tab.~\ref{tab.mlimits}~\cite{J.Phys.G33.1}. 
\begin{table}[htbp!] 
\linespread{1.5}
\begin{center}{\begin{tabular}{|ccc|}
\hline
condition & sparticle mass & lower bound/GeV \\
\hline
$\sin^2 (\alpha - \beta) > 0.95$ & $m_h$ & 114  \\
$\sin^2 (\alpha - \beta) \leq 0.95$ & $m_h$ & 89.7  \\
$m_{\tilde{\tau}_1} - m_{\tilde{\chi}^0_1} > 10$ GeV & $m_{\tilde{\tau}_1}$
& 87  \\
$m_{\tilde{\tau}_1} - m_{\tilde{\chi}^0_1} \leq 10$ GeV & $m_{\tilde{\tau}_1}$
& 73  \\
$m_{\tilde{e}_R} - m_{\tilde{\chi}^0_1} > 10$ GeV & $m_{\tilde{e}_R}$
& 100  \\
$m_{\tilde{e}_R} - m_{\tilde{\chi}^0_1} \leq 10$ GeV & $m_{\tilde{e}_R}$
& 73  \\
$m_{\tilde{\mu}_R} - m_{\tilde{\chi}^0_1} > 10$ GeV & $m_{\tilde{\mu}_R}$
& 95 \\
$m_{\tilde{\mu}_R} - m_{\tilde{\chi}^0_1} \leq 10$ GeV & $m_{\tilde{\mu}_R}$
& 73  \\
$m_{\tilde{\nu}_e} - m_{\tilde{\chi}^0_1} > 10$ GeV & $m_{\tilde{\nu}_e}$
& 94  \\
$m_{\tilde{\nu}_e} - m_{\tilde{\chi}^0_1} \leq 10$ GeV & $m_{\tilde{\nu}_e}$
& 43  \\
$m_{\tilde{\nu}_{\tau}} > 300$ GeV & $m_{\tilde{\chi}_1^{\pm}}$
& 43  \\
$m_{\tilde{\nu}_{\tau}} \leq 300$ GeV & $m_{\tilde{\chi}_1^{\pm}}$
& 92.4 \\
- & $m_{\tilde{\chi}_1^0}$ & 50 \\
$m_{\tilde{t}} - m_{\tilde{\chi}^0_1} > 10$ GeV& $m_{\tilde{t}}$
& 95 \\
$m_{\tilde{t}} - m_{\tilde{\chi}^0_1} \leq 10$ GeV& $m_{\tilde{t}}$
& 65 \\
$m_{\tilde{t}} - m_{\tilde{\chi}^0_1} > 10$ GeV& $m_{\tilde{b}}$
& 95 \\
$m_{\tilde{t}} - m_{\tilde{\chi}^0_1} \leq 10$ GeV & $m_{\tilde{b}}$
& 59 \\
- & $m_{\tilde{q}}$ & 318 \\
- & $m_{\tilde{g}}$ & 195 \\
\hline
\end{tabular}}
\end{center}
\caption{\it{The lower bounds applied to
MSSM particle masses.}} 
\label{tab.mlimits}
\end{table}

\paragraph{}
A recent random scan study of the pMSSM~\cite{Berger:2008cq} 
found that a CDF/D0
bound~\cite{Phys.Rev.D46.1889,Abazov:2008qu} was quite restrictive on
their pMSSM parameter-space random sample points when the relic
density constraint was only applied as an upper bound 
(i.e.\ allowing for additional extra-MSSM sources of dark matter). 
The bound states that, for $m_{\tilde{\chi}_1^{\pm}}
- m_{\tilde{\chi}_1^0} \leq 50$ MeV,  
$m_{\tilde{\chi}_1^{\pm}} \geq 206 \,|U_{1w}|^2 + 171 \, |U_{1h}|^2$
GeV at 95\% confidence level . Here $|U_{1w}|$, $|U_{1h}|$
are the Wino and Higgsino content of the lightest chargino, respectively. 
We did not use this bound in our {\sc MultiNest} sampling procedure,
but we have checked retrospectively that it would not have
significantly changed our fits since only less than 1$\%$ of the
posterior probability density fails this constraint. 

%%%%%%%%%%%%%%%%%%%%%%%%%%%%%%%
\subsection{Sampling procedure} \label{sub.phenoprocedure}
%%%%%%%%%%%%%%%%%%%%%%%%%%%%%%%

\paragraph{}
In this Subsection we summarise the Bayesian inference elements
(briefly reviewed in Section~\ref{sec.inference}) for the pMSSM and
the sampling procedure we employ for fitting it to the indirect
collider and cosmological data.
All of the pMSSM parameters, $\mathbf{\theta}$, listed in Tab.~\ref{tab.param}, are varied
simultaneously, our calculation being driven by {\sc MultiNest} and
the high energy physics software mentioned in the following paragraphs. 
{\sc MultiNest} is described in
  Appendix~\protect\ref{sub.nested}.
Each point is 
passed in SUSY Les Houches Accord (SLHA) format~\cite{Skands:2003cj} to the
different particle physics software we use to predict the
observables described in Subsection~\ref{sub.phenodata}. 
For each set of parameters {\sc MultiNest} selects, the following
steps are followed: 
\begin{enumerate}

\item{} The (input) parameters are  passed to
  \texttt{SOFTSUSY2.0.18}~\cite{Allanach:2001kg} which produces the
MSSM sparticle masses and couplings. 
Unphysical points are flagged by the program to have one
or some combination of the following properties: absence of electroweak
symmetry breaking, the 
presence of a tachyon, a non-perturbative point where the calculation can no
longer be trusted or the lack of numerical convergence (which usually occurs
close to a boundary of good electroweak symmetry breaking). 
If any of these properties are flagged then the
point is discarded before any further computations and 
the parameter point is given a zero likelihood. In addition
to this, sparticle spectra that violate mass limits, shown in
Tab.~\ref{tab.mlimits}, from sparticle 
searches or that have a non-neutralino LSP are also discarded. 

\item{}
Physical parameter points are passed in SLHA format to
\texttt{micrOMEGAs3.2}~\cite{Belanger:2008sj,Belanger:2006is,Belanger:2004yn,Belanger:2001fz} which
calculates the neutralino dark matter relic density, the branching ratio 
$BR(B_s \rightarrow \mu^+ \mu^-)$ and the anomalous magnetic moment of
the muon $\delta a_\mu$. 
The physical point is then
passed to the computer program \texttt{SuperIso2.0}~\cite{Mahmoudi:2007vz} 
and other subroutines. The former calculates $BR(B \rightarrow X_s
\gamma)$ and the isospin asymmetry in $B$ meson decays $\Delta_{0-}$
while the latter computes the B-physics ratios $R_{B_u \rightarrow
  \tau \nu}$ and $R_{\Delta M_{B_s}}$. 

\item{}
We then use  \texttt{SUSYPOPE}~\cite{Heinemeyer:2007bw,Heinemeyer:2006px} to
calculate the $W$-boson mass $m_W$, the effective leptonic mixing
angle variable $\sin^2 \theta^{lep}_{eff}$, the total $Z$-boson decay
width $\Gamma_Z$ and the other electroweak physics observables listed
in Tab.~\ref{tab.observ} to two loops.  
\end{enumerate}

\paragraph{}
The various physical observables
described in Subsection~\ref{sub.phenodata}, which are derived 
from the input parameters, form the data set, $D_i$, $i = 1, \ldots,
19$:
\bea \mathbf{D} &= &\{ m_W,\; \sin^2\, \theta^{lep}_{eff},\; \Gamma_Z,\; \delta
a_{\mu},\; R_l^0,\; A_{fb}^{0,l},\; A^l = A^e,\; R_{b,c}^0,\;
A_{fb}^{b,c},\; A^{b,c}, \\ \nonumber 
& &BR(B \rightarrow X_s \, \gamma),\; BR(B_s \rightarrow \mu^+ \, \mu^-),\; 
\Delta_{0-},\; R_{BR(B_u \rightarrow \tau \nu)},\; R_{\Delta M_{B_s}},\\ \nonumber 
& &\Omega_{CDM}h^2 \}. \eea  
For each predicted value $O_i$ of observable $D_i$, the corresponding
likelihood is 
\be \label{like} P(D_i|\mathbf{\Theta}, H)= \frac{1}{\sqrt{2\pi
    \sigma_i^2}}\exp\left[- \frac{(O_i - \mu_i)^2}{2\sigma_i^2}\right]
\ee   
where $\mu_i$ and $\sigma_i$ are the experimental central values and
errors given in Tab.~\ref{tab.observ}. We assume that the observables
are independent and combined the likelihoods to 
\be \label{likely} L(\mathbf{\Theta}) = P(\mathbf{D}|\mathbf{\Theta},
H) = \prod_{i=1}^{19} P(D_i|\mathbf{\Theta}, H) \, L_{BR(B_s
  \rightarrow \mu^+ \mu^-)}\ee  
where $L_{BR(B_s \rightarrow \mu^+ \mu^-)}$ is the likelihood value
for the indicated observable (see Fig.~\ref{fig.bsmumu}).
The predictions from the physical points, as enumerated above, are
checked against experimental values and the deviations from this are
quantified by the individual likelihood functions Eq.~\ref{like}.
The likelihoods from the different observables are combined into one
overall likelihood, Eq.~\ref{likely}, which is then multiplied by the
prior probability density Eq.~\ref{pro} to produce posterior
probability density Eq.~\ref{posterior}. 

%%%%%%%%%%%%%%%%%%%%%%%%%%%%%%%
\subsection{Computer resources}\label{sub.cpu}
%%%%%%%%%%%%%%%%%%%%%%%%%%%%%%%

\paragraph{}
We end this Section with the presentation of an estimate of the
computing resources used. In the {\sc MultiNest1.3}  algorithm we used
4,000 live points (see 
the Appendix and \cite{Feroz:2007kg, Feroz:2008xx} for
details) and more than $8.6 \times 10^{6}$ likelihood evaluations for the
linear prior and more than $2.1 \times 10^{7}$ for the log prior
case. The overall number of likelihood evaluations in this work is
more than $2.5 \times 10^{8}$. The computing was performed by 79
12-hour jobs at the {\it Darwin}\/ cluster of  the high performance
computing service (HPC) at the University of  Cambridge. Each run used
128 threads or CPUs.  We also used around 40 8-hour jobs at COSMOS
(the UK Cosmology supercomputer at DAMTP).
At COSMOS each run used 64 CPUs. The total run time adds to more
than 15 CPU years. However the efficiency of the more recent version of
{\sc MultiNest} has improved and these computations could take
about half of the stated time.

%SSSSSSSSSSSSSSSSSSSSSSSSSSSSS
\section{Results and Analysis} \label{sec.results}
%SSSSSSSSSSSSSSSSSSSSSSSSSSSSS

\paragraph{}
We now present the results of the global fit for the weak-scale pMSSM
parameters to indirect collider and cosmological data. We first
give the marginalised 1-dimensional posterior probability
distributions for the 25 parameters and sparticle masses in
Section~\ref{sub.maps}. We also give, see Tab.~\ref{tab.bestparam},
a good fit point in the pMSSM parameter space (i.e.\ the sampled point with
maximum likelihood) and the corresponding sparticle spectrum in
Tab.~\ref{tab.spectra}. The observables posterior PDFs are
discussed in Section~\ref{sub.observ}. The values which each of the
observables take, together with corresponding deviations from
experimental values, at the good fit point are given in
Fig.s~\ref{fig.ewwglikeN} and~\ref{fig.ewwglikeL}. In
Section~\ref{sub.signmucomp} we present an implementation of Bayesian
model selection by comparing two different pMSSM hypothesis each with
either sign of the $\mu$ parameter. We applied the `gaugino code'
prescription to the pMSSM in Section~\ref{sub.gcode}. This is
connected with the gluino-neutralino mass ratio and its role in
discriminating different SUSY-breaking phenomenology
models. We address the case of fine-tuning in the pMSSM parameters in
Section~\ref{sub.finetuning}. 

\paragraph{}
Results and analysis on neutralino dark matter and its direct
detection prospects are presented in
Section~\ref{sec.dm}. A general feature of the results is that they
exhibit some amount of prior dependence. This is clearly shown for the case of 
gluino-neutralino mass ratio in Subsection~\ref{sub.gcode}, the amount of
fine-tuning in the parameters in Subsection~\ref{sub.finetuning}, the muon
anomalous magnetic moment (see the $\delta a_\mu$ plots in
Fig.~\ref{fig.observ}), and the dominant neutralino dark matter
annihilation/co-annihilation channels addressed in in
Section~\ref{sec.dm}. It is expected that with more precise and direct data from
the Tevatron and/or the LHC would lift the prior
dependence. However interestingly enough,  some of the results,
such as the mass of the lightest CP-even Higgs boson mass, are very
similar for the different priors considered. 

%%%%%%%%%%%%%%%%%%%%%%%%%%%%%%%%%%%%%%%%%%%%%%%%%%%%%%%%%
\subsection{Parameters and sparticle mass posterior PDFs} \label{sub.maps}
%%%%%%%%%%%%%%%%%%%%%%%%%%%%%%%%%%%%%%%%%%%%%%%%%%%%%%%%%

\begin{figure}[!ht] 
  \begin{tabular}{l}
    \includegraphics[angle=0, width=1.\textwidth]{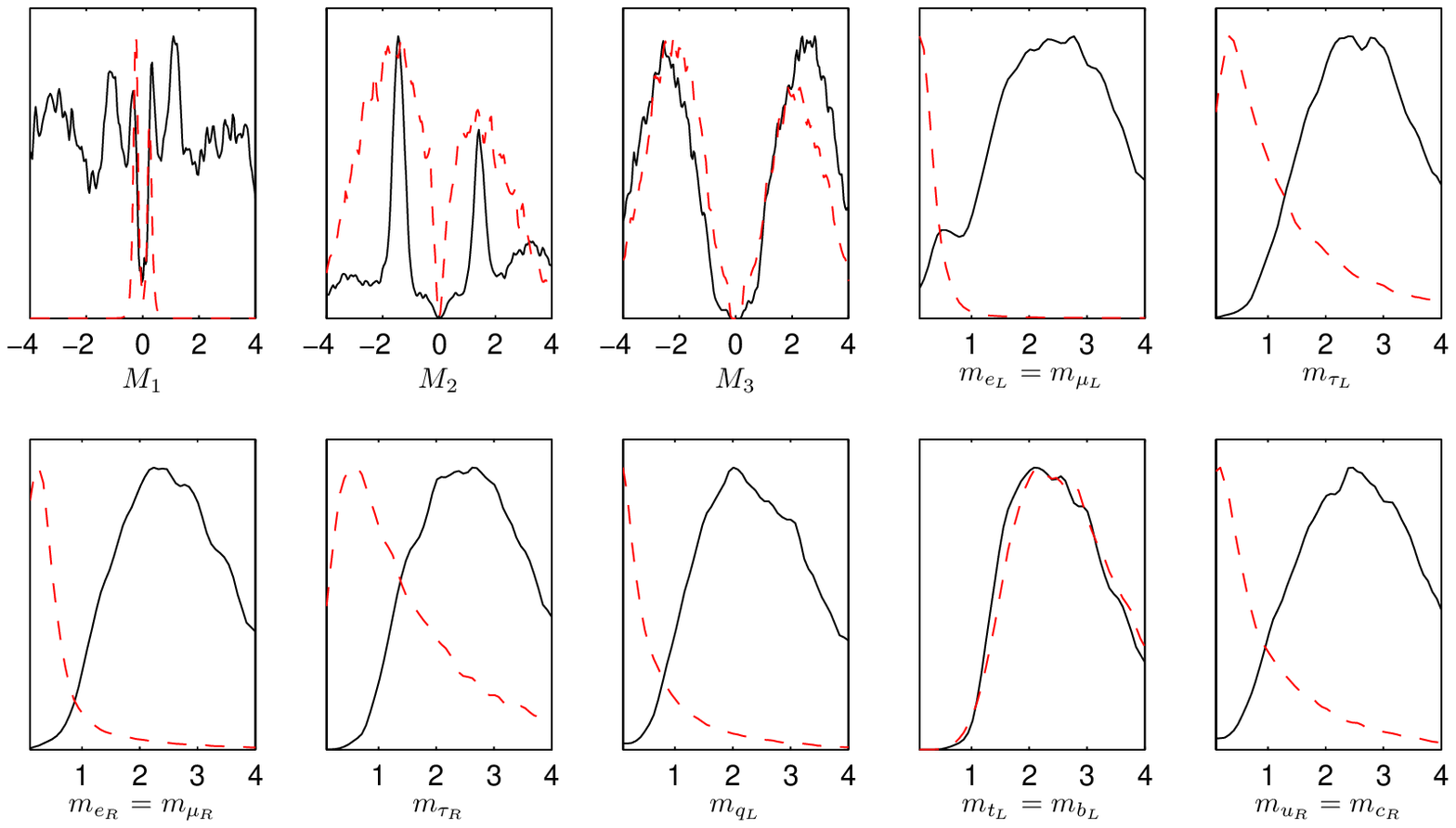} \\
    \includegraphics[angle=0, width=1.\textwidth]{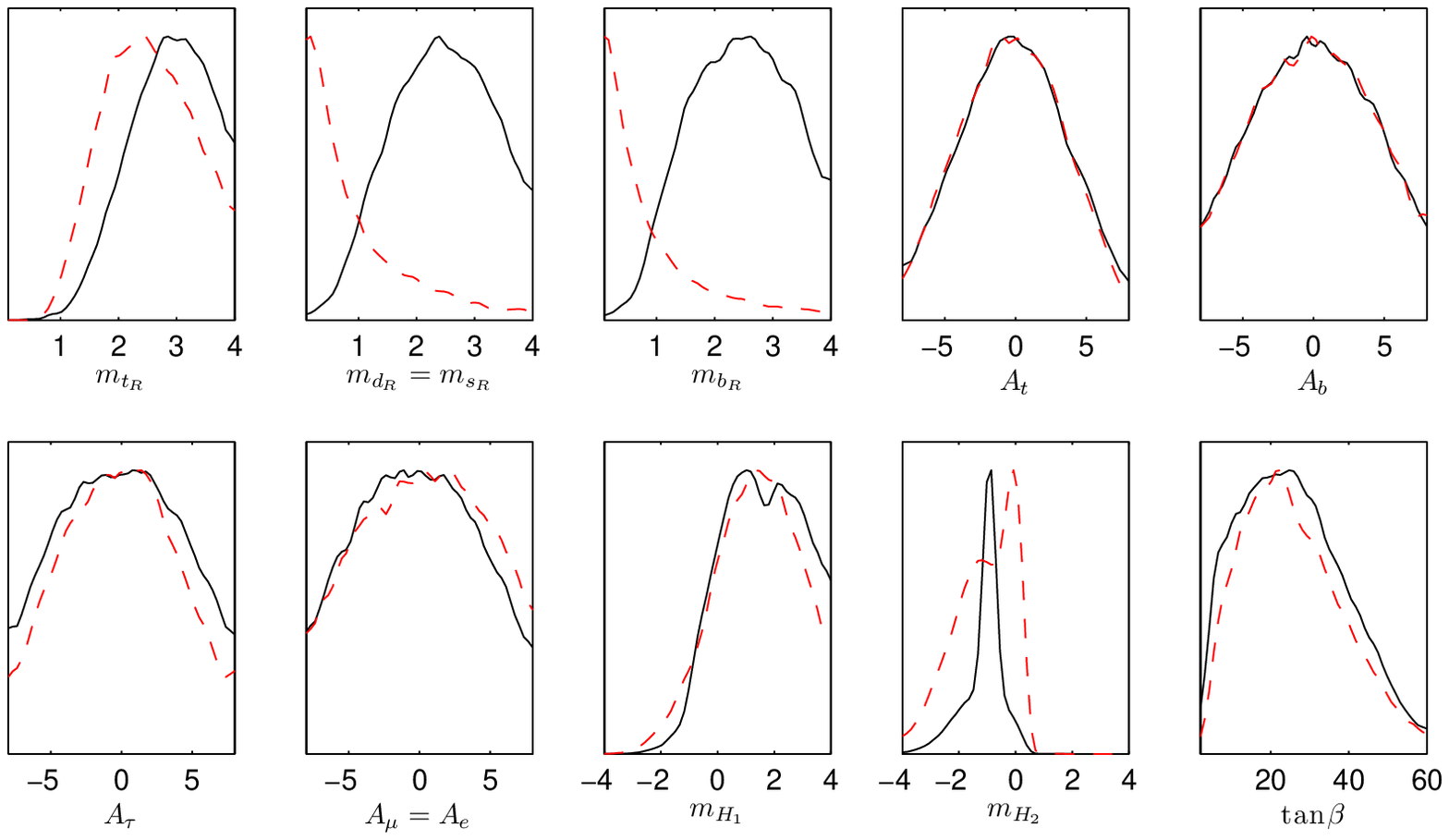}
  \end{tabular}
  \caption{\it{Marginalised 1-dimensional posterior PDFs
    for the pMSSM parameters. The soft-breaking scalar mass terms
    in the log prior scenario (broken lines) are mostly
    reduced with respect to the corresponding mass term in the linear
    prior scenario (solid lines). All masses are in TeV units and
    $m_{q_L} = m_{u_L} = m_{d_L} = m_{c_L} = m_{s_L}.$}}
 \label{fig.post}
\end{figure}

\paragraph{}
Our assumptions about the pMSSM priors are
presented in Section~\ref{sub.pars}. 
The priors are updated using
the set of different experimental measurements explained in
Section~\ref{sub.phenodata} via Bayes' theorem (see
Section~\ref{sub.bayes}). A measure of the amount of information
in the likelihood can be seen from  the results of the sampling procedure
in the form of posterior PDFs. One dimensional marginalised
posterior PDFs of the parameters are shown in
Fig.~\ref{fig.post}. 
Most of the scalar mass terms in the log prior scenario are
much reduced with respect to the corresponding mass term in the
linear prior scenario, as expected. 
Changes to the scalar mass priors (in the log prior case) do change the
posterior PDFs of $M_1$ and $M_2$. This is because the dark matter likelihood
is driving the fit, and dark matter co-annihilation requires the mass of 
the lightest neutralino (controlled by the smaller of $M_1$ and $M_2$) to be
close to that of the scalar that it is co-annihilating with
(see Subsection~\ref{sub.coann} for discussion on the co-annihilations
in the pMSSM). On the other hand, $M_3$ shows approximate prior
independence. For linear 
priors, upper bounds on the parameters come only from our constraint upon the
prior range, there not being sufficient power in the data yet to constrain the
parameters from becoming very large. 
$M_1$, $M_2$ and $M_3$
have PDFs that tend to zero as the parameter tends to zero, but a finite
bin-size means that this may not be evident in plots. 
Since $m_{H_1}$ and $m_{H_2}$ may take negative values, no logarithm was applied
to them in the log priors case. Thus the apparent prior independence of
$m_{H_1}$ is no surprise. The $A$ terms show a peak structure because if their
magnitudes are too large, the squark mass becomes tachyonic and thus
disallowed. 
Radiative corrections to the lightest CP-even higgs mass involving the
stops 
imply that they must be quite heavy, above $\sim 2$ TeV in
order to push the 
higgs mass above its direct search bound. Thus, the log prior only disfavours
points at $m_{{\tilde t}_L}=m_{{\tilde b}_L}=4$ TeV by a factor of 1/2
compared to the lower values. In contrast, other sfermion masses which may be
as low as 200 GeV obtain a log prior suppression of 1/20 at 4 TeV masses,
i.e.\ the difference between log and linear priors is much more evident in
that case. 
Thus there is no large enhancement
of the posterior for small $m_{{\tilde t}_L}=m_{{\tilde b}_L}$, which shows
approximately identical posterior PDFs for the two different prior choices.  

\begin{figure}[!ht] 
  \begin{tabular}{l}
    \includegraphics[angle=0, width=1.\textwidth]{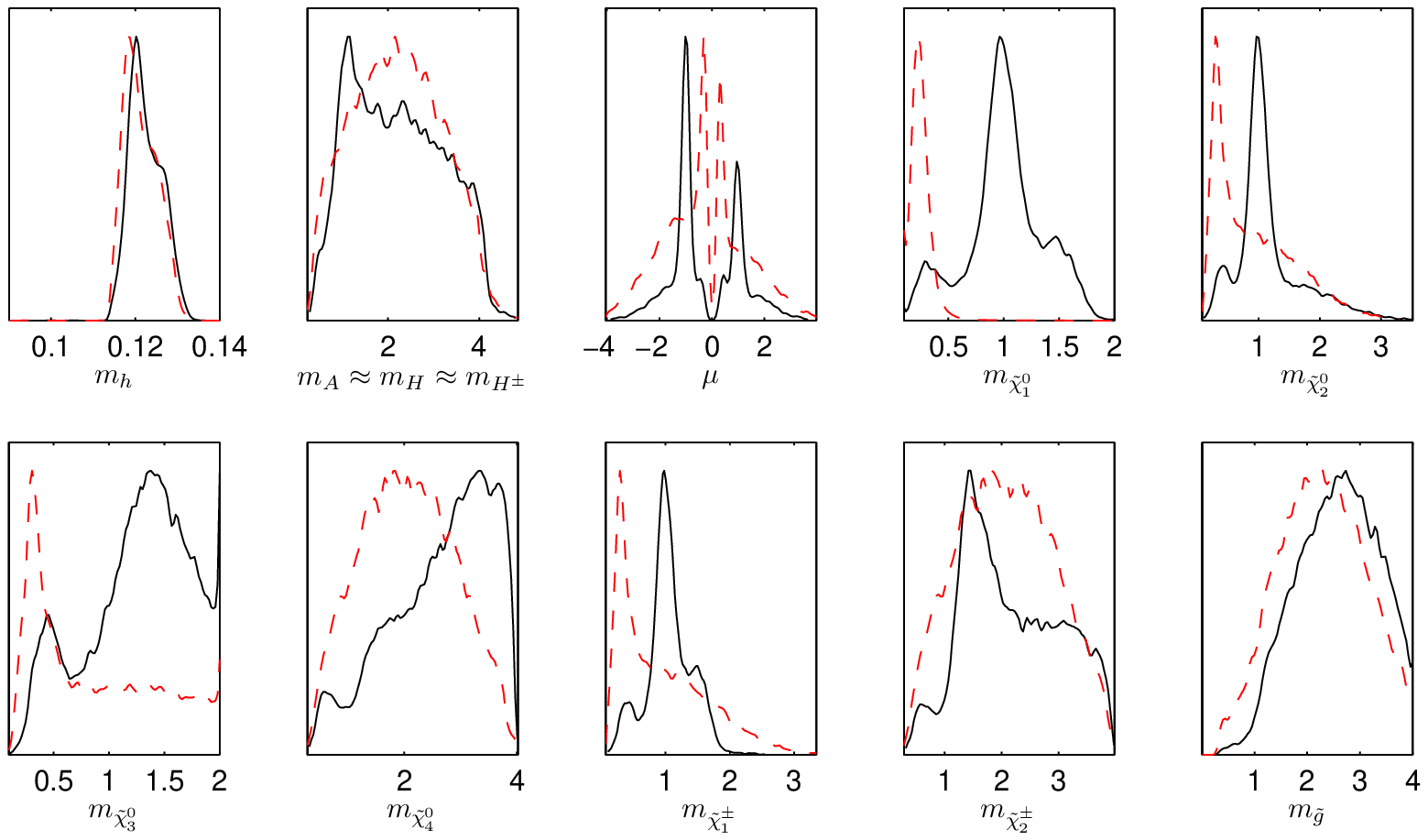} \\
    \includegraphics[angle=0, width=1.\textwidth]{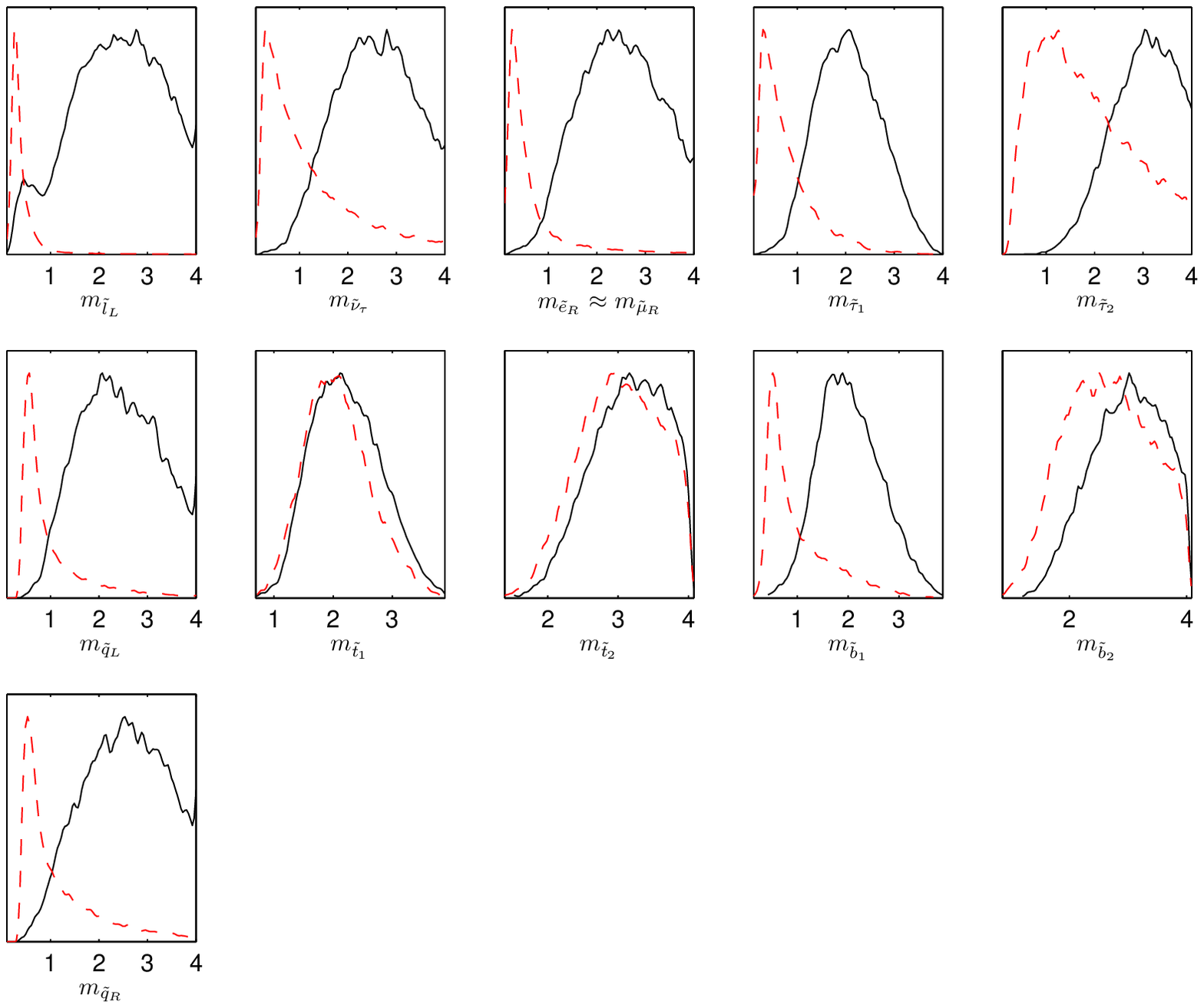}
  \end{tabular}
  \caption{\it{Marginalised 1-dimensional posterior PDFs
  for the $\mu$ parameter and pMSSM sparticle masses, all in TeV
  units, for log priors (broken lines) and linear priors (solid
  lines). 1st and 2nd generation left-handed sleptons, left-handed
  squarks and right-handed squarks are collapsed into single
  parameters, $m_{\tilde{l}_L}$, $m_{\tilde{q}_L}$ and
  $m_{\tilde{l}_R}$ respectively. 
}}  
 \label{fig.smasses}
\end{figure}

\paragraph{}
We display the posterior PDFs of the pole sparticle masses 
in Fig.~\ref{fig.smasses} 
along with the posterior PDF of the $\mu$
parameter; and the 68$\%$ and 95$\%$ Bayesian credibility ranges for the
sparticle masses in Tab.~\ref{tab.mranges}. The mass 
distributions can be understood from the 
parameters PDFs discussed above, since there is a rough one to one
correspondence between the tree-level mass and a mass parameter for many of
the sparticles. 
For the third family sparticles, the $A$-terms and $\mu$
parameter contribute via the large mixing, which is proportional to the
analogous SM fermion mass. The first and second generation sfermion masses
are approximately degenerate, 
since the degeneracy
is only broken by terms of order the second generation fermion mass, 
negligible compared to the sfermion masses. 
We see approximate
prior independence 
of $m_{\tilde{t}_1}$, $m_{\tilde{t}_2}$, $m_{\tilde{g}}$,
$m_{\tilde{b}_2}$ and $m_A \approx m_H \approx m_{H^\pm}$. Other sfermion
masses show the expected log prior suppression at high masses. 

\begin{table}[htbp!] 
\linespread{1.5}
\begin{center}{\begin{tabular}{|c|cc|cc|}
\hline
 & \multicolumn{2}{c|}{pMSSM fit (linear)}    &
 \multicolumn{2}{c|}{pMSSM fit (log)} \\
Mass [TeV] & $68$\% interval & $95$\% interval & $68$\% interval & $95$\% interval\\ 
\hline
$m_h$ & $(0.120,0.124)$ & $(0.117,0.129)$ & $(0.119,0.123)$ & $(0.118,0.128)$  \\
$m_A$ & $(0.111,0.338)$ & $(0.608,0.405)$ & $(1.129,3.284)$ & $(0.509,3.996)$\\
$m_{\tilde{g}}$ & $(2.192,4.000)$ & $(1.280,4.000)$ & $(1.844,4.000)$ & $(0.924,4.000)$\\
\hline 
$m_{\tilde{\chi}_1^0}$& $(0.899,1.125)$ & $(0.302,1.599)$ & $(0.100,0.273)$ & $(0.100,0.397)$\\
$m_{\tilde{\chi}^0_2}$& $(0.772,1.687)$ & $(0.322,2.630)$ & $(0.301,1.675)$ & $(0.176,2.491)$\\
$m_{\tilde{\chi}^0_3}$& $(0.126,2.000)$ & $(0.446,2.000)$ & $(0.539,2.000)$ & $(0.237,2.000)$\\
$m_{\tilde{\chi}^0_4}$& $(0.165,3.571)$ & $(0.649,3.877)$ & $(1171,2.967)$ & $(0.585,3.631)$\\
$m_{\tilde{\chi}_1^\pm}$ & $(0.765,1.428)$ & $(0.315,1.730)$ &$(0.290,1.692)$ & $(0.162,2.532)$\\
$m_{\tilde{\chi}_2^\pm}$ & $(1.335,3.217)$ & $(0.648,3.771)$ &$(1.174,2.981)$ & $(0.586,3.656)$\\
\hline
$m_{\tilde{l}_L}$& $(1.901,4.000)$ & $(0.664,4.000)$ & $(0.279,0.398)$ &$(0.207,0.922)$\\
$m_{\tilde{l}_R}$& $(2.039,4.000)$ & $(1.155,4.000)$ & $(1.000,0.606)$ &$(1.000,2.196)$\\
$m_{\tilde{\nu}_{\tau}}$& $(2.124,4.000)$ & $(1.230,4.000)$ & $(0.588,1.392)$ &$(0.256,3.149)$\\
$m_{\tilde{\tau}_1}$& $(1.718,2377)$ & $(1.048,3.188)$ & $(1.000,0.884)$ &$(1.000,1.965)$\\
$m_{\tilde{\tau}_2}$& $(2686,4.000)$ & $(1.832,4.000)$ & $(1.166,4.000)$ &$(0.502,4.000)$\\
 \hline
$m_{\tilde{q}}$ & $(1.965,4.000)$ & $(1.143,4.000)$ & $(0.590,0.987)$ &$(0.434,2.573)$\\
$m_{\tilde{t}_1}$ & $(1.634,2.841)$ & $(1.233,3.423)$ & $(1.529,2.633)$ &$(1.105,3.271)$\\
$m_{\tilde{t}_2}$ & $(2.591,3.710)$ & $(2.078,3.968)$ & $(2.424,3.644)$ &$(1.912,3.949)$\\
$m_{\tilde{b}_1}$ & $(1.432,2.679)$ & $(0.952,3.320)$ & $(0.478,1.695)$ &$(0.347,2.701)$\\
$m_{\tilde{b}_2}$ & $(2.319,3.637)$ & $(1.746,3.967)$ & $(1.891,3.452)$ &$(1.354,3913)$\\
 \hline
\end{tabular}}
\end{center}
\caption{\linespread{1.} \it pMSSM sparticle mass ranges corresponding to $68$\% and $95$\%
Bayesian credibility. $m_{\tilde{l}_{L(R)}}$ represents mass of 1st and 2nd
generation left-handed(right-handed) sleptons. \label{tab.mranges}} 
%\label{tab.mlimits}
\end{table}

\paragraph{}
The MSSM lightest CP-even higgs mass at tree level is,
\be
m_h = m_Z | \cos 2 \beta |,
\ee
but it
receives large radiative corrections from third generation particles, of order
30$\%$ of its mass. Since the $\tan \beta$ posterior is similar for both
priors, then the tree-level value of $m_h$ will also be similar. 
An additional small prior  dependence comes mainly from $m_{{\tilde t}_R}$
and $m_{{\tilde b}_R}$ prior dependence, but really the model itself
constrains the higgs masses to be largely prior independent.
The approximate mass degeneracy in the heavy Higgs masses and little
dependence on priors can be seen to originate from the relationships
between them. 
The tree-level MSSM pseudo-scalar CP odd higgs mass is given by
\be \label{higgsms}
m^2_A = 2|\mu|^2 + m^2_{H_1} + m^2_{H_2},
\ee
and it receives quite large radiative corrections. 
There is only a small prior dependence of $m_A$, and 
Eq.~\ref{higgsms} shows that the approximate prior independence
of $m_A$ is something of an accident since $m^2_{H_2}$ and $\mu$ show some
prior dependence, but this largely cancels in its effect on $m_A$. Notice that
\bea
m^2_{H} &=& \half \left(m^2_A + m^2_Z + \sqrt{ (m^2_A - m^2_Z)^2 + 4
  m^2_Z m^2_A \sin^2(2 \beta)} \right)  \nonumber \\ 
m^2_{H^\pm} &=& m^2_A + m^2_W.
\eea
and since $m_A$ is
usually far greater than $m_W$ and $m_Z$ then $m_A \approx m_H \approx
m_{H^\pm}$ holds, although loop corrections contribute to a small
non-degeneracy, which however is not visible to the eye. 

%%%%%%%%%%%%%%%%%%%%%%%%%%%%
\subsection{Good-fit points} \label{sub.best}
%%%%%%%%%%%%%%%%%%%%%%%%%%%%

\paragraph{}
For a given prior, the point with the highest likelihood is termed the
best-fit point. We estimate a `good-fit point' by picking the point with the
highest likelihood out of the many millions sampled by the {\sc MultiNest}
algorithm. We shall see below that our under constrained fits exhibit 
parameter degeneracies, and so we should not expect the particular values of
the good-fit parameters to be unique. 
Monte Carlo sampling methods are also ill suited to finding best-fit points,
which are much better determined by hill climbing algorithms, for example.
On the other hand, the properties of such a point are
instructive to view in order to get a feel for which observables are pulling
in which direction. 
The values  of the parameters
 at the good-fit points  are given in
Tab.~\ref{tab.bestparam}. 
The good-fit point of the linear prior has quite large SUSY breaking mass
parameters, of the order of a couple of TeV in most cases, and large $\tan
\beta$. On the other hand, the log prior sampling found a good-fit point with
a mixture of sub-TeV and several TeV masses, because it spent more time
exploring points with lower sfermion masses. It also has a moderate value of
$\tan \beta$. 
Interestingly, both good-fit
points have a negative $\mu$ parameter. This is in contrast to the CMSSM case,
where the anomalous magnetic moment of the muon prefers positive $\mu$. 
We shall discuss the sign of $\mu$ further in
Subsection~\ref{sub.signmucomp} below.  
\begin{table}[htbp!] 
{\linespread{1.5}
\begin{center}{\begin{tabular}{|clcc|}
\hline
Parameter & Description & Linear prior fit & Log prior fit\\
\hline
$M_1$               & Bino mass      & -2947 & -250\\
$M_2$               & Wino mass        & -1297 & -3017\\
$M_3$               & Gluino mass       & -2397 & -642\\
$m_{\tilde e_L} = m_{\tilde \mu_L}$    & 1st/2nd gen. $L_L$
slepton & 1040 & 174\\  
$m_{\tilde \tau_L}$  & 3rd gen. $L_L$ slepton & 2640 & 993\\   
$m_{\tilde e_R} = m_{\tilde \mu_R}$    & 1st/2nd gen. $E_R$
    slepton & 2301 & 201\\  
$m_{\tilde \tau_R}$  & 3rd gen. $E_R$ slepton & 3748 & 3530\\   
$m_{\tilde u_L} = m_{\tilde d_L} = m_{\tilde c_L} = m_{\tilde s_L}$ &
    1st/2nd gen. $Q_L$ squark   & 878 & 165 \\   
$m_{\tilde t_L} = m_{\tilde b_L}$ & 3rd gen. $Q_L$ squark
    & 2301& 2321\\
$m_{\tilde u_R} = m_{\tilde c_R}$ & 1st/2nd gen. $U_R$ squark
    & 3027 & 1515\\  
$m_{\tilde t_R}$ & 3rd gen. $U_R$ squark   & 2618  & 2905\\  
$m_{\tilde d_R} = m_{\tilde s_R}$ & 1st/2nd gen. $D_R$ squark
    & 1368 & 329\\  
$m_{\tilde b_R}$ & 3rd gen. $D_R$ squark  & 1054& 1268\\  
$A_t$ & top quark trilinear & -1963 & 651\\ 
$A_b$ & b-quark trilinear  & -3541 & 5727\\
$A_\tau$ & $\tau$-quark trilinear & 4725 & 3196\\
$A_e = A_\mu$ & $\mu$-quark trilinear & 2154 & 2951\\
$m_{H_1}$ & down-type Higgs doublet & 2548 & 3445\\
$m_{H_2}$ & up-type Higgs doublet & 882  & 669\\
$\tan \beta$ & Higgs vevs ratio & 52.0  & 21.0\\ 
$m_t$ & top quark mass & 173.4  & 175.3\\
$m_Z$ & Z-boson mass & 91.186  & 91.190\\
$m_b(m_b)^{\overline{MS}}$ & b-quark mass & 4.16  & 4.26\\
1/$\alpha_{em}(m_Z)^{\overline{MS}}$ & e-coupling constant & 127.95 & 127.91\\ 
$\alpha_{s}(m_Z)^{\overline{MS}}$ & s-coupling constant & 0.1168 & 0.1161\\ 
$\mu(M_{susy})$ & higgs parameter & -942  & -770 \\ 
\hline
\end{tabular}}\end{center}
}
\caption{\it{pMSSM input parameters for both good-fit points. Mass parameters
    are in units of GeV. The magnitude of the $\mu$ parameter is a derived
    quantity.}}\label{tab.bestparam}   
\end{table}

We see the resulting MSSM spectra in Tab.~\ref{tab.spectra}, where the log
prior good-fit point has some light sleptons and squarks. For the linear
prior, all of the sparticles have masses around the TeV scale or higher.  
\begin{table}
\begin{tabular}{ll}
\begin{minipage}[t]{6.0cm}
\begin{tabular}{|c|c|c|}
\hline
 & Linear prior& Log prior\\
\hline
$\tilde{e}_L, \tilde{\mu}_L$ & 1062  & 271 \\
\hline
$\tilde{e}_R, \tilde{\mu}_R$ & 2310  & 251 \\
\hline
$\tilde{\tau}_L$ & 2651  & 1033 \\
\hline
$\tilde{\tau}_R$ & 3740  & 3530 \\
\hline
$\tilde{u}_1, \tilde{c}_1$ & 1059  & 384\\
\hline
$\tilde{u}_2, \tilde{c}_2$ & 3067  & 1527 \\
\hline
$\tilde{t}_1$ & 2361 & 2354 \\
\hline
$\tilde{t}_2$ & 2665 & 2903 \\
\hline
$\tilde{d}_1, \tilde{s}_1$ &  1060  & 383 \\
\hline
$\tilde{d}_2, \tilde{s}_2$ &  1465  & 419 \\
\hline
$\tilde{b}_1$ &  1169  & 1296  \\
\hline
$\tilde{b}_2$ &  2367  & 2351 \\
\hline
$\chi_1^0$ &  936  & 243 \\
\hline
$\chi_2^0$ &  947  & 770 \\
\hline
$\chi_3^0$ &  1317  & 781 \\
\hline
$\chi_4^0$ &  2918  & 2864 \\
\hline
$\chi_1^{\pm}$ &  937  & 765 \\
\hline
$\chi_2^{\pm}$ &  1301  & 2916 \\
\hline
$A_0, H_0$ &  2671  & 3529 \\
\hline
$H^{\pm}$  &  2673  & 3531 \\
\hline
$\tilde{g}$ &  2470  & 735 \\
\hline
$\tilde{\nu}_{1,2}$ & 1058 & 255 \\
\hline
$\tilde{\nu}_3$ & 2645  & 1018 \\
\hline
$h$ &  121  & 119 \\
\hline
\end{tabular}
\end{minipage}
&
\begin{minipage}[t]{6.5cm}
   \includegraphics[angle=0,
   width=1.4\textwidth]{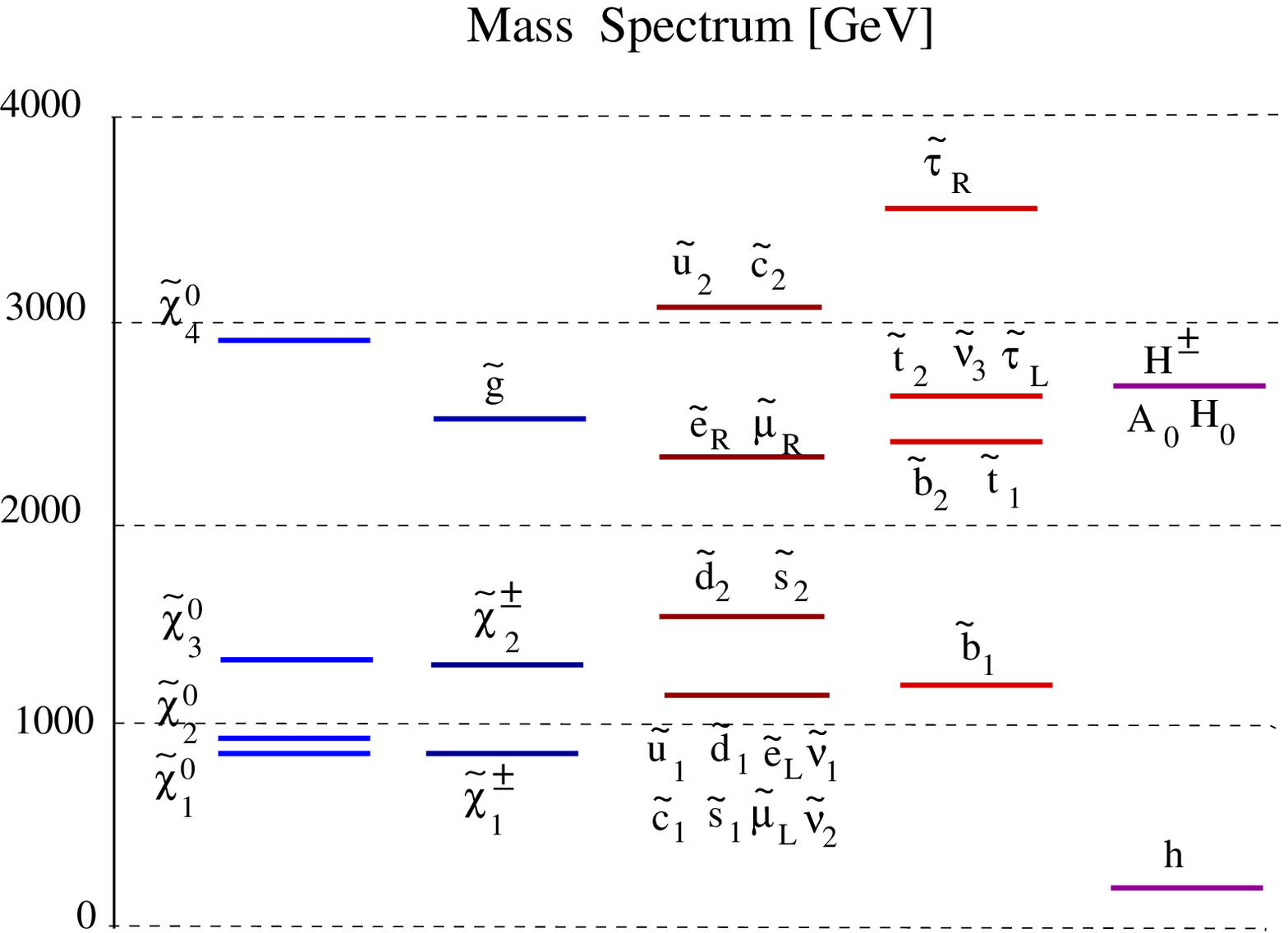}  \\  \\ \\
   \includegraphics[angle=0,
   width=1.4\textwidth]{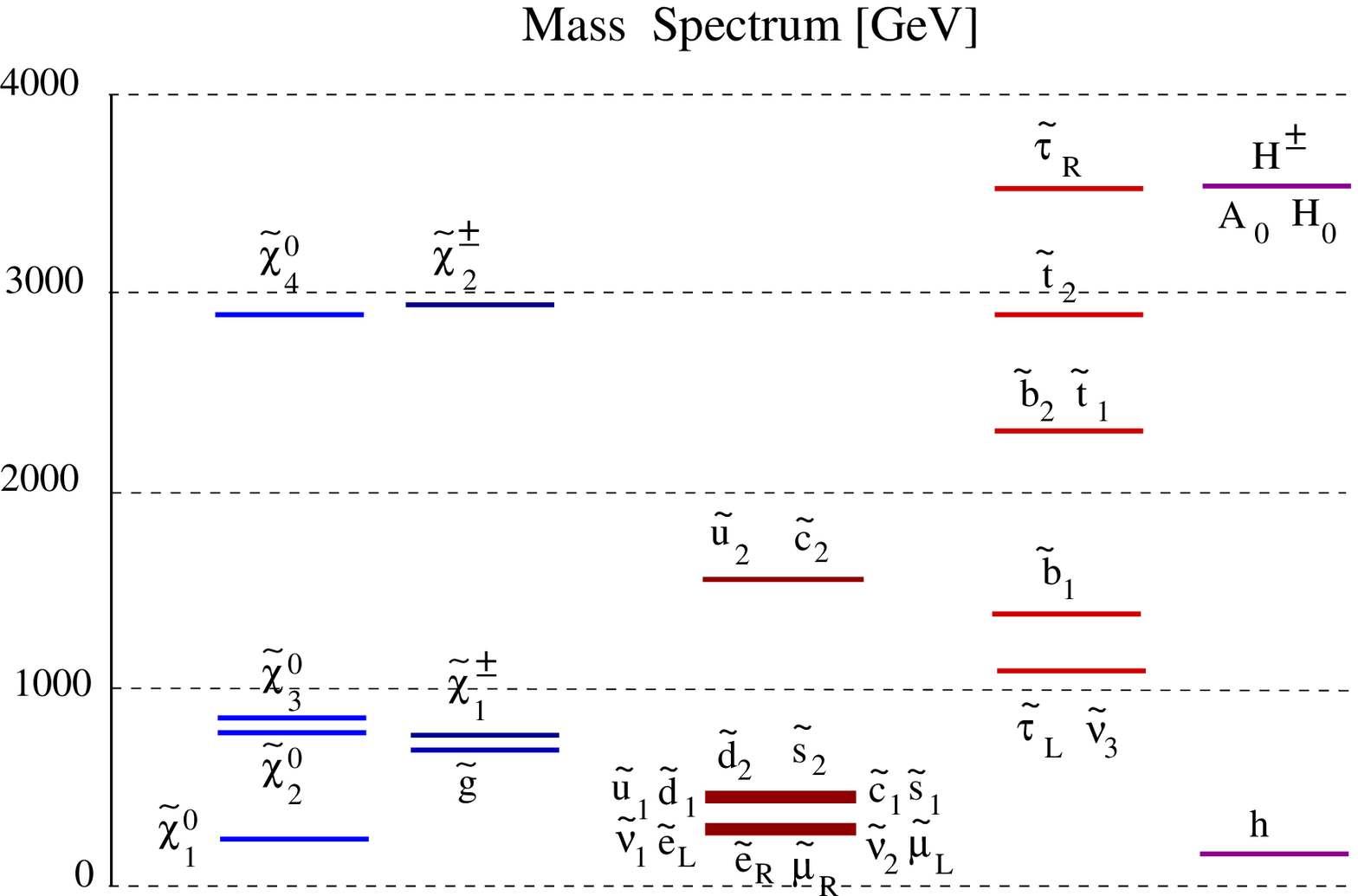}
\end{minipage}
\end{tabular}
  \label{tab.spectra}
  \caption{\it{Tabular and pictorial representation of
  the sparticle masses at the pMSSM good fit point for both linear
  and log priors. All numbers in the table are in GeV
  units. The pictures on the right hand side shows the difference in
  the pattern of sparticle masses for the linear (upper picture) and
  the logarithmic (lower picture) prior case.}} 
\end{table}

\paragraph{}

\begin{figure}[!ht] 
\begin{center}
    \includegraphics[angle=0,
      width=1.13 \textwidth]{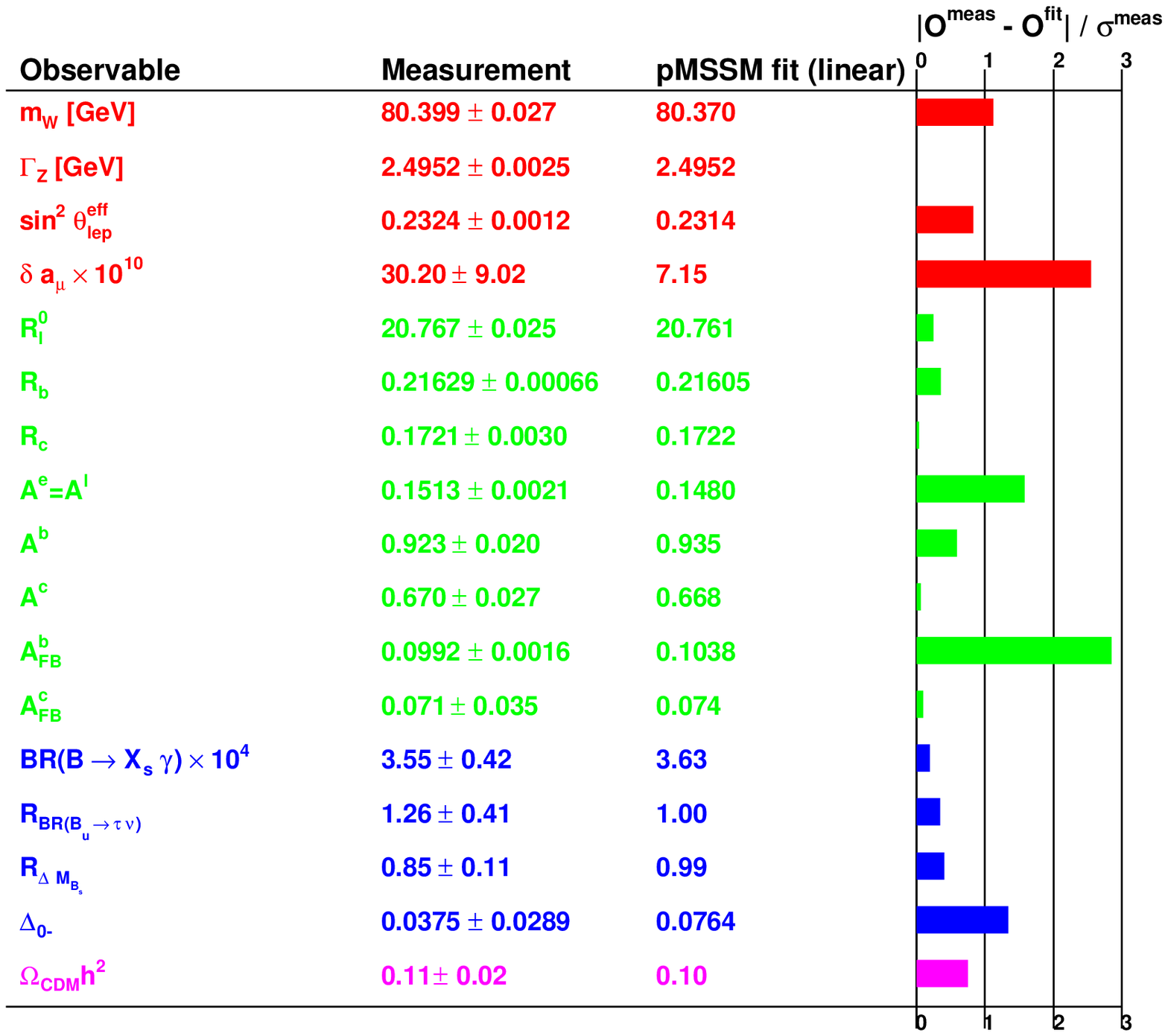} 
\end{center}
 \caption{{\it Statistical pull of various observables at the good-fit point
from the linear prior
 sampling. The good-fit value and the measurement for each observable are
 listed.  
The statistical pull, defined to be the number of sigma the point differs 
 from the experimental central value
is shown
 by the horizontal bars.}}
 \label{fig.ewwglikeN} 
\end{figure}

\begin{figure}[!ht] 
\begin{center}
    \includegraphics[angle=0,
      width=1.13 \textwidth]{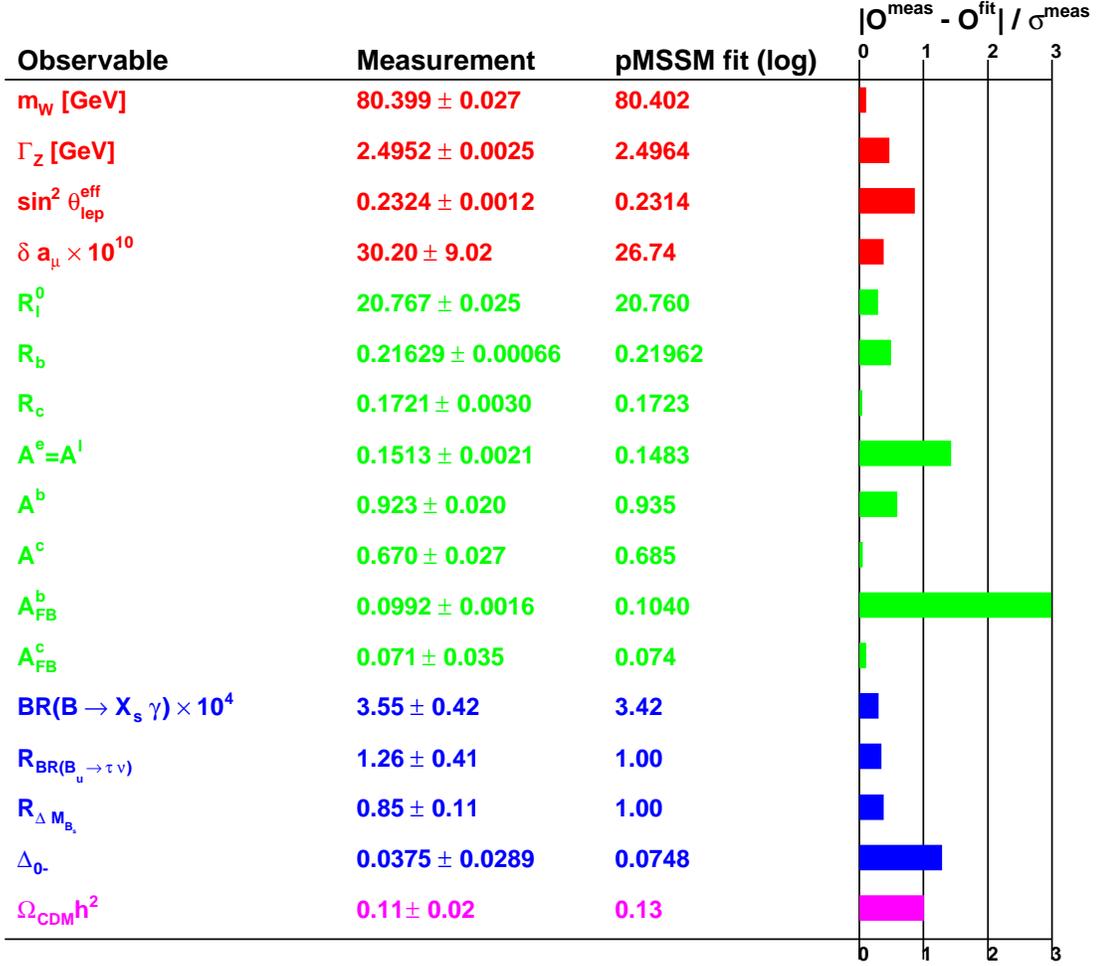} 
\end{center}
 \caption{{\it Statistical pull of various observables at the good-fit point
from the log prior
 sampling. The good-fit value and the measurement for each observable are listed. 
The statistical pull, defined to be the number of sigma the point differs 
 from the experimental central value
is shown
 by the horizontal bars.
}} 
\label{fig.ewwglikeL}
\end{figure}

The statistical pulls of the various observables are shown in
Figs.~\ref{fig.ewwglikeN} and \ref{fig.ewwglikeL}. We see that the observable
that most 
discriminates between our two good-fit points is the anomalous magnetic moment
of the muon, which is much better fit for the log prior point, where it
receives large corrections from lighter slepton and gaugino masses. $m_W$ and
$\Gamma_Z$ also show a large difference between the two points, whereas most
of the other observables display only a small difference in statistical pull
between these two points. In fact, the log prior good-fit point has a
significantly lower $\chi^2$ value (and therefore, a better likelihood). 
Were nested sampling perfect for finding the good fit point, the 
likelihood (or equivalently, the total $\chi^2$)
values would be the same for both priors. This difference is therefore due to
the finite sampling resolution of the posterior distributions. 

\FloatBarrier

%%%%%%%%%%%%%%%%%%%%%%%%%%%%%%
\subsection{Observables' PDFs} \label{sub.observ}
%%%%%%%%%%%%%%%%%%%%%%%%%%%%%%

\begin{figure}[!ht] 
  \begin{tabular}{l}
    \includegraphics[angle=0, width=1.\textwidth]{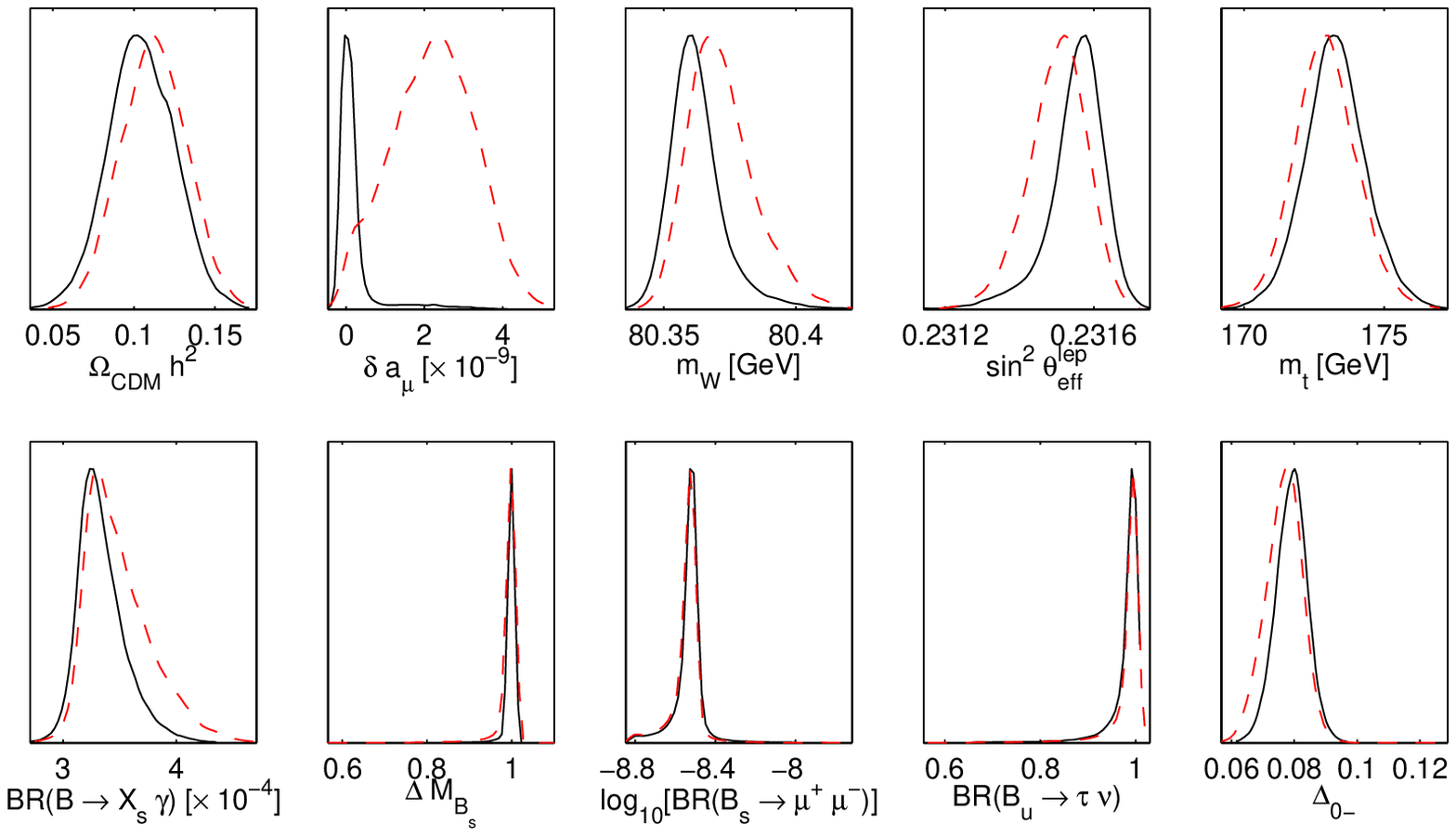} \\
    \includegraphics[angle=0, width=1.\textwidth]{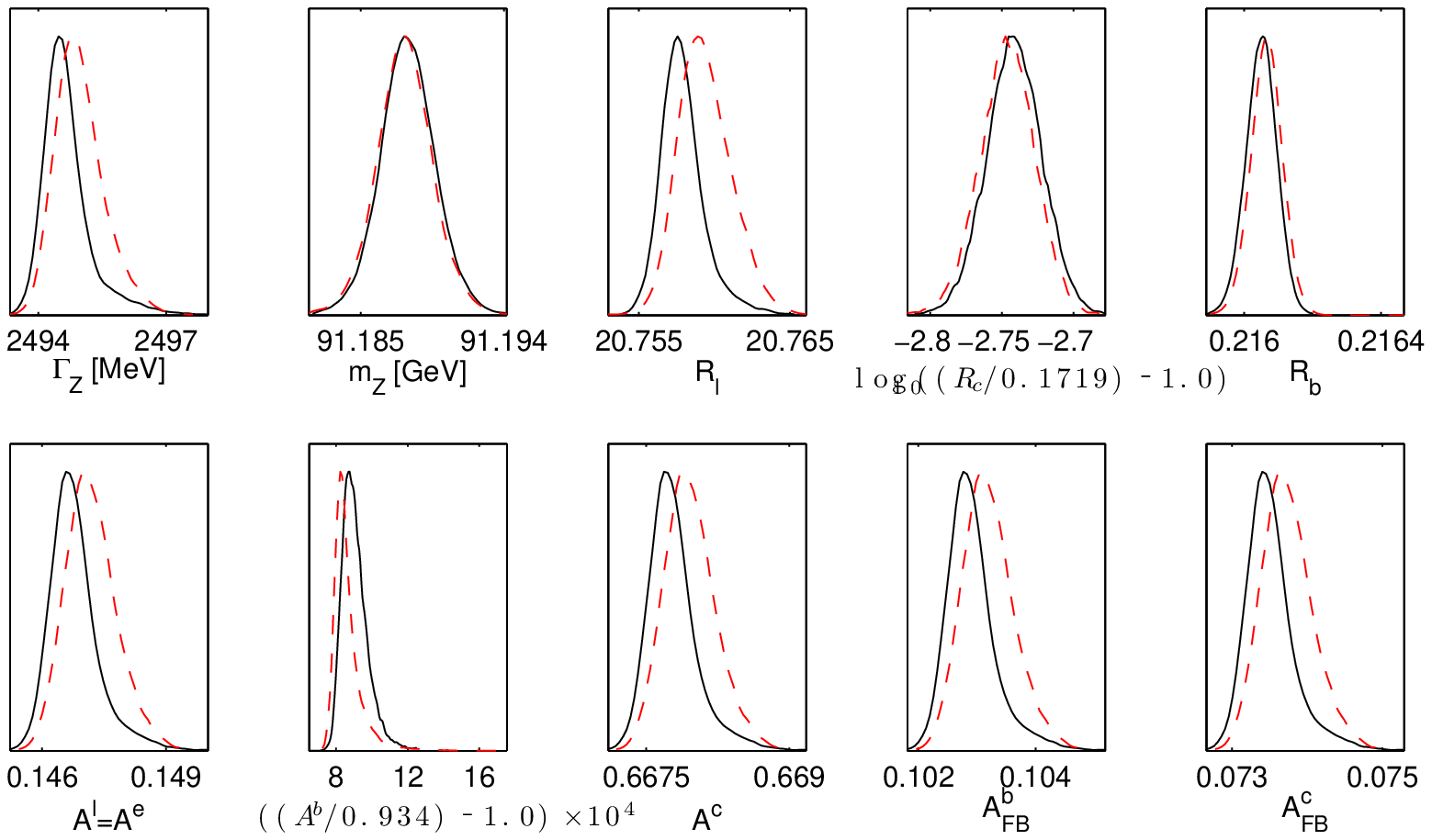}
  \end{tabular}
  \caption{\it{Posterior PDF of some of the
  observables used to constrain the pMSSM\@. Solid
  (broken) lines represents plots for the linear (log) prior
  case.}}
 \label{fig.observ}
\end{figure}

\paragraph{}
The posterior PDFs for the 
observables used to constrain the pMSSM are given in
Fig.~\ref{fig.observ}. Differences between the two prior cases are mostly
due to the fact that the sparticle mass PDFs are larger in the linear prior,
leading to a suppression of SUSY effects in the loops of most observables. 
The $B$-physics observables tend to have similar posterior PDFs in the two
prior cases. In most of the
EWPO, there are larger differences in the
posteriors. 
For $\delta a_\mu$, there is a particularly large difference between
the central values of the linear and log prior
PDFs. The leading one-loop gaugino contribution at large $\tan \beta$ is given
by~\cite{Moroi:1995yh}
\be
\delta a_{\mu} \approx
\frac{m_{\mu}^2 \mu \tan \beta}{16 \pi^2}\left( g_1^2 M_1 F_1 + 
g_2^2 M_2 F_2 \right), \label{g-2}
\ee
where $F_1$ and $F_2$ are positive loop functions proportional to
$m_{susy}^{-4}$ for the case of degenerate sparticles in the loops. 
The dominant contributions coming from
gauginos and sleptons therefore lead to an enhanced value of $\delta
a_\mu$ when they are lighter, as is evident for the log prior fits. 
As we shall see in Section~\ref{sub.coann} and Tab.~\ref{decay},
the linear prior fits prefer Higgsino exchange to be the dominant LSP
annihilation process, as opposed to 
slepton co-annihilation in the log prior case. This occurs at heavier
neutralino LSP masses, and hence heavier smuon masses (which are always
constrained to be heavier than the neutralino LSP). 
$\delta a_\mu$ is then relatively badly fit as can be
seen in the good fit point example where the statistical pull is more than
$2\sigma$.

%%%%%%%%%%%%%%%%%%%%%%%%%%%%%%%%%%%
\subsection{Sign($\mu$) comparison} \label{sub.signmucomp}
%%%%%%%%%%%%%%%%%%%%%%%%%%%%%%%%%%%

\paragraph{}
The posterior PDFs in Fig.~\ref{fig.smasses}
indicates that the pMSSM prefers $\mu < 0$ compared to $\mu > 0$. This
is interesting since in the previous studies of CMSSM $\mu > 0$
was seen to be preferred by the combination of $BR(b\rightarrow s \gamma)$ and
$\delta a_\mu$. One of the statistical tests (a predictive likelihood ratio
test) in Ref.~\cite{Feroz:2009dv} found
that the two measurements are incompatible in the CMSSM, but the other found
no strong evidence for this and so the final conclusion of the analysis
remains unclear. 
The sign of the SUSY contribution to $BR(b\rightarrow s
\gamma)$ is dependent upon the sign of $\mu$. There are two dominant SUSY
contributions to consider: 
the first comes from diagrams involving a charged Higgs boson and up-type
quarks. The second, involving
a chargino and up-type squarks, depends upon the sign of the product $A_t
\mu$. Eq.~\ref{expsg} indicates that there is a preference for a positive
total contribution at the 1-$\sigma$ level. In the CMSSM, $A_t$ is typically
negative due to RGE effects.
Eq.~\ref{g-2} shows that the sign of the non-SM contribution to the
muon anomalous magnetic moment depends upon the sign of $\mu M_1$ and $\mu
M_2$. In the CMSSM, $M_1$ and $M_2$ are positive and so the combination of the
$(g-2)_\mu$ and $BR(B \rightarrow X_s \gamma)$ constraints  implies a
preference for a definite sign of $\mu$. 
Bayesian analyses~\cite{0705.0487,0807.4512} demonstrated that the
current statistical evidence for $\mu>0$ in the CMSSM is weaker than many may
pre-suppose. 
\paragraph{}

\begin{figure}[!ht] 
  \begin{tabular}{ll}
    \includegraphics[angle=0,
      width=.5\textwidth]{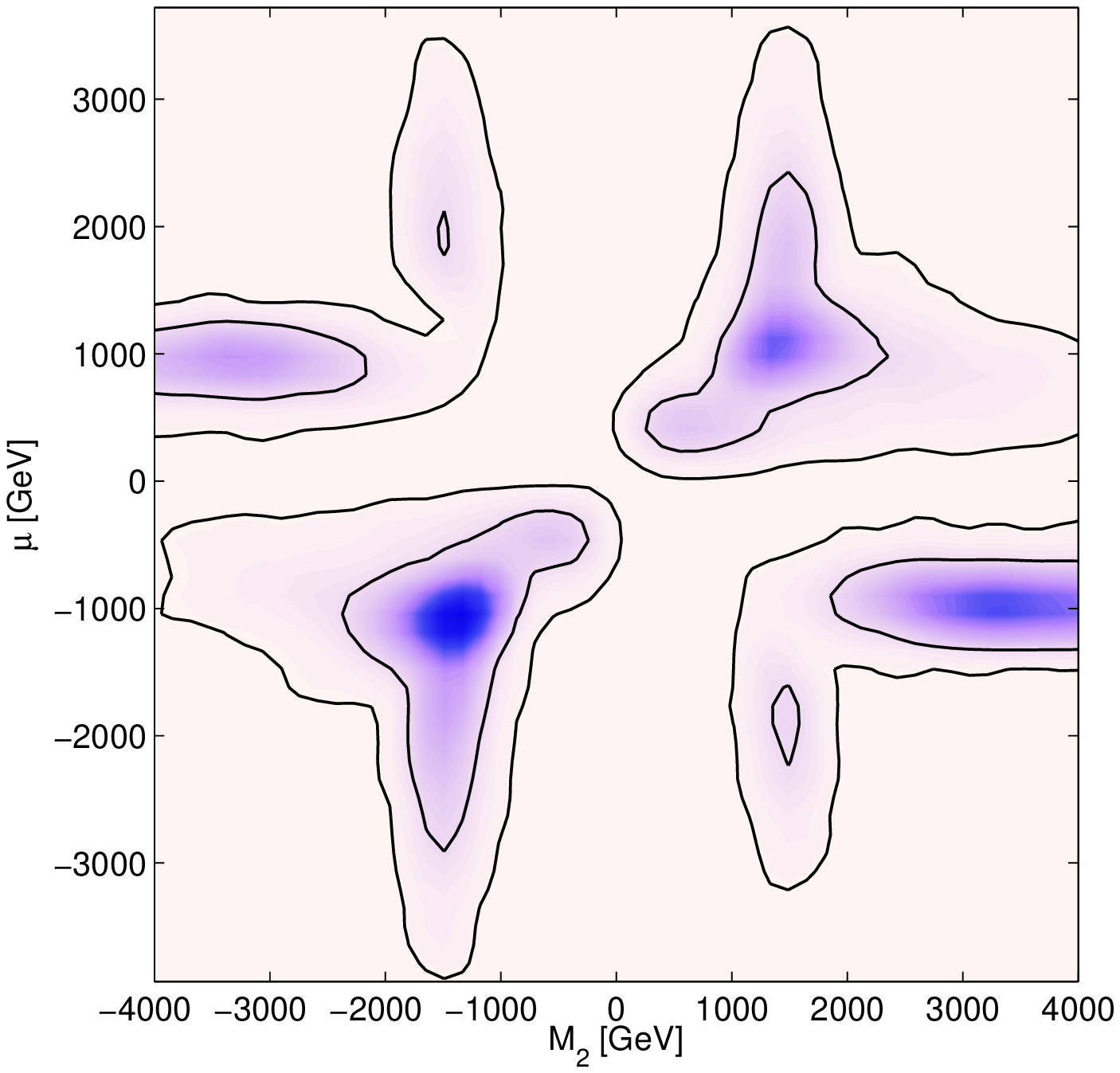} &
    \includegraphics[angle=0,
      width=.5\textwidth]{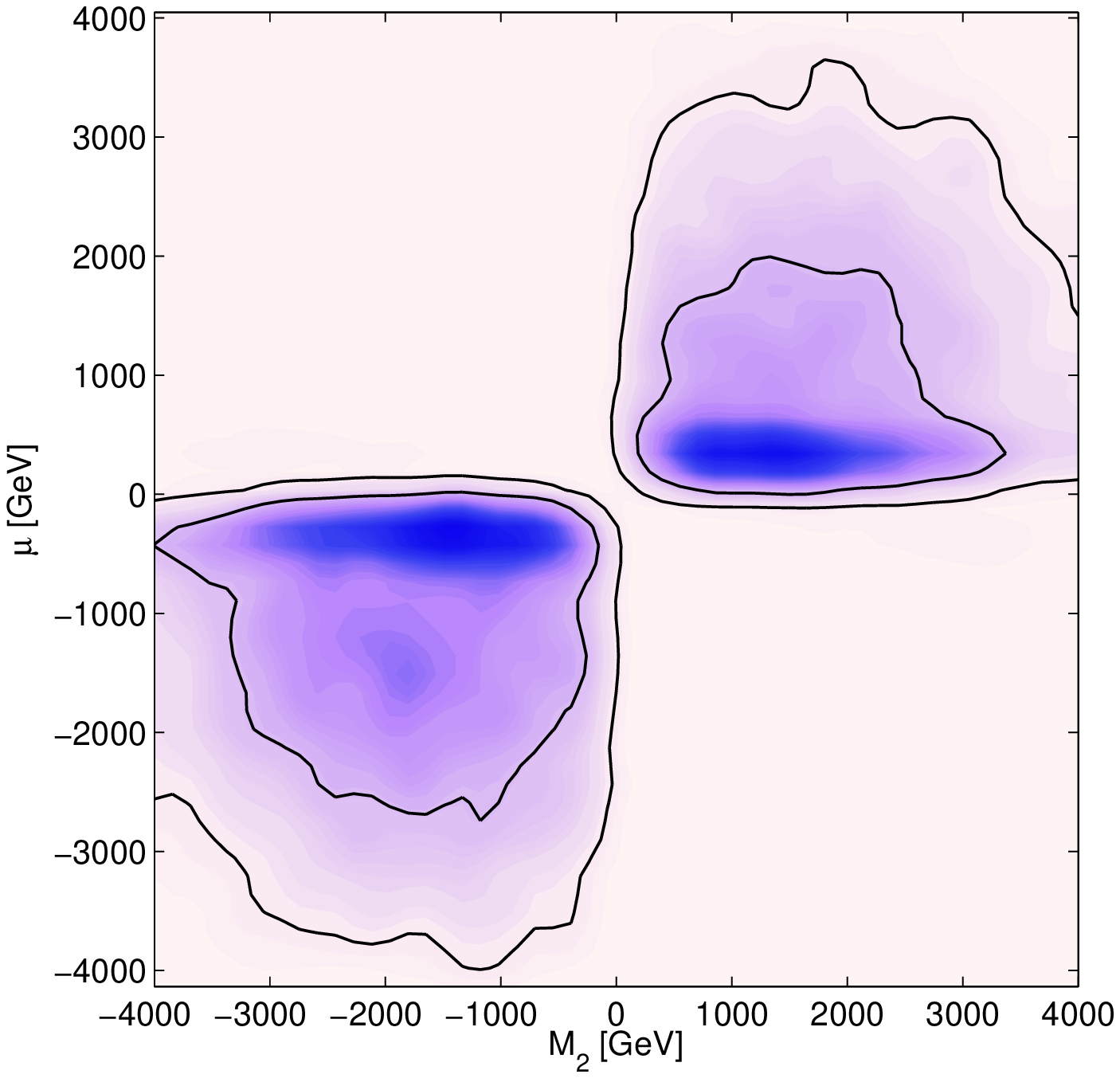} 
  \end{tabular}
 \caption{{\it Marginalised 2D posterior PDF of
 $M_2$ versus $\mu$ for a (left) linear prior and (right) log prior.
The dark contours show the 68$\%$ and 95$\%$ Bayesian credibility regions. 
\label{fig.muM1M2}}} 
\end{figure}

We have not used the freedom to re-define the phases of the fields and
make $M_2$ positive for example, and so negative $M_2$ appears in our
fits. As such, in the pMSSM both $A_t$ and $M_1$, $M_2$ may take
either sign and so the preference for $\mu>0 $ is broken and we may
expect that the dominant contributions to the observables do not
prefer either sign. On the other hand, there may be some residual
dependence from the sub-dominant contributions, as well as
sub-dominant radiative corrections to other observables. Thus it is 
still important to check the relative probabilities for $sign(\mu)$. 
For the pMSSM with a
linear prior measure we find the following probability for the
two signs of $\mu$:
\be 
P( \mu > 0 ) = 0.40 \pm 0.01 \textrm{ and }
P( \mu < 0 ) = 0.60 \pm 0.01.
\ee
Fig.~\ref{fig.muM1M2} shows the posterior PDF marginalised on the $M_2-\mu$
plane. The figure shows that, although opposite signs for $\mu$ are allowed,
it is constrained to have the {\em same sign}\/ as $M_2$, especially for log
priors where $(g-2)_\mu$ is well fit. For linear priors, the large volume of
parameter space leading to heavy sparticle
masses mean that this tendency is reduced and there is a small amount of
probability that $\mu M_2<0$, predicting a small negative $\delta
a_\mu$. Fig.~\ref{fig.muM1M2} clearly shows a symmetry of the fits
when one simultaneously flips the signs of $M_2$ and $\mu$, as should
be the case due to phase re-definition freedom.

The SM prediction of $(g-2)_\mu$ remains somewhat controversial.
The hadronic contributions that are extracted from $\tau$ and $e^+ e^-$ data
disagree, and one obtains quite different $\delta a_\mu$ constraints depending
upon which data set is used. 
To quantify the extent to which our mild preference for $\mu < 0$
depends on the $(g-2)_\mu$ observable, we made a separate sampling with all
except the $(g-2)_\mu$ constraints. We found that $\mu < 0$ is still
more probable with:
\be 
P( \mu > 0 ) = 0.46 \pm 0.02 \textrm{ and }
P( \mu < 0 ) = 0.54 \pm 0.02.
\ee
Thus, $(g-2)_\mu$ contributes around 0.06 to the probability of $\mu<0$, the
other observables including $BR(B \rightarrow X_s \gamma)$
contributing around 0.04. 
However, computing the Bayesian evidence ratios in the two scenarios 
indicate that there are no conclusive evidence, based on Jeffrey's
scale (see Tab.~\ref{tab:Jeffreys}), for one particular sign($\mu$)
over the other. The odds and logarithm of the evidence ratios are
summarised in Tab.~\ref{tab.smucomp}. 

\TABULAR{|l|l|l|l|}{\hline
Data Considered & $|\log_e \Delta E|$ & Odds, $Z_{+} / Z_{-}$ & Remark
\\  
\hline
All constraints & $-0.41 \pm 0.04$ & $0.67 \pm 0.03$ & Inconclusive \\
All minus $(g-2)_\mu$ constraints & $-0.18 \pm 0.04$ & $0.84 \pm 0.04$
& Inconclusive \\ 
\hline
}{\it The Bayesian evidence ratios for the pMSSM with $\mu > 0$ and
  $\mu < 0$ with the linear priors measure. ``All observables''
  refers to all the constraints discussed in
  Section~\ref{sub.observ}. $Z_+$ and $Z_-$ represent the evidence for
  the hypothesis for the linear prior pMSSM with $\mu > 0$ and with
  $\mu < 0$ respectively. A prior probability of 0.5 is assigned to
  each hypothesis. \label{tab.smucomp}}

\FloatBarrier
%%%%%%%%%%%%%%%%%%%%%%%%%%%%%
\subsection{The gaugino code} \label{sub.gcode}
%%%%%%%%%%%%%%%%%%%%%%%%%%%%%

\paragraph{}
The LHC being a proton-proton scattering machine, is going to be producing
a large number of strongly interacting particles. If TeV-scale SUSY
is nature's choice of physics then the LHC machine
would eventually be a gluino factory. The gluinos would cascade-decay
down to the neutralino LSP\@. Thus the visible energy of the decay products of
the gluino is determined by the gluino-LSP mass splitting.
If the ratio of these two masses is large, there will be high visible
energy and it should be easier to pick SUSY out from underneath backgrounds.
The nature of the gluino and neutralino
sparticles could also be important in discriminating
the different models of SUSY. As recently emphasised~\cite{nilles},
 patterns of gaugino masses can 
be used as an early discriminant of models of SUSY breaking at the LHC\@.  The
argument for this goes as follows. In, for instance, the CMSSM or minimal
gauge-mediated 
SUSY-breaking, the  boundary conditions of the values of the gauge
coupling constants take the form $g_1^2: g_2^2: g_3^2 \approx 1: 2: 6$
around the TeV scale. Here $g_a$, $a = 1, 2,$ or $3$ respectively
represent the electromagnetic, weak or strong interactions couplings. 
Unification of the gauge couplings at a GUT scale,
gives a
prediction on the pattern of gaugino mass terms since the ratio
$M_a^2/g_a^2$ does not run at one loop~\cite{Martin:1997ns}. If the 
neutralino is gaugino dominated, this 
translates into some gluino-to-neutralino mass ratio pattern.

\paragraph{}
Already there are distinct ratios coming from viable SUSY breaking
scenarios. For instance,  the CMSSM (and AMSB) if it has a 
predominantly bino (and wino) LSP predicts
$m_{\tilde{g}}/m_{\tilde{\chi}^0_1} \approx 6$ (and 9,
respectively)~\cite{nilles}. Mirage mediation~\cite{Lebedev:2005ge}
with predominantly bino LSP and the large volume scenario
(LVS~\cite{Balasubramanian:2005zx,Conlon:2005ki,Conlon:2006wz,Conlon:2007xv})
have a characteristic ratio less than 6 and between 
3 to 4, respectively (with the LVS having the most compact gaugino
mass pattern). 
Higgsino components of the neutralino LSP spoil a strict prediction of the mass
ratio coming from gaugino mass ratios, however.
By construction, the pMSSM set-up is the most
generic and natural approach for MSSM phenomenology. We show the
pMSSM posterior PDF for the gluino-to-neutralino mass
ratio in Fig.~\ref{fig.mgmc} which provides a clear discrimination
between the pMSSM and the other models. 
The
figure~\ref{fig.mgmc} shows a sharp dependence of the gluino-to-neutralino
mass ratio 
posterior PDF on the choice of prior. The linear prior 
predicts a compact gaugino mass ratio with\footnote{See~\cite{AbdusSalam:2008uv}
 for similar results from a pMSSM analysis done with a different set of
 observables than in this work.}
$m_{\tilde{g}}/m_{\tilde{\chi}^0_1}  
\approx 2.5 \pm 1.0$. 
For the log prior case a much broader distribution of 
mass ratios centred around $m_{\tilde{g}}/m_{\tilde{\chi}^0_1} 
\approx 10.0$ results from the fits. 

\FIGURE[!ht]{
\includegraphics[width=0.6\textwidth]{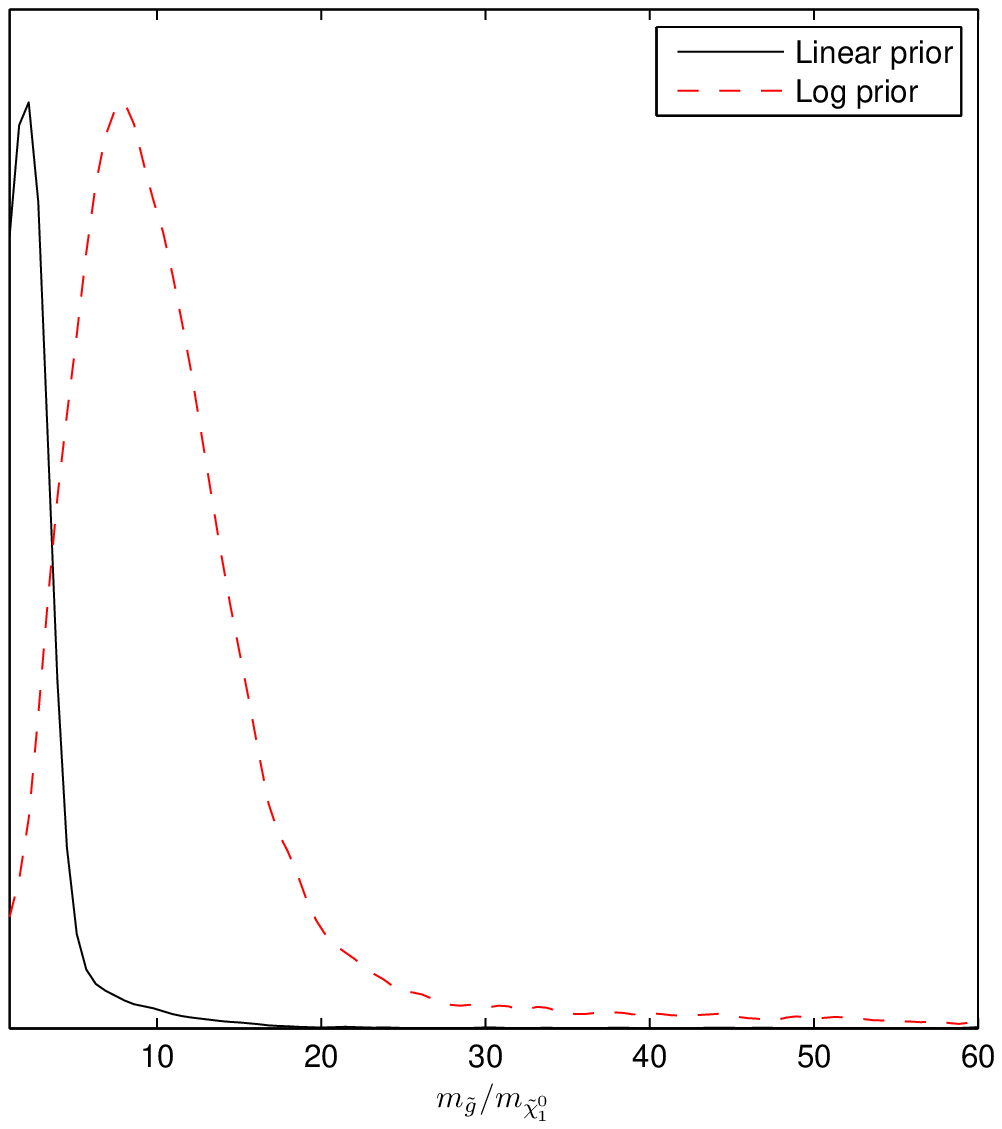}
\caption{\it{pMSSM
  gluino-to-neutralino mass 
  ratio for linear and log priors.}
  \label{fig.mgmc}}}
\paragraph{}
It is worthwhile to point out the source of the difference between
the peak positions in Fig.~\ref{fig.mgmc}. The gluino
mass, $m_{\tilde{g}}$, PDF is roughly the same for both priors. This implies that the
numerator is fixed with respect to the two different priors. So the
shift is  solely due to the different properties of
$m_{\tilde{\chi}^0_1}$ which is much lighter for the log prior
choice. 

%%%%%%%%%%%%%%%%%%%%%%%%
\subsection{Fine-tuning} \label{sub.finetuning}
%%%%%%%%%%%%%%%%%%%%%%%%

\paragraph{}
The main motivation of weak-scale SUSY is to solve the 
technical hierarchy problem, explaining why the Higgs boson 
remains at the weak scale despite quantum corrections 
which are as large as the largest mass scale in the theory
(e.g.\ the Planck scale). 
In order for softly broken SUSY to still provide a resolution of the 
technical hierarchy problem, the SUSY breaking terms should not be
much larger than the TeV scale, otherwise an a priori un-natural
cancellations between  radiative corrections are required in order to
keep the Higgs boson mass low. Direct SUSY search limits imply
lower bounds on sparticle masses, which already 
start to imply that the MSSM parameters
must cancel somewhat 
 in order to separate the electroweak and
SUSY breaking scales. This is termed the `little hierarchy problem'
\cite{hep-ph/0007265,hep-ph/0310137}, 
(see a recent discussion in~\cite{Cassel:2009ps}). We wish to
quantify the 
necessary amount of fine-tuning in the pMSSM parameters
needed to make the set-up consistent with the imposed sparticle
mass bounds. 

\paragraph{}

We follow the approach in~\cite{Kasahara:2008up}, quantifying the
amount of fine-tuning in the Z-boson mass prediction coming from 
Higgs potential minimisation conditions. We consider this as a measure
of fine-tuning in the pMSSM\@. The tree-level Z-boson mass is given by  
\be m_Z^2 = - m_{H_1}^2 \left( 1-\frac{1}{\cos 2 \beta} \right) 
+ m_{H_2}^2 \left(1 + \frac{1}{\cos 2 \beta} \right) - 2|\mu|^2,
\label{mzsq}
\ee 
where \be \sin 2 \beta = \frac{2 m_3^2}{m_{H_1}^2 + m_{H_2}^2 +
2|\mu|^2}. \label{m3sq}\ee 
The amount of fine-tuning is quantified by considering the sensitivity of
$m_Z$ to a variation of a parameter $\xi$~\cite{Nucl.Phys.B306.63}:
\be \Delta(\xi) = \left| \frac{\partial \log m_Z^2}{\partial \log
\xi}\right|, \label{defn} \ee where $\xi=m_{H_1}^2, m_{H_2}^2, m_3^2$ and $\mu$ are the
relevant parameters in the pMSSM\@. Assuming $\tan\beta>1$, from
Eqs.~\ref{mzsq},~\ref{m3sq} 
and~\ref{defn}, one derives:
\bea 
\Delta(\mu)   &=& \frac{4\mu^2}{m_Z^2}\left(1 + \frac{m_A^2+m_Z^2}{m_A^2}
           \tan^2 2\beta \right), \quad 
\Delta(m_3^2)     = \left(1 + \frac{m_A^2}{m_Z^2} \right) \tan^2 2\beta,
\nonumber \\
\Delta(m_{H_1}^2) &=& \left| \frac{1}{2}\cos2\beta +
\frac{m_A^2}{m_Z^2}\cos^2\beta - \frac{\mu^2}{m_Z^2} \right| \times
\left(1 - \frac{1}{\cos2\beta} + \frac{m_A^2+m_Z^2}{m_A^2} \tan^2
2\beta \right), \nonumber \\
\Delta(m_{H_2}^2) &=& \left| -\frac{1}{2}\cos2\beta  +
\frac{m_A^2}{m_Z^2}\sin^2\beta - \frac{\mu^2}{m_Z^2}\right| \times
\left|1 + \frac{1}{\cos2\beta} + \frac{m_A^2+m_Z^2}{m_A^2} \tan^2
2\beta \right|. \label{ghghg}
\eea 
$\Delta(\mu)$, $\Delta(m_3^2)$, $\Delta(m_{H_1}^2)$ and $\Delta(m_{H_2}^2)$
are added in quadrature to obtain an over-all measure of fine-tuning, 
$\Delta_T$:
\be \Delta_T \equiv \sqrt{\Delta(\mu)^2+\Delta(m_3^2)^2+\Delta(m_{H_1}^2)^2+\Delta(m_{H_2}^2)^2}.\ee
 Values of $\Delta_T$ far greater than unity indicate
large fine-tuning. 

\paragraph{}

\FIGURE[ht!]{
\includegraphics[width=0.6\textwidth]{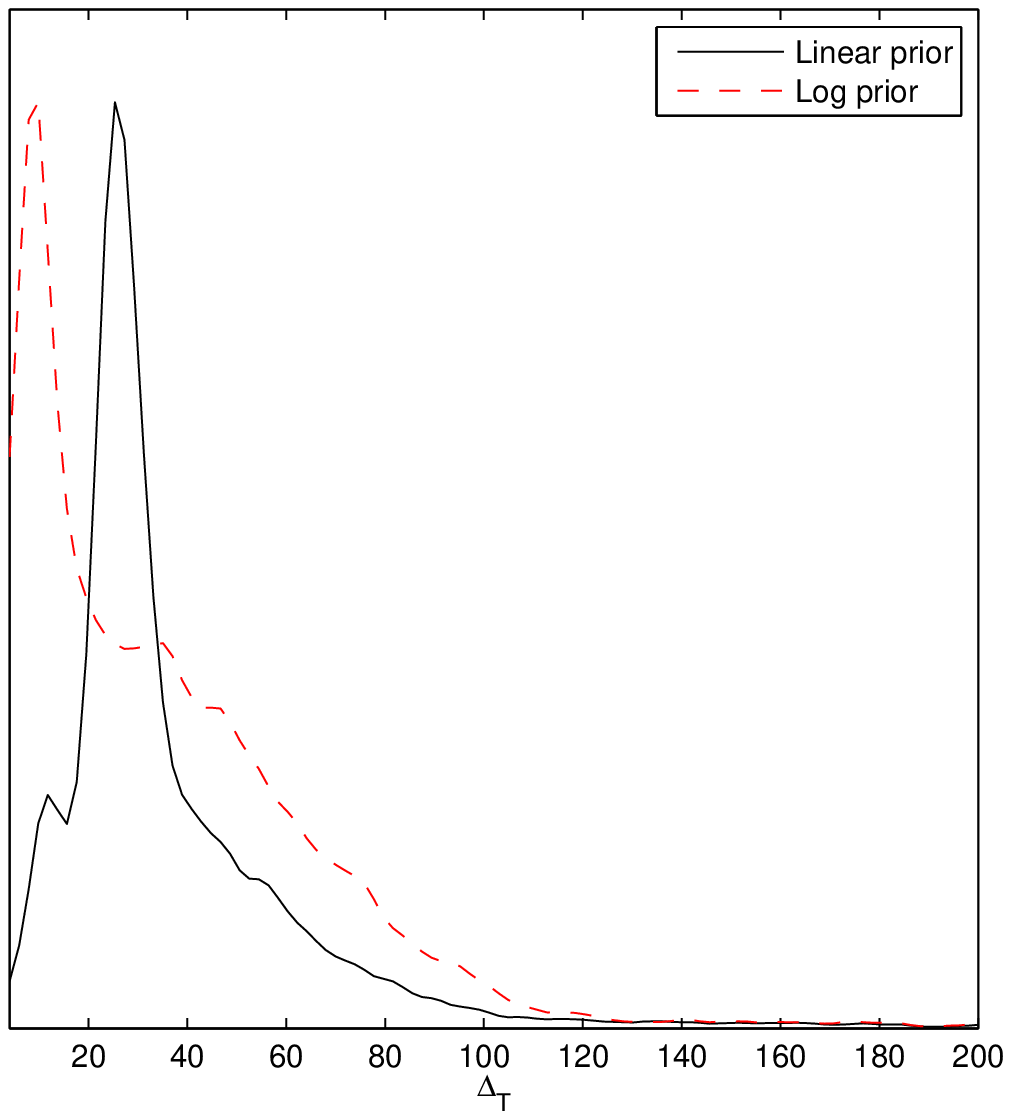}\caption{\it{Fine-tuning PDFs
  in the pMSSM.}
  \label{fig.ftune}}  }
The posterior PDF for the amount of
fine-tuning in the pMSSM is shown in Fig.~\ref{fig.ftune}. 
The logarithmic prior scenario have lower $\Delta_T$ than in the
linear prior. This is not surprising since the SUSY breaking terms are
much reduced in the former scenarios than in the latter. 
We see from the figure that the fine-tuning is most likely low at
around $\Delta_T\sim 20-30$, but there is a tail extending beyond
$\Delta_T=100$. 
One could use a prior of $1/\Delta_T$, to encode a belief in
less fine-tuned points in our global fits~\cite{Allanach:2006jc}. 
Alternatively, one could place a cut on $\Delta_T$, but the value of such a
cut is of course subjective. 
82$\%$ of the high posterior PDF points, around $4.0 \times 10^{4}$ of
samples, have $\Delta_T>10$, so a hard cut placed at 10 would have a drastic
effect on the fits. 
Here we decline to change the prior or place cuts, since we 
are content with observing that, for most of the probability mass in the fits,
it is not unacceptably large. 
For the highest likelihood models the fine-tuning is reduced
from its average. For example, for the good-fit point  in the linear prior
sample, $\Delta_T=24$, whereas for the good-fit point in the log prior
sample, $\Delta_T = 27$.

Notice that in general the amount of fine-tuning we find is small
compared to previous studies of the MSSM that start at a high scale,
running down to the TeV scale using the RG equations. 
The reasons for
this include, as explained in Refs.~\cite{Kane:1998im,hep-ph/0310137}, that the
amount of fine-tuning is a function of the cut-off scale and tends to
decrease with this scale because the interval of RG running of the
soft parameters induces EWSB at tree level and the cross-talk (through
RG running) between parameters in the Higgs sector and those from the
squarks, gluinos, etc sectors is drastically reduced such that the
latter parameters can be much heavier than $m_Z$ without disturbing
the naturalness of the electroweak scale\footnote{We thank
  J.~R. Espinosa for interesting conversations on this point and on
  fine-tuning in general.}. 
Ref.~\cite{Kane:1998im} found some approximate semi-numerical solutions of the
RGEs for the case that the boundary conditions on SUSY breaking parameters are
set at the GUT scale $\sim 10^{16}$ GeV. The dominant term in 
$\Delta \mu(M_{GUT})$ typically comes from cross-talk with the GUT scale
gluino mass $M_{3}(M_{GUT})$:
\begin{equation}
\Delta (\mu(M_{GUT}))=a(\tan \beta) \frac{M_{3}^2 (M_{GUT})}{M_Z^2},
\end{equation}
where the authors determined the coefficient $a(\tan \beta)$ numerically:
$a(2.5)=24$ and $a(10)=12$, for example. 
This is to be contrasted with the pMSSM in Eq.~\ref{ghghg}, where the terms in
$\Delta (\mu)$
are set at the SUSY breaking scale and are $\sim O(1)$, and do not involve the
gluino mass, upon which there are strong empirical lower bounds. 

%SSSSSSSSSSSSSSSSSSSSSSSSSSSSSSS
\section{Neutralino Dark Matter} \label{sec.dm}
%SSSSSSSSSSSSSSSSSSSSSSSSSSSSSSS

\begin{figure}[!ht] 
  \begin{tabular}{ll}
     \includegraphics[angle=0,width=.5\textwidth]{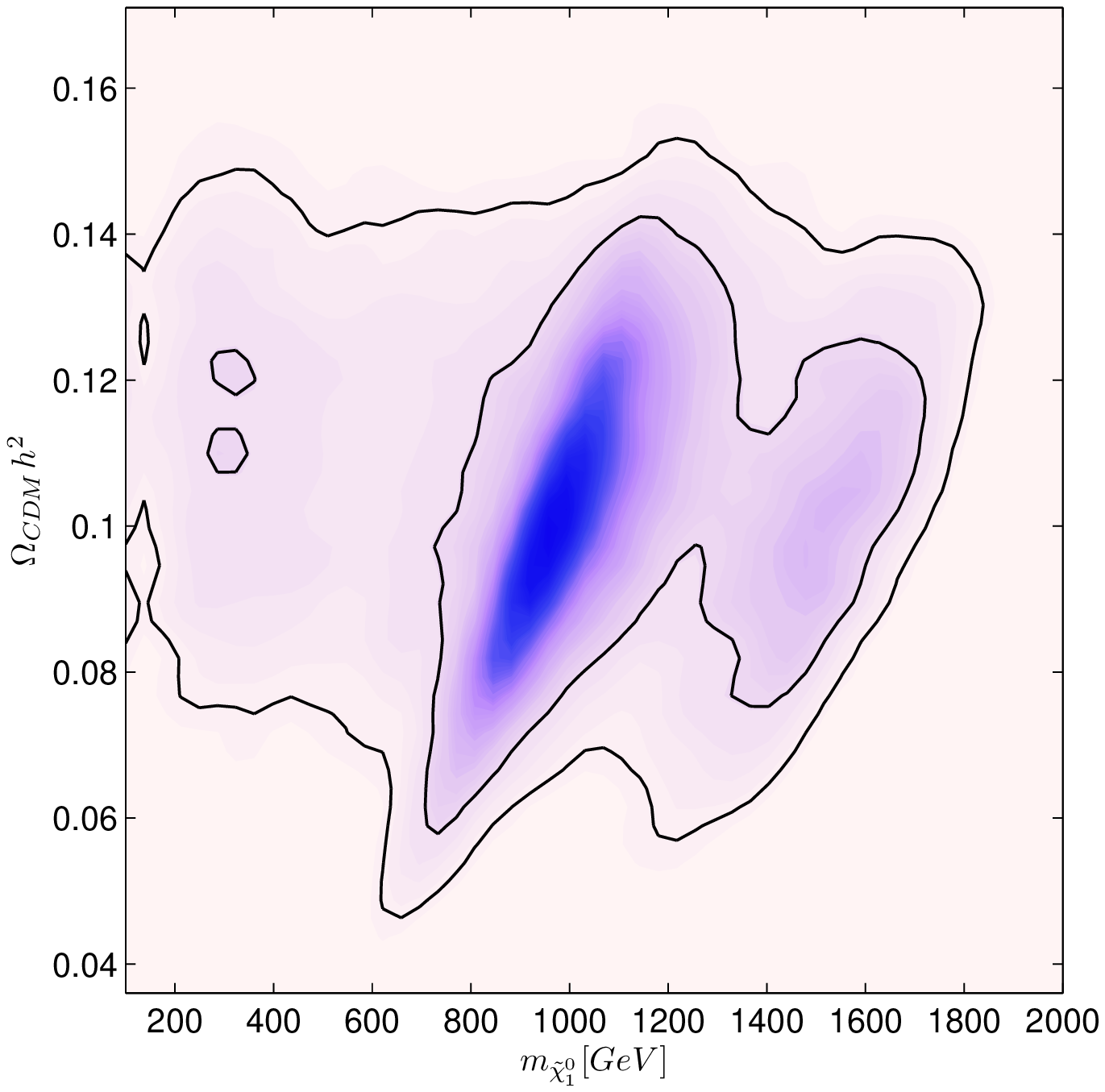} &
     \includegraphics[angle=0,width=.5\textwidth]{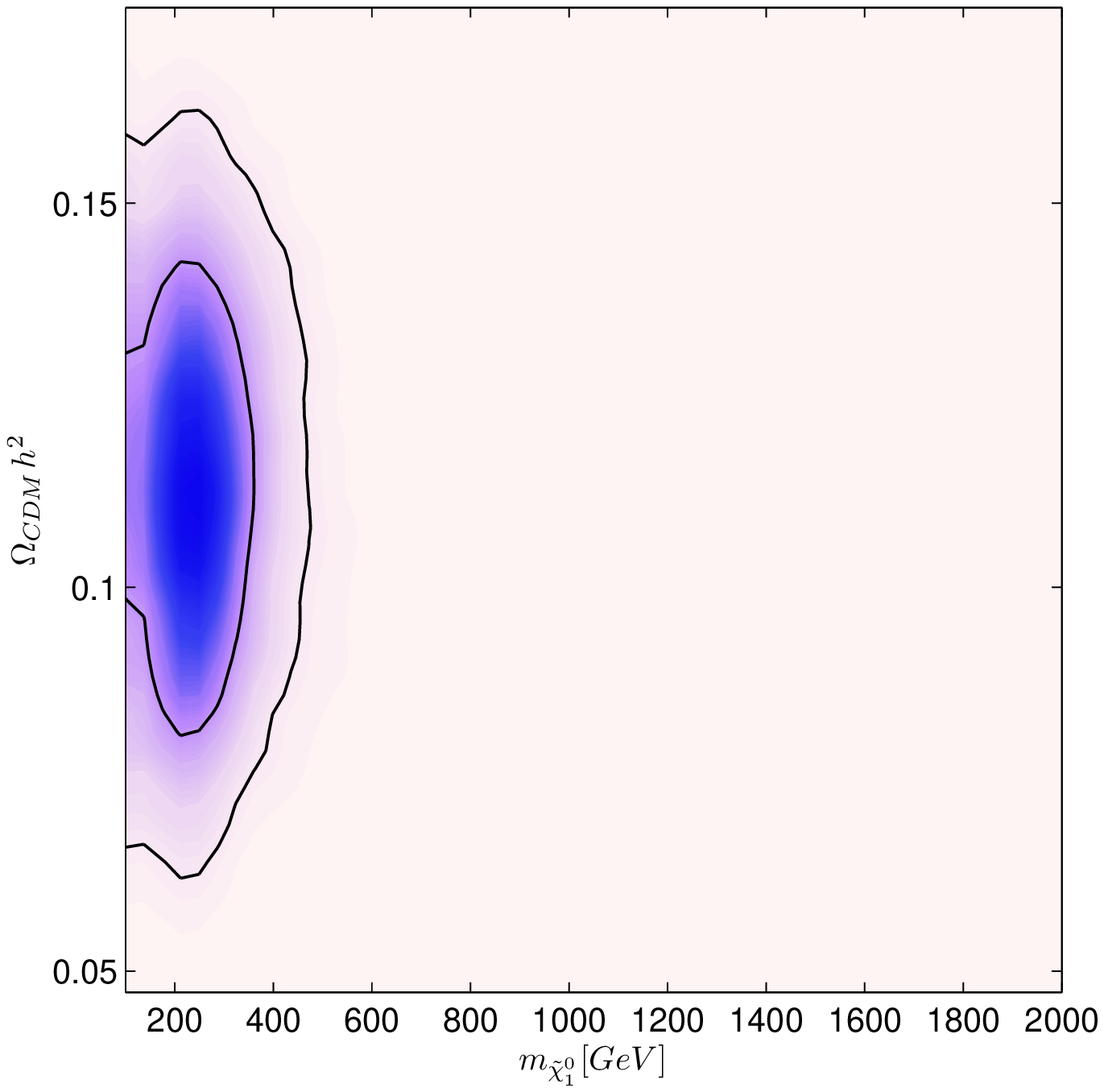} \\
   \end{tabular}
  \caption{{\it Marginalised 2D posterior PDFs of the 
  neutralino dark matter relic density against the neutralino
  mass. Linear priors are on the left hand side and 
  log priors on the right. The dark contours show the 68$\%$ and 95$\%$
  Bayesian credibility regions. Note that the feature shown towards the
  left hand side of the linear prior plot is due to the peak in the
  neutralino mass at around 200 GeV in Fig.~\protect\ref{fig.smasses} which
  here just makes it into the 68$\%$ region.}}
 \label{fig.omega-neut}
 \end{figure}

\paragraph{}
Assuming R-parity conservation the lightest supersymmetric particle (LSP) may
be a good dark matter (DM) candidate. In Ref.~\cite{Baltz:2006fm}, 
a pMSSM study of the ability of SUSY measurements at future colliders to
constrain dark matter properties was considered. 
We assume that the neutralino LSP 
constitutes the DM in 
the universe. The DM relic density then depends upon the LSP mass and, through
its composition in terms of gauginos and Higgsino, 
its interactions. 2-D 
marginalised posterior PDFs showing preferred regions in the
relic density versus LSP mass are shown in
Fig.~\ref{fig.omega-neut}. 
There is a mild positive correlation between the preferred mass and the 
dark matter relic density $\Omega_{CDM} h^2$ for linear priors which is
not evident for the log priors. It is clear that the LSP mass is not well
constrained by current data, since it is highly prior dependent. 
The nature of the neutralino LSP in the
pMSSM is addressed in Section~\ref{sub.neut}. The mass difference
between the LSP and the next to lightest supersymmetric particle
(NLSP) is important because if it is small, the LSP may efficiently
co-annihilate, in the early universe, with the 
NLSP, significantly reducing the relic density. The different
possible NLSPs in the pMSSM with the corresponding posterior probabilities are
listed in Section~\ref{sub.coann}. The dominant (co-)annihilation 
channels are presented in Section~\ref{sub.coann}. 
We present the prospects of direct dark matter detection in 
Section~\ref{sub.ddetection}.   

%%%%%%%%%%%%%%%%%%%%%%%%%%%%%%%%%%%%%%%%%%%%%%%
\subsection{Neutralino dark matter composition} \label{sub.neut}
%%%%%%%%%%%%%%%%%%%%%%%%%%%%%%%%%%%%%%%%%%%%%%%

\paragraph{}
The nature of the neutralino LSP determines the (dominant) processes
by which it (co-)annihilates into SM particles and
therefore affects its present number density. This was illustrated
using the randomly scanned pMSSM samples in~\cite{Profumo:2004at} where
it was shown that the  
nature of the neutralino LSP depends upon whether one assumes that
the LSP makes up all or only some of the DM relic density.
The neutralino mass matrix is given by $\frac{1}{2} 
   {\psi^0}^T M_N \psi^0 + H.c.$ where ${\psi^0}^T = ( - i \tilde{b},
   - i \tilde{w}^3,\tilde{H_1^0}, \tilde{H_2^0} )$ and,  
\be \label{neutmass}
M_\textrm{N} = \left( \begin{array}{cccc} 
M_1 & 0 & - m_Z c_\beta s_W & m_Z s_\beta c_W \\
0 & M_2 &  m_Z c_\beta c_W & - m_Z s_\beta c_W \\
- m_Z c_\beta s_W & m_Z c_\beta c_W &  0 & - \mu \\
 m_Z s_\beta s_W & - m_Z s_\beta c_W &  -\mu & 0 
\end{array} \right),
\ee
$c_x = \cos x$ and $s_x = \sin x$.
The neutralino mass eigenstates are $\tilde{\chi}_i^0 = N_{ij}
\psi_j^0$ where N is a unitary transformation that diagonalises
$M_N$. The LSP neutralino mass eigenstate is therefore a mixture of bino, wino
and Higgsino: 
\be \label{neut1}
\tilde{\chi}_1^0 = N_{11}\tilde{b} + N_{12}\tilde{w}^3 +
   N_{13}\tilde{H_1^0} + N_{14}\tilde{H_2^0}.
\ee
Different regions of parameter space give different neutralino LSP compositions.
When $M_1 \ll min(M_2,|\mu|)$, 
$N_{11} \sim 1$ and the LSP is dominantly bino. Bino LSPs give a relic density
that is too high for most of the parameter space unless some specific
mechanism (such as efficient co-annihilations or annihilations through a
resonance) is working.  
When $M_2 < min(M_1, |\mu|)$,  $N_{12}$ dominates, i.e.\ the LSP is dominantly
wino and is quasi mass degenerate with the lightest chargino. This leads to
strong co-annihilations between the LSP and the chargino, and typically the
relic density much smaller than the WMAP constraint for wino LSPs. 
For $|\mu| < min(M_1, M_2)$, $N_{13}$ and $
N_{14}$ are of order one and the LSP is dominantly Higgsino and there may be
efficient annihilations into top and weak gauge boson pairs. 
In the Higgsino-dominated LSP scenario, $\tilde{\chi}_1^0$,
$\tilde{\chi}_2^0$, $\tilde{\chi}_1^\pm$ are almost all mass degenerate
and are Higgsino-like. 
Of course, there exist mixed cases which include several of these limiting
behaviours. 

\paragraph{}
\begin{figure}[!ht] 
\begin{center}
\begin{minipage}{6cm}
 \includegraphics[angle=0, width=6cm]{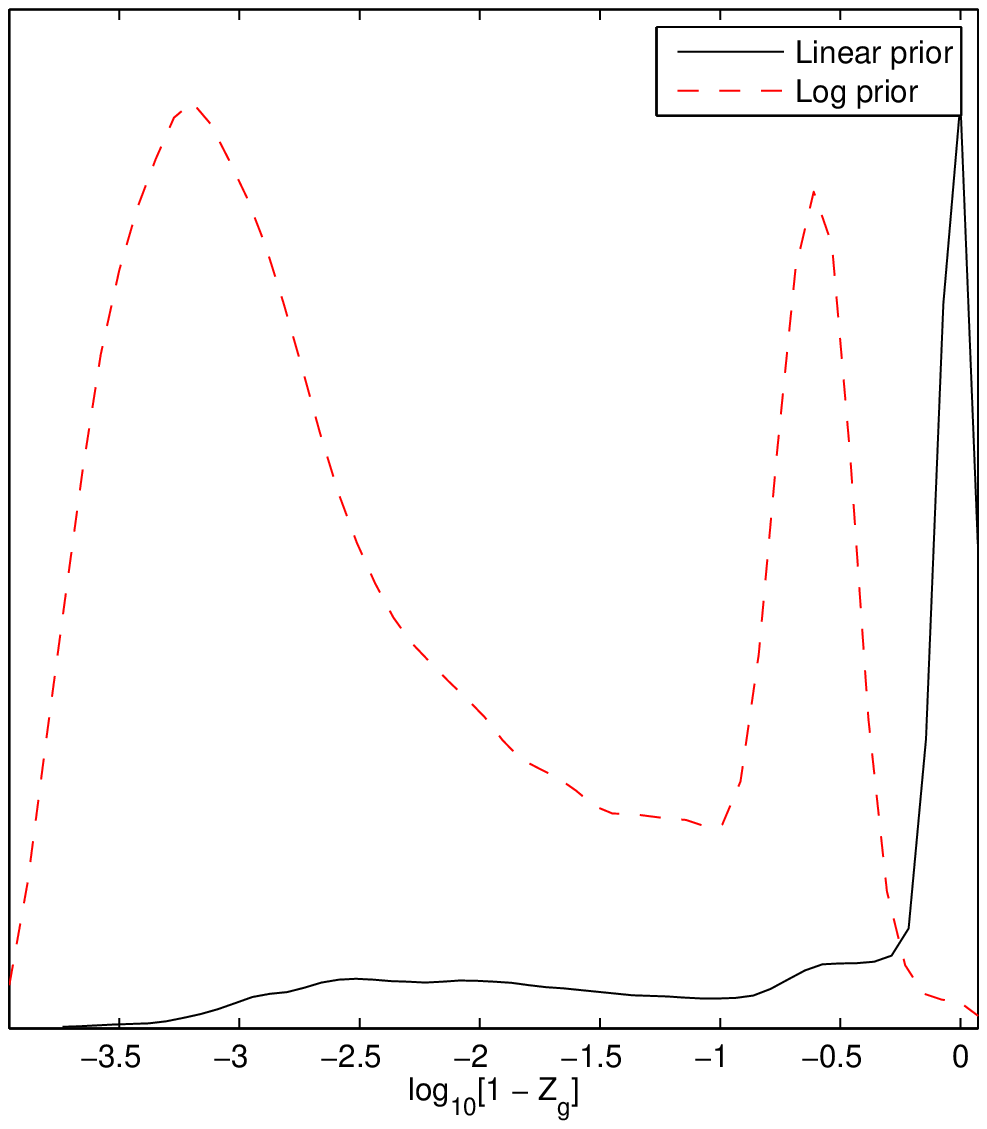}
\end{minipage}
\end{center}
 \caption{{\it pMSSM neutralino gaugino-Higgsino admixture
  fractions. Higgsino domination is   at the right-hand side of the
  plot and gaugino domination is to the   left-hand side.}}  
\label{fig.hfrac}
\end{figure}
The gaugino/Higgsino mixture PDF
of the LSP is shown in Fig.~\ref{fig.hfrac}
constructed from the fraction $$Z_g = |N_{11}|^2 + |N_{22}|^2$$ 
following~\cite{Barger:2007nv}. $1 - Z_g$ is unity if the neutralino
LSP is fully Higgsino-like and zero if fully gaugino-like.
The plot shows that LSP is mostly Higgsino in the linear prior case,
similar to the non-universal higgs mass  scenario~\cite{Roszkowski:2009sm},
and mostly gaugino 
for the log prior scenario. Thus, current data do not unambiguously
constrain the LSP content.  
Referring to Tab.~\ref{tab.bestparam}, we see that the good-fit point from
the linear prior sample 
has a mixed wino-Higgsino LSP (more precisely, the point has $|N_{13}| \sim
|N_{14}|=0.7$ and $N_{12}=0.15$). The log prior sample good-fit point 
has a bino dominated LSP ($|N_{11}|=0.998$), but there are several light
sparticles, allowing  sufficient annihilation. 

\FloatBarrier

%%%%%%%%%%%%%%%%%%%%%%%%%%%%%%%
\subsection{(Co-)Annihilations} \label{sub.coann}
%%%%%%%%%%%%%%%%%%%%%%%%%%%%%%%

\paragraph{}
At early times of the universe the LSP is in thermal
equilibrium with other particles and, ignoring for now
co-annihilations, its number density 
evolution is governed by the Boltzmann equation 
\be \label{boltz}
\frac{dn_{\widetilde{\chi}_1^0}}{dt} = -3 H n_{\widetilde{\chi}_1^0} -
<\sigma v> \{ n_{\widetilde{\chi}_1^0}^2 -
(n_{\widetilde{\chi}_1^0}^{eq})^2 \}.\ee
Here $H$ is the Hubble expansion rate of the universe,
$n_{\widetilde{\chi}_1^0}$ is the number density and $<\sigma v>$ 
is the thermally averaged annihilation 
cross sections of the neutralino LSP\@. $v$ is the relative velocity of
the annihilating pair. The LSP annihilation rate is given by
$\Gamma_{\widetilde{\chi}_1^0} = <\sigma v> \,
n_{\widetilde{\chi}_1^0}$. At a freeze-out temperature $T_f$, the
neutralino decouples, $\Gamma_{\widetilde{\chi}_1^0} = H(T_f)$.
Substituting $H(T_f)$ into Eq.~\ref{boltz} predicts that the LSP relic density
is inversely proportional to the thermally averaged annihilation
cross section, $<\sigma v>$. This means that for the LSP relic
abundance today to be in the WMAP-5 range Eq.~\ref{wmap5} there must
be a significant number of annihilations of the neutralino LSP at earlier
times. The possible processes are mostly
two-particle final states which could be fermion anti-fermion pair,
combinations of the weak gauge-bosons $(W^{\pm}, Z^0)$ or combinations
of the Higgs bosons $(h^0, H^0, A^0, H^{\pm})$  (see
e.g.\ \cite{Stefaniak:2008zj}). 
The discussion becomes much more involved once co-annihilation processes are
taken into account, since coupled Boltzmann equations
are required for each relevant SUSY particle species.

\paragraph{}
Co-annihilation processes dominate in parameter space points where
next-to-lightest supersymmetric particle (NLSP) that are almost mass
degenerate with the LSP\@. At such points the neutralino abundance
also depends strongly on the annihilations of the
NLSPs~\cite{Griest:1990kh,Edsjo:1997bg} and the number  densities of
the LSPs and NLSPs are coupled.
A review of different co-annihilation studies was presented
in~\cite{Profumo:2004at}. Here we present and analyse the outcome of
the pMSSM annihilation and co-annihilation results for our two different
prior measures. We shall only discuss processes that contribute 1$\%$ or more
of the annihilation cross section. 

\paragraph{}

In Tab.~\ref{tab.nlsps} we give a list of
possible NLSPs and corresponding posterior probabilities for each.
The probabilities indicates that neutralino-chargino
co-annihilations are most likely to be dominant in the pMSSM with a linear prior
measure. For the log prior measure, neutralino-slepton
co-annihilations are the most probable. 
We illustrate the abundance and complexity of the annihilation channels by
presenting the dominant ones at the good fit points in
Tab.~\ref{decay}. The dominant channels for the linear prior sample good-fit
point are direct neutralino-chargino co-annihilation and neutralino
annihilation via chargino exchange into $Z$- and $W$-boson pairs. 
For the 
log prior measure good-fit point, the dominating channels are neutralino
co-annihilations with various sleptons. Many different processes contribute at
the percent level. 
We present the identities of the dominant channel, along with its posterior
probability in Tab.~\ref{tab.coannprob}.
The most likely channel is neutralino-chargino co-annihilation for the linear
prior and neutralino annihilation for the log prior. 
From the large prior dependence in the results, we deduce that current
data are not sufficient to constrain the dark matter annihilation
properties of the LSP\@. 

\begin{table}[htbp!] 
\begin{center}{\begin{tabular}{|c|cc|}
\hline
NLSP & $P(NLSP)_{Linear}$ & $P(NLSP)_{Log}$ \\
\hline
$\tilde{\chi}^0_2$        & 14\%  & 1\%  \\
$\tilde{\chi}_1^{\pm}$    & 77\%  & 15\%  \\
$\tilde{g}$               & 1\%   & 0\% \\
$\tilde{\nu}_e$           & 2\%   & 39\% \\
$\tilde{\nu}_\tau$        & 0\%   & 4\% \\
$\tilde{e}_L$             & 0\%   & 2\% \\
$\tilde{e}_R$             & 0\%   & 27\% \\
%$\tilde{\mu}_L$           & 0\%   & 0.115\% \\
$\tilde{\tau}_1$          & 0\%   & 7\% \\
$\tilde{u}_L$             & 1\%   & 1\% \\
$\tilde{u}_R$             & 1\%   & 1\% \\
%$\tilde{s}_L$             & 0.122\%   & 0.252\% \\
$\tilde{s}_R$             & 1\%   & 1\% \\
$\tilde{t}_1$             & 1\%   & 0\% \\
$\tilde{b}_1$             & 1\%   & 1\% \\
\hline
\end{tabular}}
\end{center}
\caption{\it{pMSSM NLSP identity probabilities for linear and log
    priors. }}   
\label{tab.nlsps}
\end{table}

\begin{table}[htbp!] 
{\begin{tabular}{|l|c|l|c||l|c|l|c|}
\hline
\multicolumn{4}{|c||}{\bf Linear Prior} & \multicolumn{4}{|c|}{\bf Log Prior}\\
\hline\hline
Event & $\%$ & Event & $\%$ & Event & $\%$ & Event & $\%$ \\
\hline
 ${\tilde\chi}_1^0 + \tilde{\chi}_1^0 \rightarrow W^+ W^-$ & 3
&${\tilde\chi}_1^0 + \tilde{\chi}_1^0 \to ZZ$ & 2 & ${\tilde\chi}_1^0
+ \tilde{\chi}_1^0\to \mu\bar{\mu}$ & 2 &  ${\tilde\chi}_1^0 +
\tilde{\chi}_1^0\to e\bar{e}$ &  2 \\ 
$\tilde{\chi}_1^0 + \tilde{\chi}_1^+ \to W^+h $ & 3 &
$\tilde{\chi}_1^0 + \tilde{\chi}_1^+\to t\bar{b}$ & 5 & ${\tilde\chi}_1^0 +
\tilde{e}_R\to Ze$ & 4 & ${\tilde\chi}_1^0 + \tilde{e}_R\to Ae$ & 14
\\  
 ${\tilde\chi}_1^0 + \tilde{\chi}_1^+\to u\bar{d}$ & 10 &  ${\tilde\chi}_1^0 + \tilde{\chi}_1^+\to \bar{s}c$ & 10 & ${\tilde\chi}_1^0 + \tilde{\mu}_R\to Z\mu$ & 4 & ${\tilde\chi}_1^0 + \tilde{\mu}_R\to A\mu$ & 14  \\
${\tilde\chi}_1^0 + \tilde{\chi}_1^+\to \nu_\tau \tau$ & 3 & ${\tilde\chi}_1^0 + \tilde{\chi}_1^+\to\bar{\mu} \nu_\mu$ & 3 & ${\tilde\chi}_1^0 + \tilde{\nu}_e\to Z\nu_e$ & 2 &${\tilde\chi}_1^0 + \tilde{\nu}_e\to W^+ e$ & 3 \\
${\tilde\chi}_1^0 + \tilde{\chi}_1^+\to \bar{e}\nu_e $ & 3 & ${\tilde\chi}_1^0 + \tilde{\chi}_1^+\to ZW^+$ & 2 & ${\tilde\chi}_1^0 + \tilde{\nu}_\mu\to Z\nu_\mu$ & 2 & ${\tilde\chi}_1^0 + \tilde{\nu}_\mu\to W^+\mu $ & 3 \\
  ${\tilde\chi}_1^0 + \tilde{\chi}_1^+\to AW^+ $ & 2 &  ${\tilde\chi}_1^0 +\tilde{\chi}_2^0\to d\bar{d} $ & 1 & $\tilde{e}_R+ \tilde{e}_R \to ee$ & 9 & $\tilde{e}_R+ \tilde{\bar{e}}_R \to AZ$ & 1 \\
${\tilde\chi}_1^0 +\tilde{\chi}_2^0\to u\bar{u} $ & 1 & ${\tilde\chi}_1^0 +\tilde{\chi}_2^0\to c\bar{c} $ & 1 & $\tilde{e}_R+ \tilde{\bar{e}}_R \to AA$ & 2 & $\tilde{e}_R+ \tilde{\mu}_R \to e\mu$ & 9 \\
${\tilde\chi}_1^0 +\tilde{\chi}_2^0\to s\bar{s} $ & 1 & $\tilde{\chi}_1^+ + \tilde{\chi}_1^- \to Zh $ & 2 & $ \tilde{e}_R+ \tilde{\bar{\nu}}_e\to e\bar{\nu}_e $ & 1 & $ \tilde{e}_R+ \tilde{\bar{\nu}}_\mu\to e\bar{\nu}_\mu $  & 1 \\ 
 $\tilde{\chi}_1^+ + \tilde{\chi}_1^- \to t\bar{t} $ & 1 & $\tilde{\chi}_1^+ + \tilde{\chi}_1^- \to d\bar{d} $ & 2 & $ \tilde{\mu}_R + \tilde{\mu}_R \to \mu\mu $ & 9 &  $ 
\tilde{\mu}_R + \tilde{\bar{\mu}}_R \to AZ$ & 1 \\
$\tilde{\chi}_1^+ + \tilde{\chi}_1^- \to u\bar{u} $ & 3 & $\tilde{\chi}_1^+ + \tilde{\chi}_1^- \to c\bar{c} $ & 3 & $ \tilde{\mu}_R + \tilde{\bar{\mu}}_R \to AA$ & 2 & $ \tilde{\mu}_R + \tilde{\bar{\nu}}_e \to \mu \nu_e $ & 1 \\
$\tilde{\chi}_1^+ + \tilde{\chi}_1^- \to s\bar{s} $ & 2 & $\tilde{\chi}_1^+ + \tilde{\chi}_1^- \to \tau\bar{\tau} $ & 1 &  $ \tilde{\mu}_R + \tilde{\bar{\nu}}_\mu \to \mu \nu_\mu $ & 1 & $\tilde{\nu}_e + \tilde{\bar{\nu}}_e \to W^+W^-$ & 1 \\
$\tilde{\chi}_1^+ + \tilde{\chi}_1^- \to \mu\bar{\mu} $ & 1 & $\tilde{\chi}_1^+ + \tilde{\chi}_1^- \to e\bar{e} $ & 1 & $\tilde{\nu}_e + \tilde{\bar{\nu}}_e \to ZZ$ & 1 &
$\tilde{\nu}_\mu + \tilde{\bar{\nu}}_\mu \to W^+W^- $ & 1 \\
$\tilde{\chi}_1^+ + \tilde{\chi}_1^- \to W^+W^- $ & 3 & $\tilde{\chi}_1^+ + \tilde{\chi}_1^- \to AZ $ & 1 & $\tilde{\nu}_\mu + \tilde{\bar{\nu}}_\mu \to ZZ $ & 1 & -  & -  \\
$\tilde{\chi}_1^+ + \tilde{\chi}_1^- \to AA $ & 1 & $\tilde{\chi}_2^0 + \tilde{\chi}_1^+ \to t\bar{b} $ & 2 & - & - & - & - \\
$\tilde{\chi}_2^0 + \tilde{\chi}_1^+ \to u\bar{d} $ & 4 & $\tilde{\chi}_2^0 + \tilde{\chi}_1^+ \to c\bar{s} $ & 4 & - & - & - & -\\
$\tilde{\chi}_2^0 + \tilde{\chi}_1^+ \to \tau\nu_\tau $ & 1 & $\tilde{\chi}_2^0 + \tilde{\chi}_1^+ \to \bar{\mu} \nu_\mu $ & 1 & - & - & - & -\\
$\tilde{\chi}_2^0 + \tilde{\chi}_1^+ \to e\nu_e $ & 1 & $\tilde{\chi}_2^0 + \tilde{\chi}_1^+ \to ZW^+ $ & 1 & - & - & - & -\\
\hline
\end{tabular}}
\caption{\it (Co-)annihilation channels  
at the good fit points. We display the percentage contribution to the
annihilation cross section for each channel. Channels which contribute 
less than $1\%$ to the annihilation cross sections are not listed.}
\label{decay}
\end{table}

\begin{table}[!ht] 
\begin{center}
\begin{tabular}{|c|c|c|}
\hline
(Co-)annihilation & Linear prior case & Log prior case\\
\hline \hline
$\tilde{\chi}_1^0 \tilde{\chi}^{\pm}$ &  35\%  & 5\%\\
\hline
$\tilde{\chi}_1^0 \tilde{\chi}_1^0$ &  20\%  & 28\%\\
\hline
$\tilde{\chi}_1^0 \tilde{\chi}_2^0$ &  0\%  & 7\%\\
\hline
$\tilde{\chi}_1^0$ sleptons &  0\%  & 23\% \\
\hline
\end{tabular}
\end{center}
\caption{\it Posterior probabilities for dominant annihilation and
  co-annihilation channels. \label{tab.coannprob}}  
\end{table}

%%%%%%%%%%%%%%%%%%%%%%%%%%%%%
\subsection{Direct detection} \label{sub.ddetection}
%%%%%%%%%%%%%%%%%%%%%%%%%%%%%

\paragraph{}
Many different experiments search for
the nature of dark matter (e.g.\ see \cite{Hooper:2008sn} and
references therein). 
Indirect detection
experiments are designed to observe the annihilation
products of dark matter particles. We do not address indirect detection
here and save it for future consideration. 
Here, we consider direct detection
experiments such as XENON~\cite{0706.0039},
CDMS~\cite{astro-ph/0509259,astro-ph/0509269}, ZEPLIN~\cite{astro-ph/0701858,APhys.23.444},
Edelweiss~\cite{astro-ph/0503265}, CRESST~\cite{astro-ph/0408006}, WARP~\cite{astro-ph/0701286,astro-ph/0405342} or
COUPP~\cite{J.Phys.Conf.Ser.39.126}. Such experiments
are designed to observe the elastic scattering of
dark matter particles with nuclei. The LSP may
interact with quarks in target nuclei via
$t$-channel CP-even Higgs exchange or $s$-channel squark exchange and
with gluons via squark loop contributions. DM direct detection rates also depend
on the local neighbourhood DM density and velocity distribution. The
density, which is estimated to lie between $4 \times 10^{-25}
g/cm^{-3}$ and $13 \times 10^{-25} g/cm^{-3}$ ($0.22 - 0.73$
GeV/$cm^3$), is inferred by fitting observations to models of galactic
halo \cite{astro-ph/9508071,astro-ph/9704253}. The velocity is expected to be around $230 \pm
20$ km/s \cite{Phys.Rev.D33.3495}. 

\paragraph{}
The elastic scattering cross section is partitioned into
spin-dependent and spin-independent components.
The spin independent part is currently the most constraining, and
we concentrate on it. 
It is
proportional to the square of the target nucleus atomic number,
$A^2$. This enhancement is because the dark matter wavelength
is of same order as the size of a nucleus and hence the scattering
amplitudes on individual nucleons add coherently.
There is one experimental claim in
direct detection experiments of a signal in 
the annual modulation rate~\cite{{Bernabei:2003wy}}.
This result has not been confirmed by other experiments and would be
incompatible with a neutralino LSP candidate, so we do not use it to constrain 
the pMSSM\@.
Aside from this, no positive signal has been detected
to date in dark matter detection experiments. A positive signal would
constrain SUSY parameter space if one assumed a particular local neighbourhood
DM density and velocity distribution. 

\paragraph{}
The spin-independent neutralino-nucleus elastic scattering
cross section is given by   
\be
\sigma \approx \frac{4 m^2_{\tilde \chi^0_1} m^2_{T}}{\pi (m_{\tilde \chi^0_1}
  + m_T)^2} 
       [Z f_p + (A-Z) f_n]^2, 
\ee
where $m_T$ is the mass of the target nucleus and $Z$ and $A$ are the
atomic number and atomic mass of the nucleus, respectively.  $f_p$ and
$f_n$ are the neutralino's couplings to protons and neutrons, given
by~\cite{hep-ph/9506380} 
\be 
f_{p,n} = \sum_{q=u,d,s} f^{(p,n)}_{T_q} a_q \frac{m_{p,n}}{m_q} +
\frac{2}{27} f^{(p,n)}_{TG} \sum_{q=c,b,t} a_q  \frac{m_{p,n}}{m_q}, 
\label{quds}
\ee
where $a_q$ are the neutralino-quark
couplings~\cite{hep-ph/9506380,Nucl.Phys.B351.623,Phys.Lett.B225.140,hep-ph/9307208,hep-ph/9210272,hep-ph/0001005} and $f^{(p,n)}_{T_q}$ denote
the quark content of the nucleon. They have been experimentally bounded to be:
$f^{(p)}_{T_u} \approx 0.020\pm0.004$, $f^{(p)}_{T_d} \approx
0.026\pm0.005$,  $f^{(p)}_{T_s} \approx 0.118\pm0.062$, $f^{(n)}_{T_u}
\approx 0.014\pm0.003$,  $f^{(n)}_{T_d} \approx 0.036\pm0.008$ and
$f^{(n)}_{T_s} \approx 0.118\pm0.062$~\cite{hep-ph/0111229,hep-ph/9909228,hep-ph/0502001}.  The first term in
Eq.~\ref{quds} corresponds to interactions with the quarks in the
target, which can occur through either $t$-channel CP-even Higgs
exchange, or $s$-channel squark exchange. The second term corresponds
to interactions with gluons in the target through a quark/squark
loop. $f^{(p)}_{TG}$ is given by $1
-f^{(p)}_{T_u}-f^{(p)}_{T_d}-f^{(p)}_{T_s} \approx 0.84$, and
analogously, $f^{(n)}_{TG} \approx 0.83$.  

\paragraph{}

\begin{figure}[!ht] 
  \begin{tabular}{ll}
    \includegraphics[angle=0,
      width=.5\textwidth]{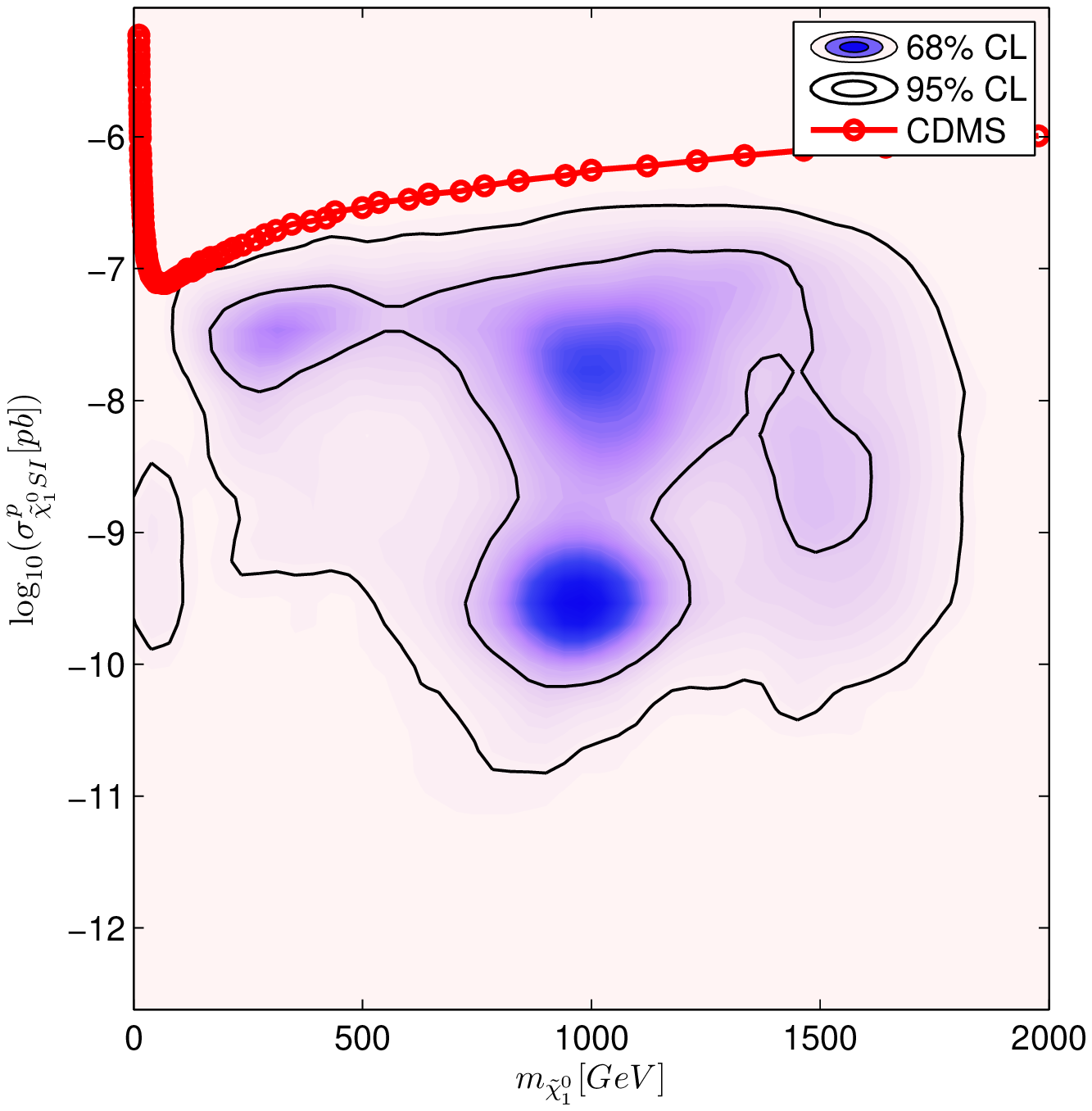} 
    \includegraphics[angle=0,
      width=.5\textwidth]{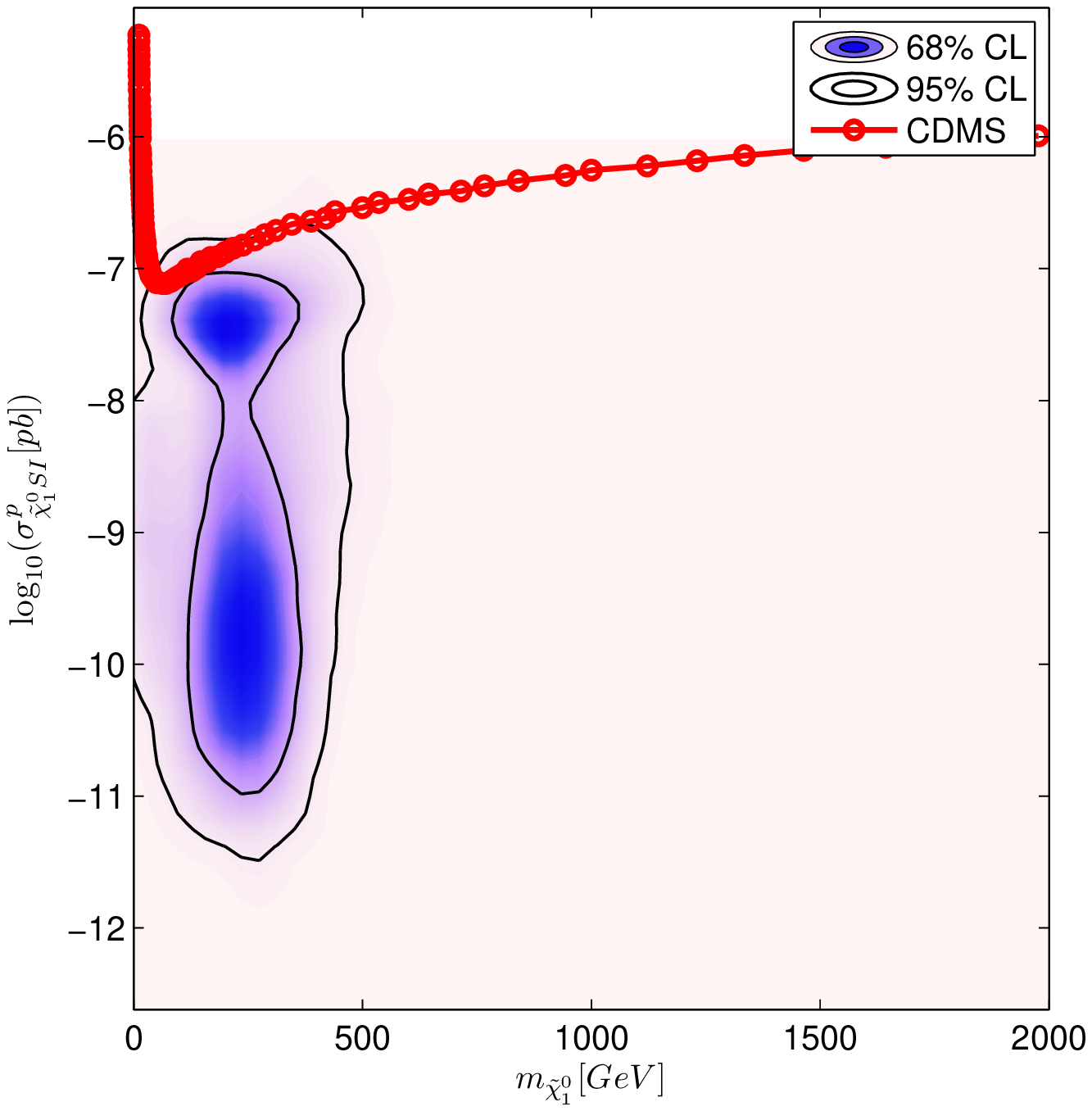} 
  \end{tabular}
 \caption{{\it Posterior PDF of the
 neutralino-proton spin-independent scattering cross section for the pMSSM
 with linear (left) and log (right) prior measures. The CDMS
 90\% confidence level upper bound is also shown, assuming a local DM density
 of 0.3 GeV/cm$^{3}$. The dark contours show the 68$\%$ and 95$\%$ Bayesian
 credibility regions.}}  
\label{fig.ddetect}
\end{figure}
The direct detection constraints from the cryogenic cold dark matter
search (CDMS) experiments on the pMSSM is shown in
Fig.~\ref{fig.ddetect}. The large prior dependence of the results indicates
that current data are insufficient to constrain the direct detection
cross sections. One can say that there is clearly a wide allowed range for the
direct detection cross sections. 

%%%%%%%%%%%%%%%%%%%%%%%%%%%%%%%%%%%%%%%%%%%%%%%%%%%%%%%%%%%
\subsection{Relaxing the purely LSP dark matter assumption} \label{sub.lspDMfrac}
%%%%%%%%%%%%%%%%%%%%%%%%%%%%%%%%%%%%%%%%%%%%%%%%%%%%%%%%%%%

\EPSFIGURE{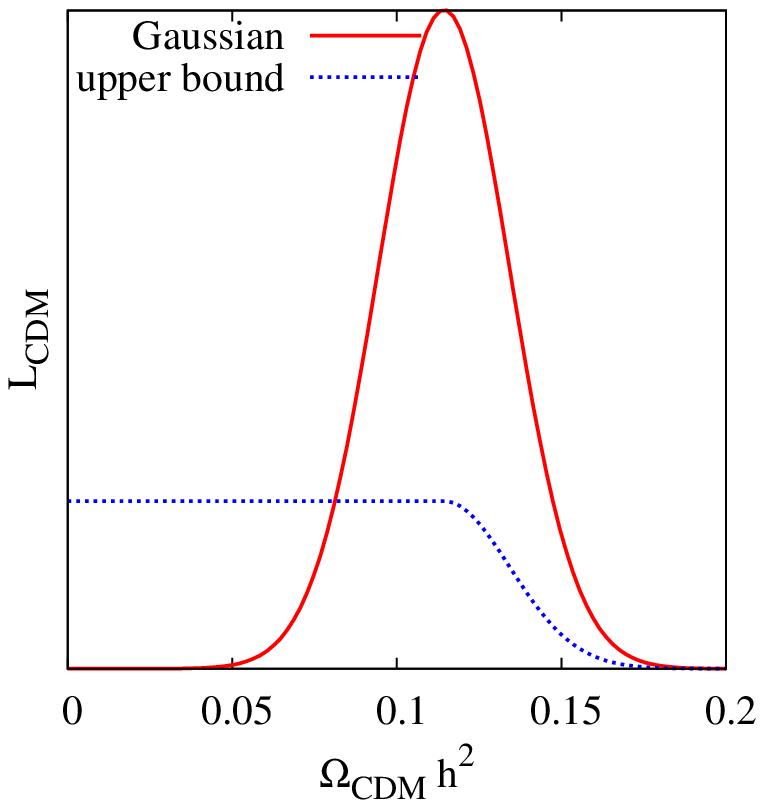}{\it {Depiction of the likelihood constraint on the
  predicted value of $\Omega_{\mathrm{DM}}h^2$ due to lightest
  neutralinos, compared to a simple Gaussian with WMAP5 central value
  and a 1$\sigma$ uncertainty of 0.02 used in the rest of the paper. \label{fig.omega}}}

\paragraph{}
The analysis presented above was done assuming that the neutralino
LSP  is the only source of dark matter. It is known that the LSP relic
density is sensitive to the assumed cosmology. For example, Big Bang
Nucleosynthesis (BBN) expansion rates can enhance the calculated relic
density without affecting other important cosmological quantities
\cite{Arbey:2008kv}. The inclusion of right-handed neutrinos could
change the relic density prediction, see for instance
\cite{Barger:2008nd}. One could also allow for additional
non-neutralino dark matter components. In order to see the potential
effect of such model changes, we relax the constraint from the DM
relic density to the case where only $\Omega_{CDM} h^2$ predictions
larger than the central values are penalised according to the likelihoods: 
\be \label{halfull}
L_{CDM}(\Omega_{CDM}) = 
\begin{cases}
 \frac{1}{\mu \quad +  \quad \sqrt{\pi s^2/2}}, & \textrm{if $\Omega_{CDM} < \mu$} \\
 \frac{1}{\mu \quad + \quad \sqrt{\pi
 s^2/2}}\exp\left[-\frac{(\Omega_{CDM}-\mu)^2}{2s^2}\right], &  
 \textrm{if $\Omega_{CDM} \geq \mu$} \\
\end{cases} 
\ee
where $\mu = 0.1143$ is the experimental central value quoted above and
$s$ is an inflated error on relic density that includes theoretical
uncertainties in its prediction.

\begin{figure}[!ht] 
\begin{center}
    \includegraphics[angle=0,
      width=0.5\textwidth]{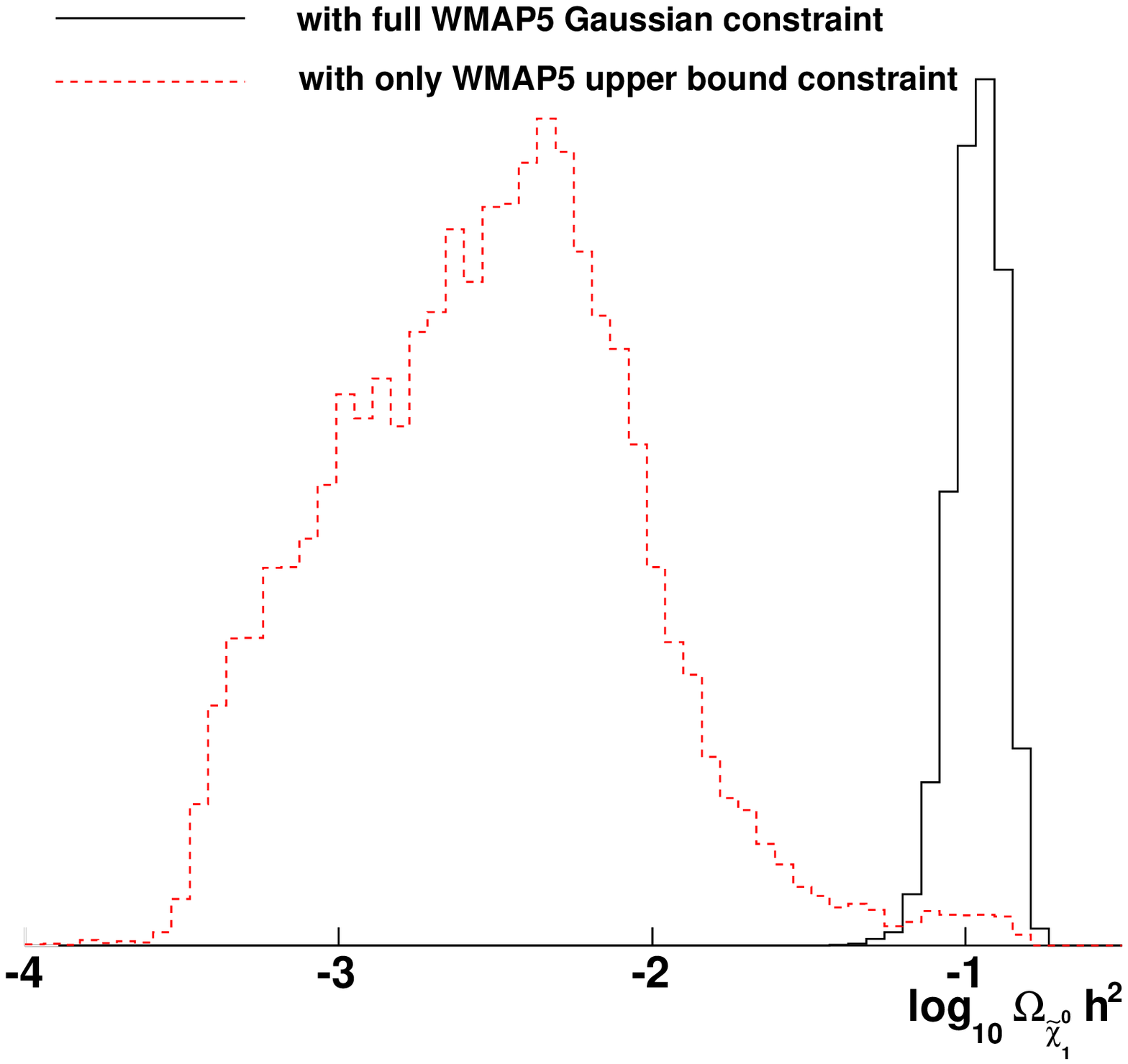} 
\end{center}
 \caption{{\it Neutralino relic density assuming WMAP5 as a Gaussian
likelihood constraint or as an upper bound. For this plot only, a 2 TeV
range linear pMSSM with other parameters as in
Ref.~\protect\cite{AbdusSalam:2008uv} was taken. \label{fig.non}}} 
\end{figure}

\paragraph{}
We have made an independent run using Eq.~\ref{halfull}, i.e.\ relaxing the
purely LSP DM assumption, i.e.\ implicitly assuming some other component of
dark matter. We wish to examine the amount of DM that
comes from the LSP. These runs were performed in
Ref.~\cite{AbdusSalam:2008uv}, where the relevant constraints can be
found. Linear priors were used on the parameters, which had a 2 TeV
upper bound. We find in Fig.~\ref{fig.non} that the preferred relic
density is low compared with the purely LSP DM assumption: around $\Omega_{CDM}
h^2=10^{-2}-10^{-3}$. Thus
once one allows for an additional component of DM to the LSP, the model
prefers the additional component to dominate the relic density
\footnote{Some of us hope to return to this issue in future work. We
  thank Bryan Webber for suggesting this comparison.}. 

%SSSSSSSSSSSSSSSSSSSSSSSSSSSSSSSS
\section{Conclusions and Outlook} \label{sub.conc}
%SSSSSSSSSSSSSSSSSSSSSSSSSSSSSSSS

\paragraph{}
We have presented the first statistically convergent global fit of the
pMSSM model with its 25 independent continuous parameters plus a
discrete parameter, $sign(\mu) = \pm$, to the dark matter relic
density, indirect observables and direct sparticle search
constraints. We have used the entire set of relevant electroweak
precision observables and $B$-physics data, as well as the anomalous
magnetic moment of the muon as indirect observables. The evidence for
the linear and log prior measure pMSSM in light of the data is $\log_e
Z = 63.211 \pm 0.033$ and $\log_e Z = 65.043 \pm 0.042$ respectively. 
We have presented a good-fit point in
the pMSSM parameter-space which is not unique since the $\chi^2$
minimum is degenerate. All the different parameter-space points with
maximal likelihood are equally interesting for phenomenological
studies.  

\paragraph{}
This work constitutes a technical demonstration that statistically
convergent global fits in high dimensions involving curving
degeneracies and several modes are now feasible. This feasibility is
due to new sampling algorithms and improvements in the speed of
computation and access to it. It allows more complete phenomenological
studies of multi-parameter models beyond the standard model that could
not have been completed in the past. In particular, the set-up and
techniques employed here provide an unbiased approach to MSSM
phenomenology -- independent of the underlying theory, the mechanism
to break SUSY or its mediation -- hence it could lead to more
robust SUSY phenomenological studies and guides for LHC SUSY searches
and for dark matter search experiments. As expected the results of
this exercise differs significantly from those of the more studied
CMSSM/mSUGRA and thus the pMSSM parameter space provides a much richer
and appropriate arena for LHC studies of the MSSM.

\paragraph{}
For the analysis we consider prior measures flat in either the
parameters (a linear prior) or flat in the logarithm of the parameters
(a log prior) in order to check robustness of inferences. Given the
large number of (the pMSSM) parameters compared with the (weak
constraining power of the) data available at the moment it is very
interesting that indeed there are some prior-independent results or
inferences. The lightest CP-even Higgs boson mass and the stop masses
fall into this category of robust and prior independent results. The
other sparticle masses and all dark matter properties exhibit
significant prior dependence and require more direct and precise data
(or more constrained models) to make their prediction robust. We
emphasise that prior dependence is present and it is a positive
feature of Bayesian methods, since its absence signals when there is
enough data to make the fits robust. This can be used as a guideline
for future studies of the experimental implications of the MSSM. In
particular, if SUSY is discovered via sparticle production at collider
and many sparticle properties are precisely measured, it will be
possible to use the techniques show-cased in the present work to
extract pMSSM information. Such information can then be checked for
consistency with more constrained models. 

\paragraph{}
We now contrast our methods with the recent random points scan
analysis of the pMSSM in Ref.~\cite{Berger:2008cq} and a similar
earlier work in Ref.~\cite{Profumo:2004at}. These works perform a
pMSSM random parameter scan to find points which  pass direct search
and dark matter constraints while being within 2 $\sigma$ of the
central values of some indirect constraints. All such points are
considered on an equal footing, and as such (as emphasised by the
authors) are not a statistical global fit. We, on the other hand,
allow a trade-off between different observables in a statistically
correct fashion; one may tolerate a moderately bad fit in one
observable if it fits the other observables particularly well. Our use
of Gaussian distributions for the likelihoods (instead of the
2$\sigma$ top-hat function in ~\cite{Profumo:2004at,Berger:2008cq})
is justified by the central limit theorem and the maximum entropy
principle~\cite{sivia}. Interesting points with LHC phenomenology not
covered by previous studies of constrained models were found in 
Ref.~\cite{Berger:2008cq}, which was the main aim of the approach (a few
thousand points passed the constraints, out of $\sim 10^7$ scanned). 
We aim to perform a complete and statistically convergent global
fit of the pMSSM. To achieve this we take advantage of the power of the
{\sc MultiNest} algorithm, which provides samples in moderately 
high dimensional parameter spaces (with curving degeneracies and
different modes) much more efficiently than in random/grid parameter
scans. In \cite{Berger:2008cq} there 
was more emphasis on direct search limits, which are more sophisticated than
the ones employed in the present paper. Since the sparticle masses implied by
our fits are large, our results are insensitive to the exact form of
the direct search limits. 
The density of points in Ref.~\cite{Berger:2008cq} also shows prior
dependence although the results were not interpreted statistically
(and thus they were not Bayesian nor frequentist). Another major difference
to our approach is that in Ref.~\cite{Berger:2008cq}, the WMAP
constraint is used only as an upper bound (making viable points much
easier to find) and so the existence of another non-MSSM
dark matter particle is assumed. This changes the character of the
points: allowing MSSM points which predict approximately zero LSP
relic density means that a large number of sampled points have a wino
dominated LSP. We, however, assume in most of our analysis that the
neutralino constitutes all of the dark matter. In Section
(\ref{sub.lspDMfrac}) we presented an independent run made using the
WMAP constraint only as a lower bound and our results (as expected)
agree with those of ~\cite{Profumo:2004at,Berger:2008cq} in the sense
that, in that case, the LSP contribution tends to be only  a small
fraction of the total dark matter.  

\paragraph{}
There are many directions in which this research could be
extended. For each of the tens of thousands of preferred points in our
sample, a detailed calculation  of LHC observables such as inclusive counts
of opposite sign dilepton and trilepton events, could be made using
standard event 
generators and detector simulators, as has been done for more constrained
models such as the CMSSM~\cite{Feldman:2008hs}. This would provide
a portrait of the signature space that may eventually be
useful in direct SUSY searches.  
On a simpler level, one could compute relative
probabilities of various sparticle mass hierarchies. 
 The indirect dark matter detection
prospects could also be evaluated, although with current data, they are likely
to be 
highly prior dependent. The impact of the inclusion of the fine-tuning into
the prior could be analysed. 
Assuming a particular parameter point, the impact of LHC SUSY measurements on
our fits 
could be evaluated and an estimate of how much luminosity would be required in
order to make inferences approximately prior independent\footnote{A Bayesian
  approach has recently been used to ameliorate the LHC inverse problem in the
MSSM by combining LHC data with indirect
observables~\cite{Balazs:2009it}.}. In this case,  
model comparison
between more constrained models and the pMSSM could be informative. 
One could determine, for a given LHC luminosity, which models could be made
prior independent. 
The extension of the analysis to the full 124 MSSM parameter space may be
still out of reach at the moment. Algorithms improving the {\sc MultiNest}
algorithm may be required before attempting it.
Adding a small number of 
minimal flavor violating parameters could however be feasible. 
An extension of this work to include 
reasonable generalisations of the minimal flavour violation scenario
adding a few extra parameters should be possible. Also, including
R-parity violation to the pMSSM as well as a phenomenological NMSSM
are within reach of the techniques we used here. 
Some of us hope to return to
these issues in future work. 

\appendix
%%%%%%%%%%%%%%%%%%%%%%%%%%%%%%%%%%%%%%%%%%%%%%%%%%%%%%%%%%%
\section{Nested sampling and the {\sc MultiNest} algorithm}\label{sub.nested}
\begin{figure}[tb]
    \includegraphics[angle=0, width=.9\textwidth]{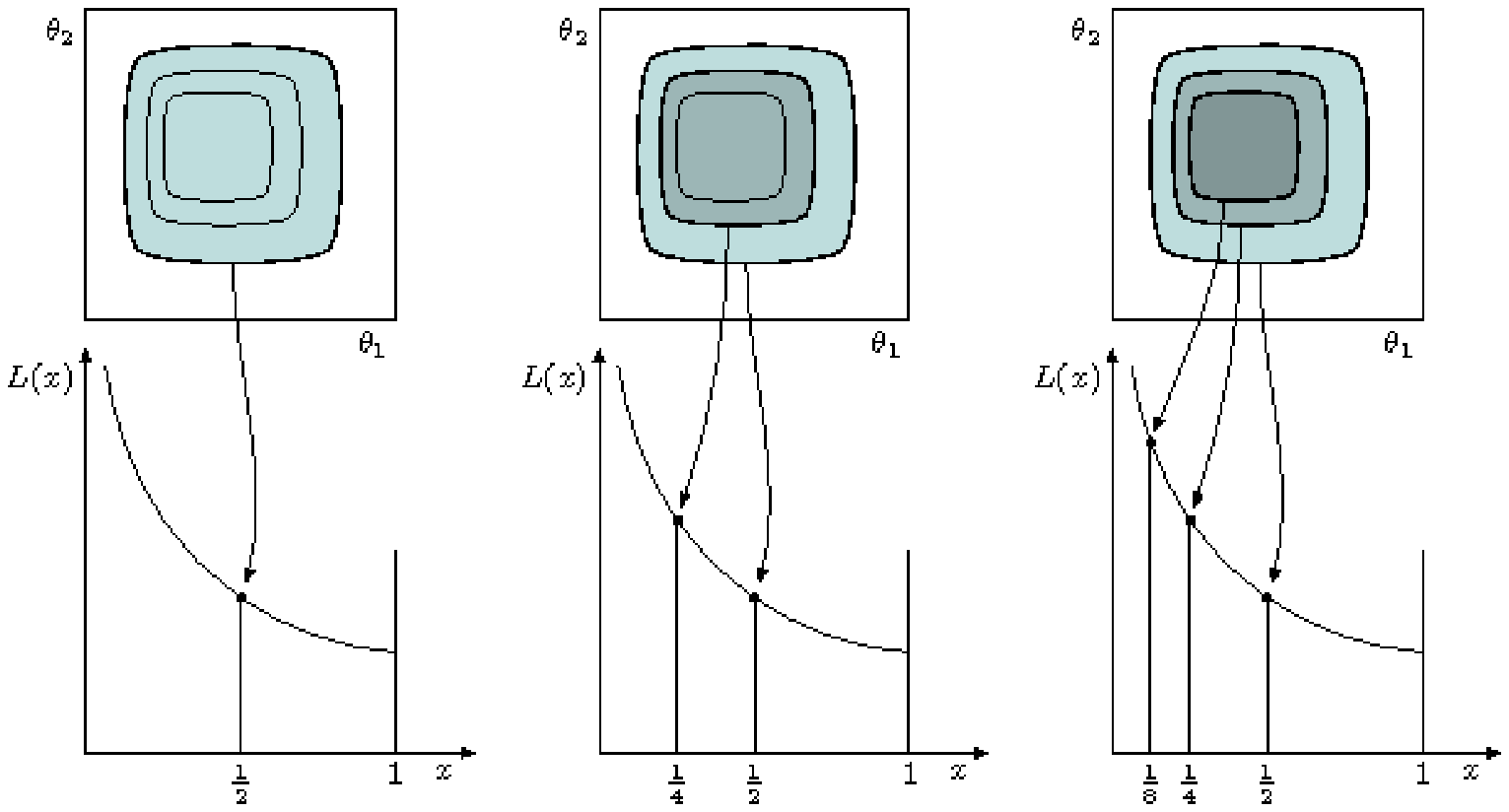}
  \caption{\it Likelihood contours over
  a toy two-parameter space showing the enclosed volume mapped to
  corresponding prior mass. A prior mass $x=\half$ is mapped to the
  likelihood contour that encloses $\half$ of the prior volume. Note
  the nested nature of the contour lines. Figures are
  from~\cite{Mackay}.} \label{fig.Lcont}   
\end{figure}

%%%%%%%%%%%%%%%%%%%%%%%%%%%%%%%%%%%%%%%%%%%%%%%%%%%%%%%%%%%
\paragraph{}
For scanning parameter spaces of large dimensionality
we have to use efficient modern approaches for sampling the posterior. In such
problems, interesting parameter regions are 
often tiny in some directions and many directions are orthogonal to ones along
which the likelihood is degenerately high. In this Section we present the
procedure, in the context 
of the pMSSM, for the Monte Carlo technique called nested
sampling developed by John Skilling~\cite{Skilling} and implemented in
{\sc MultiNest}. It is a general 
method for evaluating the integral Eq.~\ref{eq:3} from which
representative samples from the posterior distribution
Eq.~\ref{posterior} are obtained as by-product. The method differs from
the traditional approach to inference dating back to Metropolis {\it et. al.}
(1953) where the emphasis is more on evaluating the posterior density than in
calculating the evidence. Skilling's method goes as follows. Exploring the
25-dimensional co-ordinate $\mathbf{\Theta}$ to evaluate the evidence integral
is 
impractical. Instead, the prior mass $dX = \pi(\mathbf{\Theta})\,
d\mathbf{\Theta}$ can be
used directly to convert the 25-dimensional into a 1-dimensional
integral over a unit interval. Let $X(L)$ be the prior mass enclosed
within the likelihood contour, $L(\mathbf{\Theta}) = L$ in the
parameter space. That is,
\be \label{pmass}
X(L) = \int_{L(\mathbf{\Theta}) > L} \pi(\mathbf{\Theta}) \,
d^{25}\mathbf{\Theta}.\ee 
As $L$ increases from zero to infinity the enclosed prior mass decreases
from $X(0) = 1$ to 
$X(\infty) = 0$. The inverse function $L(X) \equiv L$ is the
contour value (a likelihood value) such that the volume enclosed is
$X$ (see Fig.~\ref{fig.Lcont} for an illustration).
Eq.~\ref{pmass} and the definition of its inverse implies that the
evidence Eq.~\ref{eq:3} can be expressed as 
\be 
\label{1d-evidence} Z = \int L(\mathbf{\Theta}) \pi(\mathbf{\Theta})\,
d^{25}\mathbf{\Theta} = \int_0^1 L(X) \, dX. \ee   
Given the likelihood values $L_i = L(X_i)$ at a sequence of $m$ points
$ 0 < X_{m} < \ldots < X_{2} < X_{1} < X_0=1$ the evidence
is estimated as a weighted sum,
\be \label{zsum} Z = \sum_{i=1}^m L_i w_i, \ee 
where for the trapezoidal rule $w_i = \half (X_{i-1} - X_{i+1})$. 

%%%%%%%%%%%%%%%%%%%%%%%%%%%%%%%%
\subsection{Evidence evaluation} \label{sub.eveva}
%%%%%%%%%%%%%%%%%%%%%%%%%%%%%%%%
 
\begin{figure}[!t]
    \includegraphics[angle=0, width=.9\textwidth]{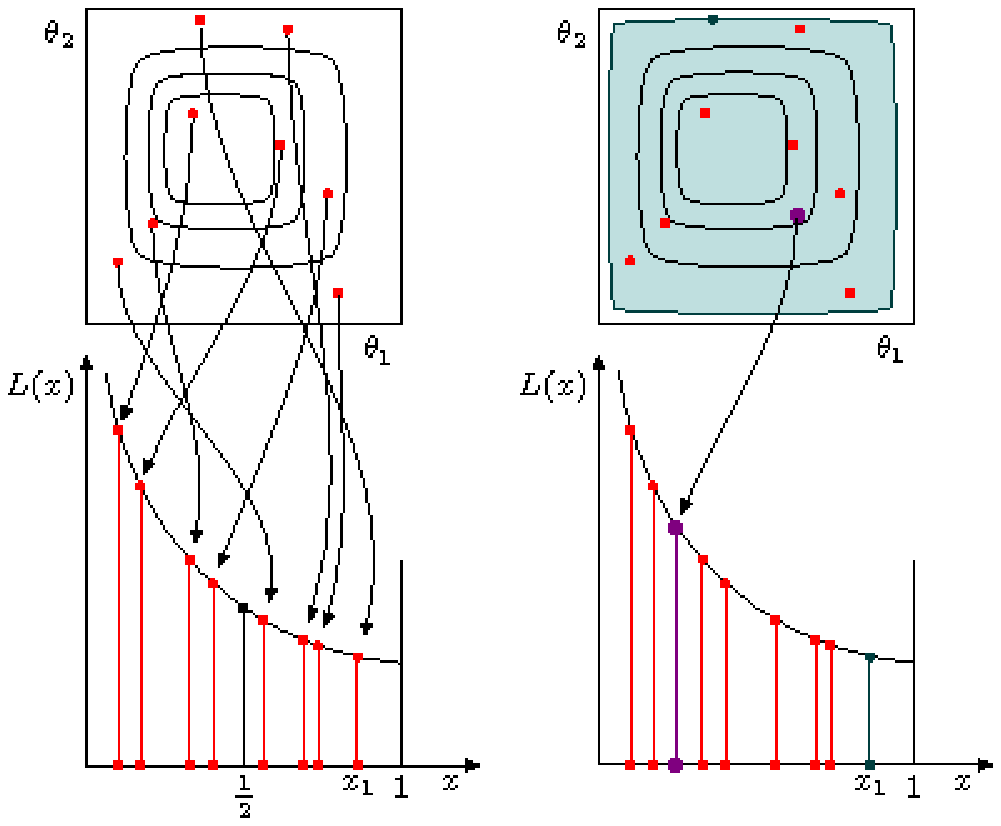}
  \caption{\it Figure on the left shows
  $N=8$ live points uniformly distributed in parameter space (or prior
  volume space $(0, 1)$) and sorted according to corresponding 
  likelihood values. On the right is a picture illustrating the
  sampling of a new point (big purple) dot from the live points
  uniformly distributed in $(0, x_1)$. Figures are from
  \cite{Mackay}. \label{fig.npoint} }
\end{figure}

\paragraph{}
The nested sampling procedure for evaluating the evidence starts with
the accumulation of $N$ points uniformly drawn from the prior, the
initialisation of the evidence, $Z = 0$, and the initialisation of the
prior volume, $X_0 = 1$. The number, $N$,  of ``live'' points,
$\mathbf{\Theta}_1, \ldots, \mathbf{\Theta}_N$ is preserved 
throughout the procedure and every point is associated with its
corresponding likelihood value: $L(\mathbf{\Theta}_1), \ldots,
L(\mathbf{\Theta}_N)$. Each step $i = 1, 2, 3, \ldots$ over the iterations is
associated with the lowest likelihood $L_i$ (or the largest prior
mass, $X_i$) that defines the contour line (or shell)
$L(\mathbf{\Theta}) = L_i$ over parameter space. For moving from the 
$(i-1)$th to the $i$th iteration a new point is drawn
from the set of points uniformly distributed in $(0, X_{i-1})$, the
parameter space region with likelihoods $L \geq L(X_i) = L_i$. This is
illustrated in Fig.~\ref{fig.npoint}. The new point replaces the one
with lowest likelihood. $X_i$ is set to $X_i = exp(-i/N)$, the weight
$w_i$ to $\half (X_{i-1} - X_{i+1})$ and the evidence $Z$ incremented
by $L_i w_i$. This procedure is repeated for the subsequent
iterations. 

\paragraph{}
The prior volume shrinkage ratios $t_i = X_i/X_{i-1}$ are distributed
according to $Pr(t_i) = N t_i^{N-1}$ in $(0,1)$ where $t_i$ is the
largest of $N$ random numbers uniformly distributed in $(0,
1)$. Sampling over $t$ represents a geometrical exploration of the parameter
space. The mean and standard 
deviation of $t$  are 
\be \label{xsigma} E(\log t)=-\frac{1}{N} \textrm{ and } \sigma[\log
  t]=\frac{1}{N}\ee
respectively. This justifies the assignment $X_i = exp(-i/N)$ since
each draw of $\log t$ is independent and after $i$ iterations of the
sampling procedure the prior volume will shrink down according to \be
\log X_i \approx -(i \pm \sqrt{i})/N. \ee 

%%%%%%%%%%%%%%%%%%%%%%%%%%%%%%%
\subsection{Stopping criterion}
%%%%%%%%%%%%%%%%%%%%%%%%%%%%%%%
\paragraph{}
The nested sampling procedure is terminated after a preset number of
the iterations (as described in Subsection~\ref{sub.eveva}) or when the
largest likelihood taken over the whole currently (at the instance of
check for termination, say the $j$th iteration) available prior mass
would not increase the evidence value by more than some preset
fraction $f$ (we use 0.5 in log-evidence). That is, the procedure is
terminated if $$max(L(\mathbf{\Theta}_1), \ldots,
L(\mathbf{\Theta}_N)) X_j < f Z_j.$$  

\paragraph{}
The integration, $Z$, is dominated around the region $X \approx
e^{-H}$, wherever the bulk of the posterior mass is to be found. Here 
\be H = \textrm{information} = - \int \log(dX/dP) \, dP \approx \sum_i
\log \left(\frac{L_i}{Z} \right) \frac{L_i w_i}{Z}\ee and  
$dX/dP$ is the compression ratio representing the fraction of the
prior mass that contains the bulk of the posterior. $dP =
p(\mathbf{\Theta})\, d\mathbf{\Theta} =
Z^{-1}L(\mathbf{\Theta})\pi(\mathbf{\Theta}), d\mathbf{\Theta}$. 
Recalling that $X_i \approx e^{-i/N}$ we expect the integration
procedure to take $NH \pm \sqrt{NH}$ steps (iterations) before
reaching covering the bulk of the posterior. Hence another termination
condition could be to continue 
iterating until the count $i$ is significantly greater than $NH$. 

\paragraph{}
The uncertainty in $X$ translates to a geometrical uncertainty factor
$exp(\pm\sqrt{H/N})$ in the weights $w_i$ of the dominating
iterates. This in turn gives the uncertainty in the evidence via
Eq.~\ref{zsum} as $dev(\log Z) \approx \sqrt{H/N}$ so that
\be \label{zevidence} \log Z = \log \left(\sum_{i=1}^m L_i w_i \right)
\pm \sqrt{\frac{H}{N}}. \ee 

%%%%%%%%%%%%%%%%%%%%%%%%%%%%%%%%%
\subsection{Posterior inferences}
%%%%%%%%%%%%%%%%%%%%%%%%%%%%%%%%%

\paragraph{}
The posterior distribution $p(\mathbf{\Theta})$ is simply the prior
distribution weighed by 
the likelihood. This can be trivially extracted from the evidence
calculation since the set of sampled points $\mathbf{\Theta}_1, \ldots,
\mathbf{\Theta}_N$ is already a posterior representative provided it is
assigned the appropriate importance weight and normalised by the
evidence, $Z$, to
produce probability density with unit total. That is at the $i$th
iteration the posterior probability density is
\be \label{postp} p_i= \frac{L_i w_i}{Z}. \ee 
These are generated from the sequences of discarded points (the points
with the lowest likelihood value at each iteration) during the
sampling procedure. From these posterior sequence properties such as
the mean $\mu$ and standard deviation $\sigma$ of some
$Q(\mathbf{\Theta})$ are easily computable: 
\be \mu = \sum_i p_i Q(\mathbf{\Theta}_i) \textrm{ and } \mu^2 + \sigma^2 =
\sum_i p_i Q(\mathbf{\Theta}_i)^2. \ee  
Equally-weighed samples selected proportionally to $p_i$ can be used to construct marginalised posterior
distributions in $\mathbf{\Theta}$.  

\paragraph{}
For completeness, it is worth mentioning that there are alternative
methods for evaluating the evidence with other advanced MCMC
algorithms like thermodynamic integration and it is not  
clear yet which method is best for high dimensional problems:
dimensions greater than  \footnote{We thank David Mackay for
  interesting discussions about 
    this. See for example \\{\texttt
    http://www.inference.phy.cam.ac.uk/mackay/presentations/nested06/}} 10.
However, for this paper we implement the nested sampling algorithm for
our purpose using the {\sc MultiNest} code~\cite{Feroz:2007kg} which has
additional quality of being efficient in sampling multi-modal
posteriors exhibiting curving degeneracies (see a summarised account
in Subsection~\ref{multinest}). 

%%%%%%%%%%%%%%%%%%%%%%%%%%%%
\subsection{{\sc MultiNest}} \label{multinest}
%%%%%%%%%%%%%%%%%%%%%%%%%%%%

\begin{figure}
\twographshuge{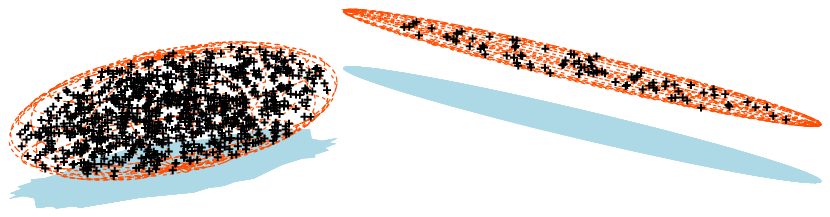}{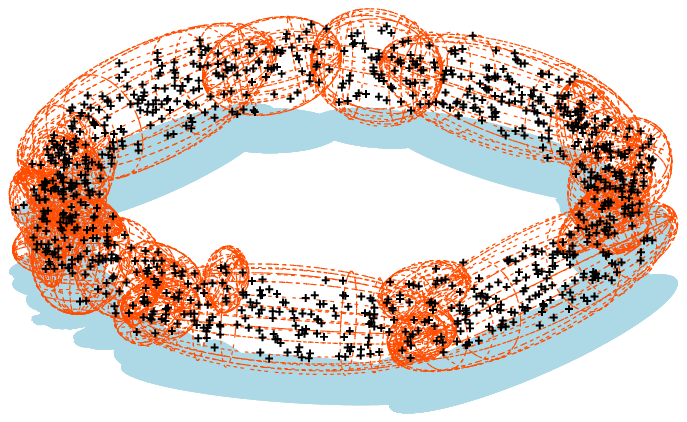}
\caption{Illustrations of the ellipsoidal decompositions performed by
  {\sc MultiNest}. The points given as input
  are overlaid on the resulting ellipsoids. 1000 points were sampled
  uniformly from: (a) two non-intersecting
  ellipsoids; and (b) a torus.}
\label{fig:dino}
\end{figure}

\begin{figure}[!ht] 
  \begin{tabular}{c}
    \includegraphics[angle=0, width=.5\textwidth]{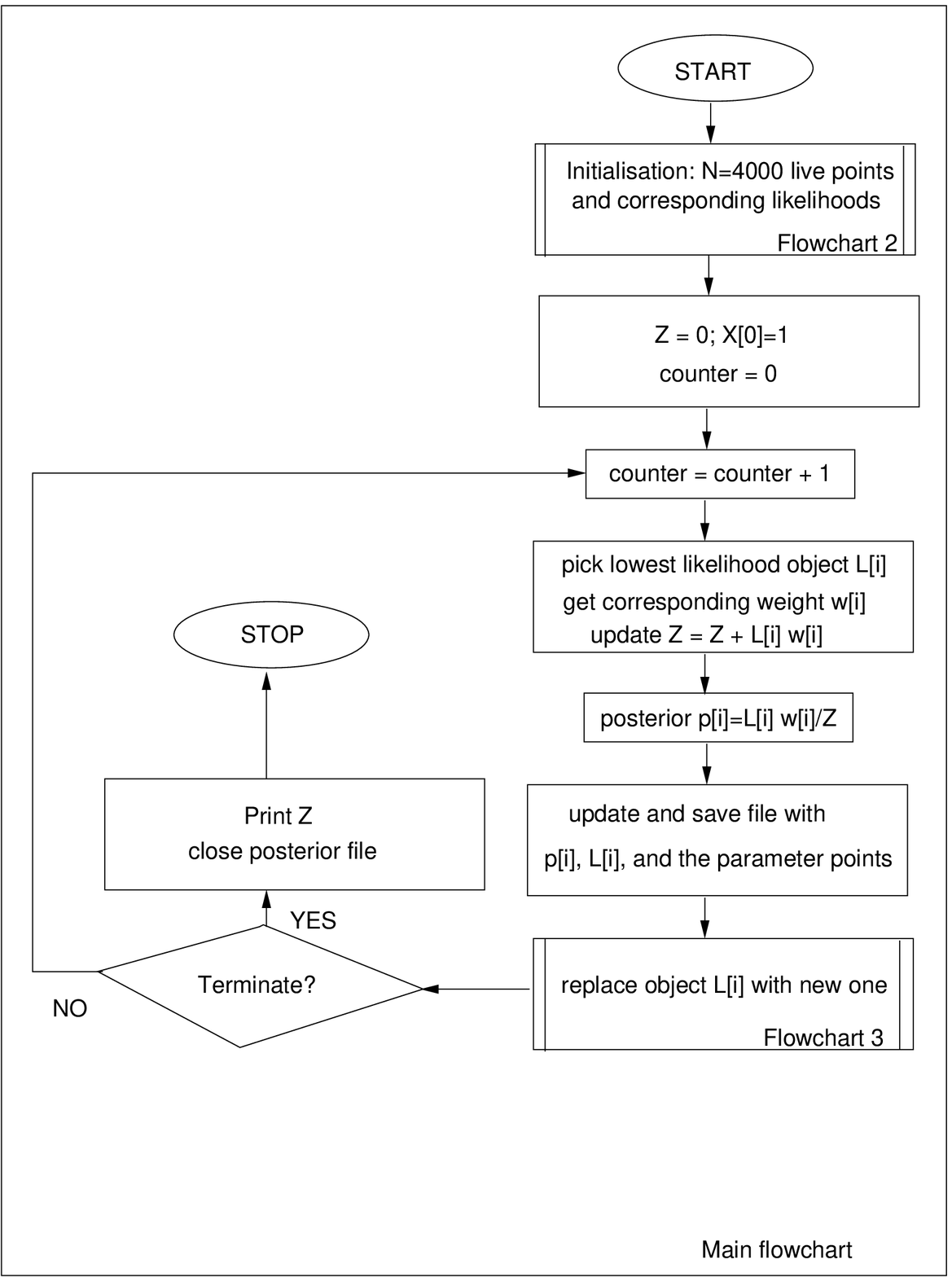} 
    \includegraphics[angle=0, width=.5\textwidth]{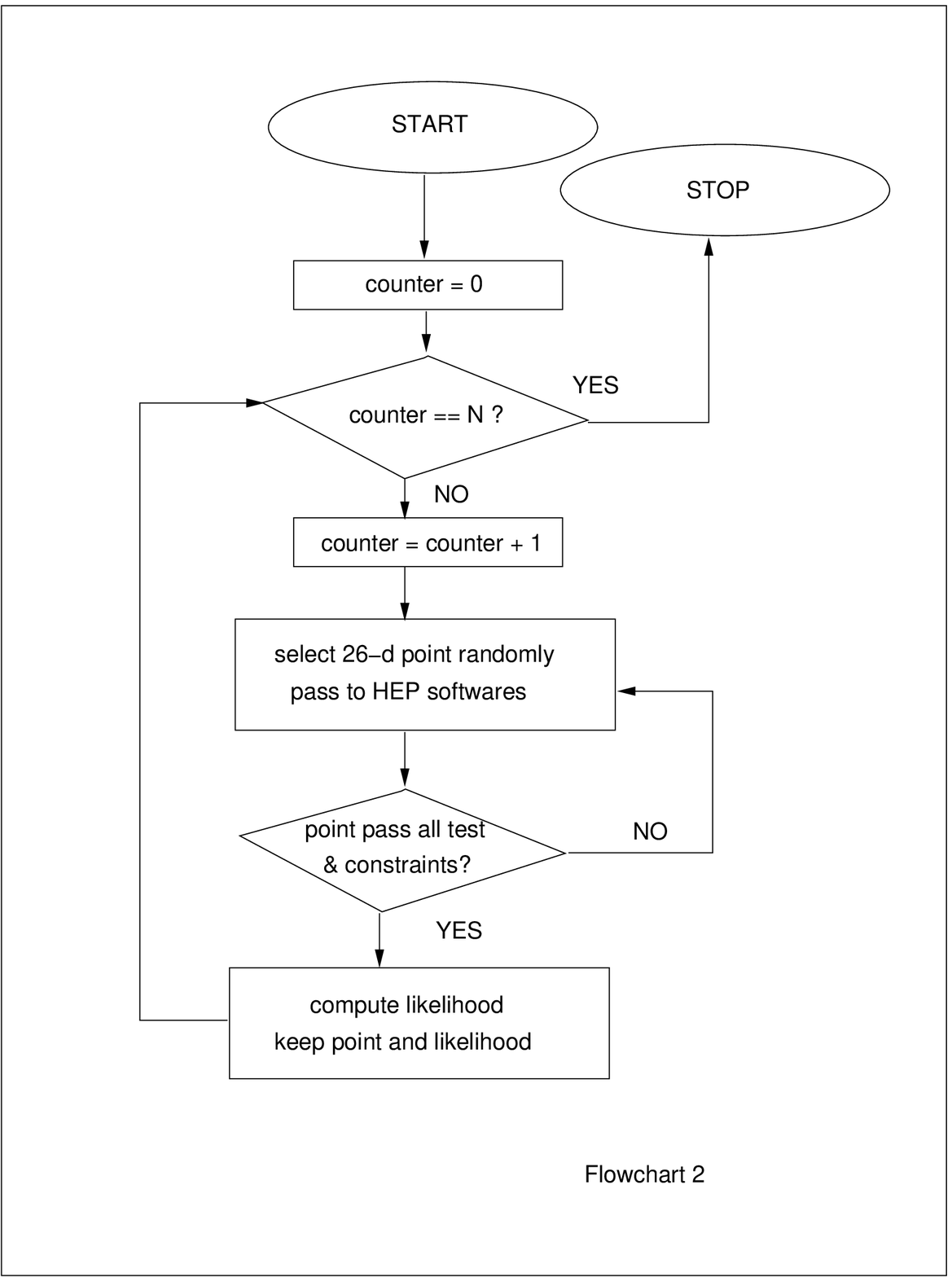}\\
    \includegraphics[angle=0, width=.5\textwidth]{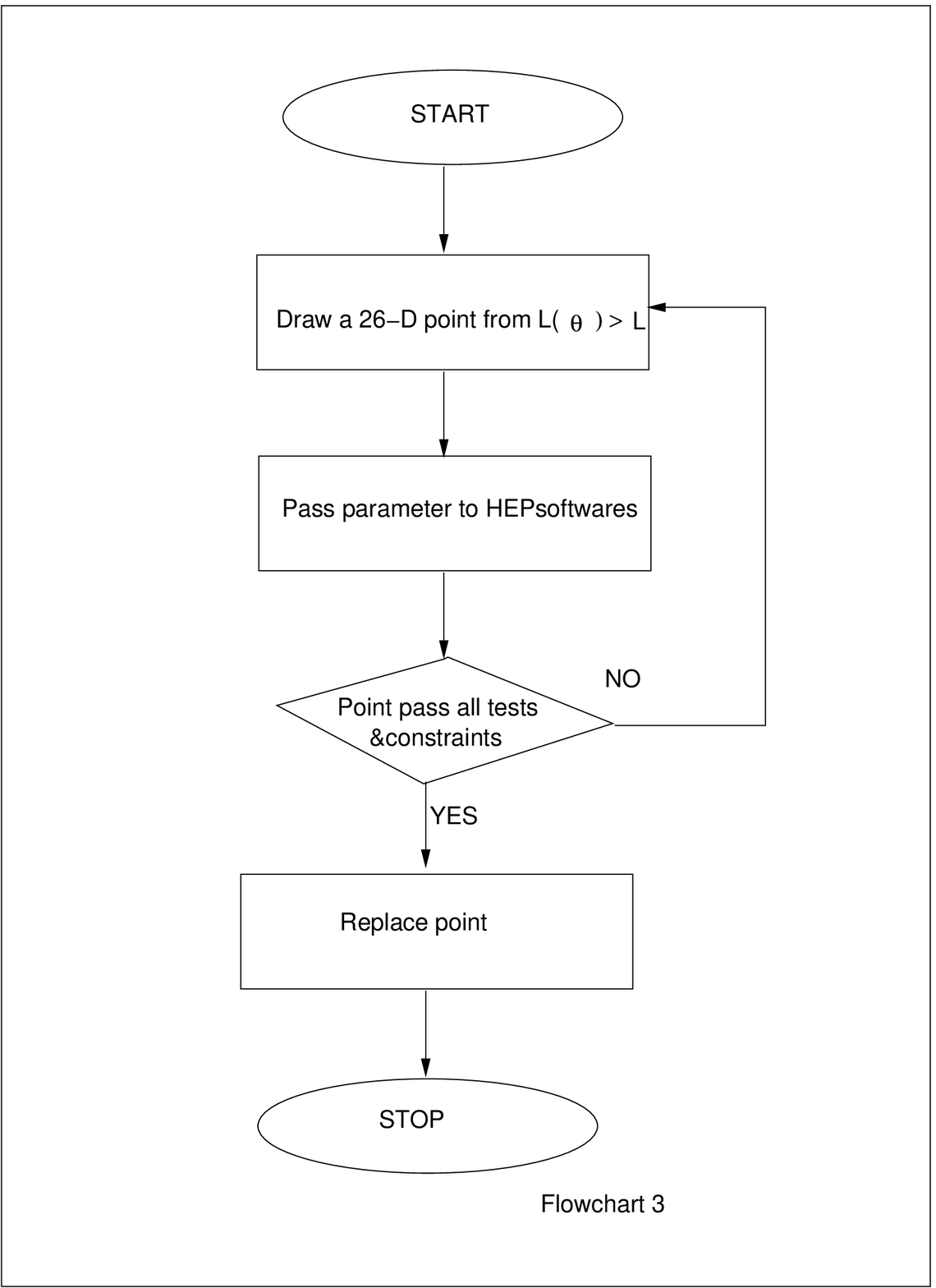}
  \end{tabular}
  \caption{\it{Flow charts summarising the sampling procedure: Z
  refers to the Bayesian evidence, 
  Eqs.~(\protect\ref{eq:3},~(\protect\ref{1d-evidence}),~(\protect\ref{zsum});
  $X_i$ is the prior mass, Eq.~\protect\ref{pmass}; and $p_i$ is
  the   posterior probability, Eq.~\protect\ref{postp}. HEP software
  refers to the different computer programs described in
  Section~(\protect\ref{sub.phenoprocedure}).}}
 \label{fig.charts}
\end{figure}

\paragraph{}
The most challenging task in implementing the nested sampling
algorithm is drawing samples from the prior within
the hard constraint ${L}> {L}_i$ at each iteration $i$. Employing a
naive approach that draws
blindly from the prior would result in a steady decrease in the
acceptance rate of new samples with decreasing
prior volume (and increasing likelihood). {\sc MultiNest} algorithm
\cite{Feroz:2007kg,Feroz:2008xx} tackles this
problem through an ellipsoidal rejection sampling scheme by enclosing
the live point set into a set of (possibly
overlapping)  ellipsoids and a new point is then drawn uniformly from
the region enclosed by these ellipsoids.
The number of points in an individual ellipsoid and the total number
of ellipsoids is  decided by a an
`expectation--maximisation' algorithm so that the total sampling
volume, which is equal to the sum of volumes of
the ellipsoids, is minimised. This allows maximum flexibility and
efficiency by breaking up a mode resembling a
Gaussian into  relatively fewer number of ellipsoids, and if the
posterior mode possesses a pronounced curving 
degeneracy so that it more closely resembles a (multi--dimensional)
`banana' then it is broken into  a relatively
large number of small `overlapping' ellipsoids (see
Fig.~\ref{fig:dino}). With enough live points, this approach
allows the detection of all the modes simultaneously resulting in
typically two orders of magnitude improvement
in efficiency and accuracy over standard methods for inference
problems in cosmology and particle physics
phenomenology (see
e.g.\cite{Feroz:2008rm,Feroz:2008dt,0807.4512,AbdusSalam:2008uv,Trotta:2008bp}).
The {\sc MultiNest} procedure as applied to our pMSSM fits
is summarised by the flow charts in Fig.~\ref{fig.charts}.

\FloatBarrier

\acknowledgments{}
We would like to thank Arne Weber for help with fixes to the code
\texttt{SUSYPOPE}, A. Pukhov for help with \texttt{micrOMEGAs},
N. Mahmoudi for her help with \texttt{SuperIso}, and M. Dolan for
checks with the likelihood code. We 
thank S. Abel, J. Conlon, D.~J.~C. MacKay, J.~R. Espinosa,
S. Krippendorf, C. Lester, M. Gomez-Reino, F. Marchesano,  P. Slavich, K. Suruliz,
A. Uranga, B. Webber and E. Witten for useful discussions. The
calculations performed in this paper were done using the 
Cambridge High Performance Computing Cluster (HPC) {\it Darwin}\/ and
COSMOS, the UK National Cosmology Supercomputer. We thank Andrey
Kaliazin, Stuart Rankin and Victor Travieso for assistance in setting
and running the codes on these computing facilities and John Turner and
the HPC group for important assistance regarding the rights to
use HPC\@. This research was partially funded by STFC\@. SSA is supported
by The Gates Cambridge Trust and thank the African Institute for
Mathematical Sciences (AIMS) for hospitality during the early stages of
this work. FQ thanks the organisers of the Cooks Branch 2009 meeting
for hospitality during the last stages of this work. 

\bibliographystyle{JHEP}
\bibliography{references}

\end{document}